\documentclass[aoas]{imsart}

\RequirePackage{amsthm,amsmath,amsfonts,amssymb}
\RequirePackage[authoryear]{natbib}
\RequirePackage[colorlinks,citecolor=blue,urlcolor=blue]{hyperref}
\RequirePackage{graphicx}
\RequirePackage{multirow}
\RequirePackage{afterpage}

\startlocaldefs

\endlocaldefs

\begin{document}

\begin{frontmatter}
\title{Joint identification of spatially variable genes via a network-assisted Bayesian regularization approach}
\runtitle{SV-network}

\begin{aug}
\author[A]{\fnms{Mingcong}~\snm{Wu} \thanksref{t2}\ead[label=e1]{wumingcong@ruc.edu.cn}},
\author[A]{\fnms{Yang}~\snm{Li} \thanksref{t2}\ead[label=e2]{yang.li@ruc.edu.cn}},
\author[B]{\fnms{Shuangge} \snm{Ma}\ead[label=e3]{shuangge.ma@yale.edu}}
\and
\author[C]{\fnms{Mengyun}~\snm{Wu} \thanksref{t1}\ead[label=e4]{wu.mengyun@mail.shufe.edu.cn}},
\address[A]{Center for Applied Statistics and School of Statistics,
Renmin University of China, Beijing, China\printead[presep={,\ }]{e1,e2}}
\address[B]{Department of Biostatistics,
Yale School of Public Health, New Haven, USA\printead[presep={,\ }]{e3}}
\address[C]{School of Statistics and Data Science,
Shanghai University of Finance and Economics, Shanghai, China\printead[presep={,\ }]{e4}}

\thankstext{t1}{corresponding author: wu.mengyun@mail.shufe.edu.cn}
 \thankstext{t2}{the co-first authors}
\end{aug}

\begin{abstract}
Identifying genes that display spatial patterns is critical to investigating expression interactions within a spatial context and further dissecting biological understanding of complex mechanistic functionality. Despite the increase in statistical methods designed to identify spatially variable genes, they are mostly based on marginal analysis and share the limitation that the dependence (network) structures among genes are not well accommodated, where a biological process usually involves changes in multiple genes that interact in a complex network. In addition, the latent cellular composition within the spots can introduce confounding variations, negatively affecting the accuracy of the identification. In this study, we develop a novel Bayesian regularization approach for spatial transcriptomic data, with confounding variations induced by varying cellular distributions effectively corrected. Significantly advancing from existing studies, a thresholded graph Laplacian regularization is proposed to simultaneously identify spatially variable genes and accommodate the network structure among genes. The proposed method is based on a zero-inflated negative binomial distribution, effectively accommodating the count nature, zero inflation, and overdispersion of spatial transcriptomic data. Extensive simulations and applications to real data demonstrate the competitive performance of the proposed method.
\end{abstract}


\end{frontmatter}


\section{Introduction}

Recently developed and rapidly advancing spatial transcriptomics (ST) technology enables gene expression profiling across numerous spatial locations within a tissue, offering biological insights into various contexts \citep{rao2021exploring}, with particularly prominent applications in cancer research. One crucial task in ST data analysis is to identify genes with varying expressions across space, termed as spatially variable (SV) genes \citep{svensson2018spatialde}. SV genes have been shown to be associated with disease characteristics such as immune cell infiltration and tumor cell proliferation \citep{zuo2024dissecting}, thus facilitating the discovery of tumorigenesis mechanisms and the development of therapeutic strategies.

Many approaches have recently been proposed for the detection of SV genes. The great majority of them are based on the Gaussian process (GP). For example, SpatialDE \citep{svensson2018spatialde} models normalized expression data using GP regression and tests the significance of the spatial covariance matrix for each gene separately. SPARK \citep{sun2020statistical}, BOOST-GP \citep{li2021bayesian}, and GPcounts \citep{bintayyash2021non} also take advantage of GP regression but directly model raw count data with Poisson, Negative Binomial (NB), and zero-inflated NB (ZINB) distributions, respectively. CTSV \citep{yu2022identification}, on the other hand, implements a slightly different technique based on ZINB regression, where the mean expression level is parameterized as a linear combination of functions of spatial coordinates. There are some non-parametric approaches with more computational efficiency, such as SPARK-X \citep{zhu2021spark} based on covariance-based testing, MERINGUE \citep{miller2021characterizing} based on Voronoi tessellation and classical Moran's I score, and HEARTSVG \citep{yuan2024heartsvg} based on constant variance testing. Despite considerable successes, the results of the aforementioned works are still sometimes unsatisfactory due to the high dimensionality of genes, high levels of noise and sparsity, and low resolution of spots. In addition to the transcriptomic data, other biological information, such as cellular phenotypes and genetic interactions, is often available and potentially provides a valuable complement to the present analysis. The integration of such assisted information into transcriptomic analysis is a promising direction to mitigate the aforementioned challenges in ST data and further advance existing ST research.

Specifically, first, most existing approaches for SV gene detection rely on marginal analysis, which models each gene separately and has the limitation that the dependence among genes is not well utilized. Increasing evidence has shown that diseases are mostly a result of a combination of multiple genetic alterations, and genetic factors usually interact with each other and are involved in a biological network \citep{barrio2023network} (Figure 1). Specifically, genes connected within a network are believed to have similar biological functions, leading to potentially similar contributions to cellular organizations and functional mechanisms. In the context of SV gene, many detected SV genes are confirmed to be associated with certain common pathways or networks. For example, the SV genes detected in a human colorectal cancer (CRC) dataset have been found to be enriched in immune-associated GO terms and KEGG pathways \citep{yuan2024heartsvg}. Similarly, SV genes identified in human lung and kidney cancer datasets have been confirmed to share certain common biological functionalities \citep{shang2025statistical}. Recently, an increasing number of biological networks have been amassed, such as protein-protein interaction (PPI) networks, metabolic networks, and regulatory networks. Curated network information has been widely adopted as a powerful supplement to gene expression analysis, particularly in bulk and single-cell sequencing analysis \citep{li2010variable,elyanow2020netnmf, qin2023two}. However, the integration of network information for SV gene detection remains limited. 

Second, despite prosperous developments in recent years,  measurements obtained using the sequencing-based ST technologies, such as Slide-seq and 10X Genomics Visium, are still ``spot''-based, where the gene expression measurement at a single spot is usually a mixture of diverse cells from heterogeneous types rather than at a single-cell resolution \citep{yu2022identification} (Figure 1). This cellular composition diversity among spots has been demonstrated to probably contribute to expression variability \citep{cable2022robust}, potentially confounding the detection of biologically relevant SV genes. In the published studies, there are very few approaches that conduct simultaneous cellular composition accommodation and SV gene detection. Limited existing studies include SPARK-X, which converts the inferred major cell types into binary indicators to mitigate the confounding effect caused by latent cell type distributions. Alternatively, CTSV concentrates on the detection of cell-type-specific SV genes whose expressions are affected by the spatial coordinates of cells of the same type. 

\begin{figure}[htbp]
	\centering
		\includegraphics[width=5.5in]{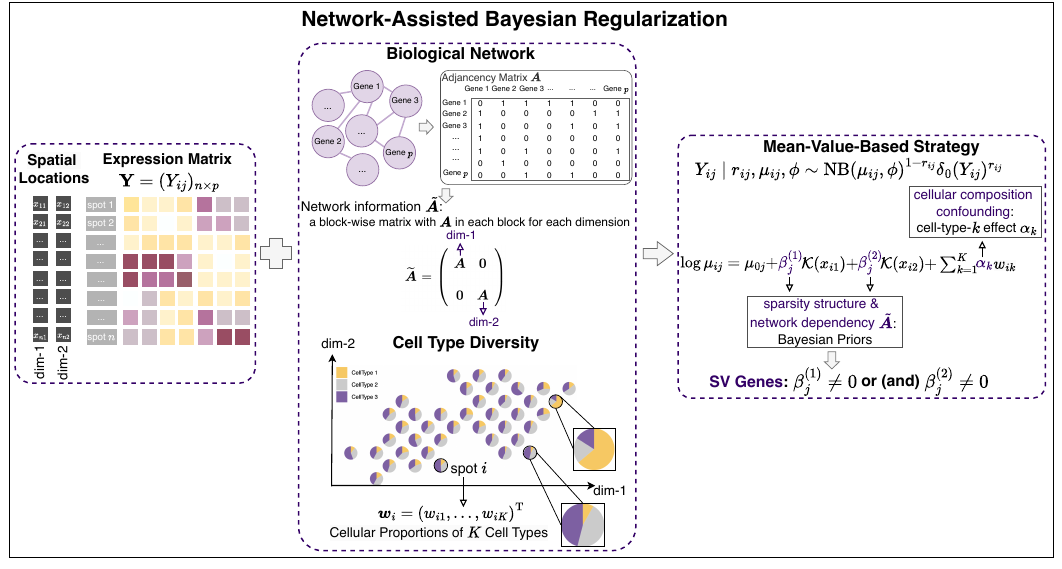}
	\caption{
   Overview of the proposed network-assisted Bayesian regularization framework. Left: ST data consisting of spatial coordinates and corresponding gene expression matrix. Middle: Biological network and cellular composition information for integration. 
    Right: Mean-value-based sparsity-inducing strategy for joint identification of SV genes.
	\label{fig:overview}}
\end{figure}

Motivated by the aforementioned challenges, we propose a Bayesian regularization approach for the joint identification of SV genes as shown in Figure \ref{fig:overview}. Specifically, the ZINB distribution is adopted to account for the count nature, sparsity, and overdispersion of raw expression measurements. The mean-value-based strategy is applied for SV gene detection, which enjoys the advantages of simplicity, interpretability, and computational efficiency. The most substantial advancement is that the proposed approach conducts joint detection with the network dependency structure among genes well incorporated, making it a big step forward from the existing marginal analysis. Moreover, the proposed approach introduces a series of cell-type-specific parameters to effectively correct for the confounding variations induced by varying cell-type compositions across spots. Overall, this study provides a practically useful and biologically meaningful approach for SV gene identification in ST analysis, with improved performance over alternatives as demonstrated in both simulation studies and application to multiple real-world ST datasets.

\section{Methods}
Consider a tissue section consisting of $n$ spots, $p$ genes, and $K$ cell types. Let $\mathbf{Y} \in \mathcal{R}^{n\times p}$ be the ST expression matrix composed of $\mathbf{Y}_{j}$'s, where $\mathbf{Y}_j = \left( Y_{1j}, \dots, Y_{nj}\right)^{\mathrm{T}}$ is the vector of $n$ observed raw counts of the $j$th gene. $\mathbf{Y}$ usually has a very high level of sparsity because of a low capture rate. Each spot $i$ is associated with a 2-dimensional coordinate $\mathbf{x}_i = \left(x_{i1}, x_{i2}\right)$ which represents the location of the corresponding spot center.

\subsection{Network and Cellular Composition Information for Integration}

Consider an undirected network $G(V, E)$ that is constructed using biological information, where $V$ is the node set consisting of $p$ genes and $E = \left\{e(j,l), j,l \in\{1,\right. $ $\left.\cdots,p\}\right\}$ is the set of edges between nodes. For genes connected within the network, it is expected that they have similar biological functionalities, leading to potentially similar spatial variability, and thus are more likely to be \textit{SV} or \textit{non-SV} genes simultaneously. To induce such network-assisted identification, an adjacency matrix $\boldsymbol{A}=(a_{jl})_{p \times p}$ is first constructed based on $G(V, E)$, with $a_{jl}=1$ if there is an edge $e(j, l)$ between the $j$th and $l$th genes and $a_{jl}=0$ otherwise, and ${a}_{jj}=0$, for $j,l=1,2,\ldots,p$.

As for cellular compositions, denote $\boldsymbol{w}_{i} = \left(w_{i1}, \dots, w_{iK}\right)^{\mathrm{T}}$ as the vector of cellular proportions for the $i$th spot, which satisfies the constraint that $ 0 \leq w_{ik} \leq 1, k = 1,\dots, K$ and $ \sum_{k=1}^{K} w_{ik} = 1$. Such information is typically available as ground truth or can be obtained using deconvolution methods such as RCTD \citep{cable2022robust}, Redeconve \citep{zhou2023spatial}, and SONAR \citep{liu2023sonar}. 
Since the distributions of cellular proportions usually display spatial relatedness, which may confound SV gene detection, we correct for this potential confounding in SV gene identification.

\subsection{Network-assisted Bayesian Modeling}
We introduce latent binary variables to accommodate the zero-inflation in $\mathbf{Y}$ and consider the following zero-inflated negative binomial (ZINB) model:
\begin{equation}
  Y_{ij} \mid r_{ij}, \mu_{i j}, \phi \sim \mathrm{NB}\left(Y_{ij}\mid\mu_{ij},\phi\right)^{(1-r_{ij})} \delta_0(Y_{ij})^{r_{ij}},
  \label{eq:2}
\end{equation}
where $r_{ij} = 1$ indicates that $Y_{ij}$ is from a Dirac probability measure $\delta_0(\cdot)$ with a point mass at zero, and otherwise $Y_{ij}$ is from a NB distribution $\mathrm{NB}\left(Y_{ij}\mid \mu_{i j},\phi\right)$ with expectation $\mu_{ij}$ and dispersion $1/\phi$. The NB distribution has variance $\mu_{ij} + \mu_{ij}^2/\phi$, thus allowing modeling extra variation.

To accommodate spatial differential expression and cell-type-specific confounding, the logarithm of $\mu_{ij}$ is modeled as:
\begin{equation}\label{mu}
     \log \mu_{i j}  = \mu_{0 j} + \beta^{(1)}_{j}  \mathcal{K}\left(x_{i 1}\right)+\beta^{(2)}_{j} \mathcal{K}\left(x_{i 2}\right)  +
        \sum_{k=1}^{K}\alpha_k w_{ik},
\end{equation}
where $\mu_{0j}$ is the baseline expression parameter, $\alpha_k$ is the cell-type-specific coefficient to accommodate the potential influence of cellular distributions, and $\beta^{(1)}_{j}$ and $\beta^{(2)}_{j}$ reflect the degree of spatial differential expression, with $\mathcal{K}(\cdot)$ being the pre-specified spatialization function to measure the specified trend of spatial gene expression variation.

\begin{figure}[htbp]
\centering
\includegraphics[width=5.5in]{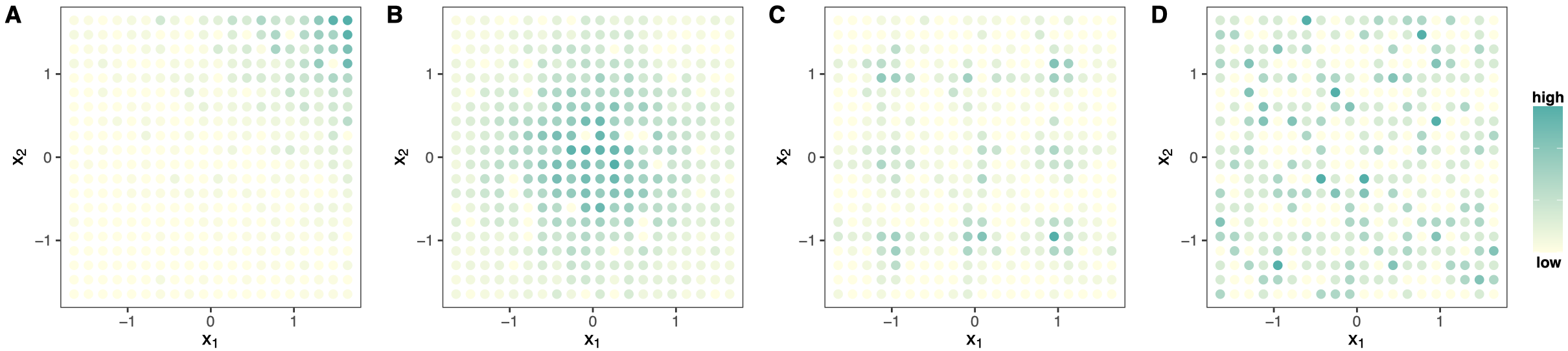}
\caption{Illustrative spatial expression patterns of SV gene with (A) linear, (B) exponential, and (C) periodic pattern with $\beta^{(1)}_{j}\neq 0$ or $\beta^{(2)}_{j}\neq 0$, and (D) non-SV gene with $\beta^{(1)}_{j}=0$ and $\beta^{(2)}_{j}=0$, which are simulated based on the proposed ZINB models (\ref{eq:2}) and (\ref{mu}). 
\label{fig:sv_pattern}
}
\end{figure}

We adopt the ZINB model due to its superiority for simultaneously accommodating the count measure, over-dispersion, and excessive zeros caused by dropouts.  For SV gene identification, significantly different from the previous studies that introduce spatial differential expression via a covariance matrix, 
we adopt the mean-value-based formulation. Specifically, we introduce the spatialization functions $\mathcal{K}(x_{id})$'s and the corresponding coefficients $\beta^{(d)}_{j}$'s in the mean parameter of the NB distribution to describe the spatial variability. When $\beta^{(1)}_{j}\neq 0$ or $\beta^{(2)}_{j}\neq 0$, the $j$th gene has the expression pattern associated with its spatial location $(x_{i1}, x_{i2})$ and is regarded as SV. Such a mean-value based strategy was first introduced by SPARK-X \citep{zhu2021spark} and has demonstrated strong empirical performance in \cite{yu2022identification, cable2022cell} and \cite{yu2024spvc}. Compared to covariance-based approaches, our mean-based formulation is more intuitive and interpretable, enabling insights into axis-specific spatial effects about whether expression varies along one specified spatial direction while remaining constant along the other, as shown in  Supplementary Figure S13 \citep{wu2025supp}. More importantly, the covariance-based strategy always involves constructing and inverting large covariance matrices, which makes the covariance-based strategy usually space- and time- consuming, and thus it lacks efficiency and scalability for large-scale analysis. Here, we follow the published studies \citep{svensson2018spatialde,sun2020statistical,yu2022identification, seal2023smash} and consider three of the most popular and widely adopted kernel functions to account for a variety of potential spatial patterns that have been discovered in common biological datasets. Illustrative examples are shown in Figure \ref{fig:sv_pattern} and the detailed settings are provided in Section 2.4. Specifically, a linear function is adopted to recognize a linearly varying expression pattern (Figure \ref{fig:sv_pattern}(A)), while an exponential function is adopted for the expression pattern clustered in focal areas (Figure \ref{fig:sv_pattern}(B)), and a periodic function is particularly useful to detect an expression pattern that is periodically expressed across the spatial area (Figure \ref{fig:sv_pattern}(C)). For the non-SV gene with $\beta^{(1)}_{j}=0$ and $\beta^{(2)}_{j}=0$ (Figure \ref{fig:sv_pattern}(D)), it is observed that the expression pattern is random without association with the spatial location. This mean-value-based strategy requires less storage space and also involves a simpler estimation procedure.

Moreover, we innovatively utilize a set of cell-type-specific $\alpha_k$'s for eliminating the confounding impact of the latent cellular composition, where spots that are spatially closer are often observed to have similar cell-type proportions \citep{cable2022robust}. Different from the mean-value-based strategy adopted by \cite{yu2022identification} for marginal cell-type-specific SV gene detection, we focus on the detection of global SV genes while accommodating the cell-type proportion confounding.

\subsection{Priors Specification}

The proposed priors are defined as follows:
\begin{equation}\label{priors}
   \begin{aligned}
   & \boldsymbol{\beta}=\boldsymbol{\gamma} \circ {\mathbf{t}_{\lambda, \boldsymbol{\rho}}(\boldsymbol{\gamma})}, \boldsymbol{\gamma} \sim \mathrm{N}\left(\mathbf{0}_{2 p}, \sigma_\gamma^2\left(\mathbf{L}+\varepsilon \mathbf{I}_{2 p}\right)^{-1}\right),   \sigma_\gamma^2 \sim \operatorname{IG}\left(a_\gamma, b_\gamma\right), \lambda \sim \operatorname{Unif}\left(\lambda_l, \lambda_u\right), \\ &
    r_{i j} \sim \operatorname{Bern}\left(\pi_j\right), \pi_j \sim \operatorname{Beta}\left(a_\pi, b_\pi\right), \phi \sim \operatorname{Ga}\left(a_\phi, b_\phi\right), \\ &
   \mu_{0 j} \sim N\left(0, \sigma_{0 j}^2\right), \alpha_{k} \sim N\left(0, \sigma_{\alpha_{k}}^2\right),
   \end{aligned}
\end{equation}
where $\boldsymbol{\beta} = \left(( \boldsymbol{\beta}^{(1)}) ^\mathrm{T}, ( \boldsymbol{\beta}^{(2)}) ^\mathrm{T} \right)_{(2p)}^{\mathrm{T}} = \left( \beta^{(1)}_{1}, \ldots, \beta^{(1)}_{p}, \beta^{(2)}_{1}, \ldots, \beta^{(2)}_{p} \right)^{\mathrm{T}}$,  $\boldsymbol{\gamma}   = \left(( \boldsymbol{\gamma}^{(1)}) ^\mathrm{T}, ( \boldsymbol{\gamma}^{(2)}) ^\mathrm{T} \right)_{(2p)}^{\mathrm{T}} $ represents the effect size of the genes, and $\circ$ denotes the element-wise product.
 $\mathbf{t}_{\lambda,\boldsymbol{\rho}}(\boldsymbol{\gamma})=\left\{\operatorname{I}\left(\left|\gamma^{(1)}_{1}\right|>\lambda \cdot \rho^{(1)}_{1} \right), \ldots, \operatorname{I}\left(\left|\gamma^{(2)}_{p}\right|>\lambda  \cdot \rho^{(2)}_{p} \right)\right\}^{\mathrm{T}} $ is a vector thresholding function with $\lambda$ being a parameter controlling model sparsity and $\boldsymbol{\rho} = \left(( \boldsymbol{\rho}^{(1)}) ^\mathrm{T}, ( \boldsymbol{\rho}^{(2)}) ^\mathrm{T} \right)_{(2p)}^{\mathrm{T}}$ being the adaptive weights. Moreover, $\mathbf{L}=\left(\operatorname{sgn}(\tilde{\boldsymbol{\beta}}) \operatorname{sgn}(\tilde{\boldsymbol{\beta}})^{\mathrm{T}}\right) \, \circ \, \widetilde{\mathbf{L}}$ is a block diagonal adaptive Laplacian matrix, where $\operatorname{sgn}(\tilde{\boldsymbol{\beta}})=\left({\operatorname {sgn}}\left(\tilde{\beta}^{(1)}_{1}\right), \ldots, {\operatorname {sgn }}\left(\tilde{\beta}^{(2)}_{p}\right)\right)^{\mathrm{T}}$ with $\tilde{\boldsymbol{\beta}}$ being a rough estimate of $\boldsymbol{\beta}$ and $\widetilde{\mathbf{L}}=\mathbf{I}-\widetilde{\boldsymbol{D}}^{-1 / 2} \widetilde{\boldsymbol{A}} \widetilde{\boldsymbol{D}}^{-1 / 2}$ with $ \widetilde{\boldsymbol{A}}=\left(\begin{array}{cc}
 \boldsymbol{A}& \mathbf{0} \\
 \mathbf{0} & \boldsymbol{A}
 \end{array}\right)=(\widetilde{a}_{jl})_{2p\times 2p}$ and $\widetilde{\boldsymbol{D}}=\operatorname{diag}\left(d_1,\cdots,d_p,d_1,\cdots,d_p\right)$ with $d_j=\sum_{l=1}^{p} a_{jl}$. $\varepsilon$ is a small constant to make ${\mathbf{L}} + \varepsilon \mathbf{I}_{2p}$ strictly positive-definite, which is set as $10^{-3}$ in our numerical work.

The proposed priors have been motivated by the following considerations. The identification of SV genes is achieved using the hard-thresholded Gaussian prior, where $\beta^{(d)}_j$ is shrunk to zero when {$\mathrm{t}_{\lambda,\rho^{(d)}_j}\left(\gamma^{(d)}_j\right)=\operatorname{I}\left(\left|\gamma^{(d)}_{j}\right|>\lambda \cdot \rho^{(d)}_j \right)=0$ ($ d = 1, 2$).} The thresholded Gaussian prior has been shown as a useful alternative to shrinkage priors in Bayesian sparse analysis \citep{wu2024bayesian}, and is favored here for its simplicity and flexibility as well as its appealing interpretability as a minimal detectable signal. Here, we further introduce a series of weights $\rho_j^{(d)}$'s in $\mathbf{t}_{\lambda,\boldsymbol{\rho}}(\boldsymbol{\gamma})$ to adjust the shrinkage of various $\gamma^{(d)}_j$'s to improve selection efficiency, where the genes with strong spatial variability are potentially assigned small weights and thus are more likely to have nonzero $\beta^{(d)}_j$'s.

Moreover, the network dependency is introduced via the graph Laplacian matrix in the covariance matrix of the hard-thresholded Gaussian prior. The proposed network-assisted strategy is motivated by the successes of Laplacian shrinkage in high-dimensional regression analysis \citep{chakraborty2019graph, cai2020bayesian}. Different from these studies, we innovatively conduct SV gene detection with the network structures among genes incorporated. In particular, the Laplacian matrix $\widetilde{\mathbf{L}}$ is further modified with a pre-defined sign matrix $\operatorname{sgn}(\tilde{\boldsymbol{\beta}}) \operatorname{sgn}(\tilde{\boldsymbol{\beta}})^{\mathrm{T}}$ to accommodate the scenario where two neighborhood genes are negatively correlated and have opposite directions of spatial variability. With the proposed priors, we have $\boldsymbol{\gamma}^{\mathrm{T}}\left(\mathbf{L}+\varepsilon \mathbf{I}_{2p}\right)\boldsymbol{\gamma}=\sum_{j\sim l}\left(\frac{\text{sgn}\left(\tilde{\beta}_{j}^{(1)}\right)\gamma_{j}^{(1)}}{\sqrt{d_j}}-\frac{\text{sgn}\left(\tilde{\beta}_{l}^{(1)}\right)\gamma_{l}^{(1)}}{\sqrt{d_l}} \right)^2+\sum_{j\sim l}\left(\frac{\text{sgn}\left(\tilde{\beta}_{j}^{(2)}\right)\gamma_{j}^{(2)}}{\sqrt{d_{j}}}-\frac{\text{sgn}\left(\tilde{\beta}_{l}^{(2)}\right)\gamma_{l}^{(2)}}{\sqrt{d_{l}}} \right)^2$, where
for genes $j$ and $l$ with an edge $e(j,l)$, the absolute values of $\gamma^{(d)}_{j} $ and $\gamma^{(d)}_{l}$ are promoted to be similar, further inducing simultaneous \textit{SV} or \textit{non-SV}.

We assign a Bernoulli prior for the latent variable $r_{ij}$ with the hyperparameter $\pi_j \sim \text{Beta}(a_{\pi},b_{\pi})$, where $\pi_j$ is the probability that $Y_{ij}$ is a dropout zero.
A Gamma distribution $\text{Ga}(a_\phi, b_\phi)$ is assumed for the dispersion parameter $\phi$.  For the variance term $\sigma_{\boldsymbol{\gamma}}^2$, we use the conjugate prior by assigning the Inverse-Gamma distribution $\operatorname{IG}\left(a_\gamma, b_\gamma\right)$. The non-negative uniform prior $\text{Unif} \left(\lambda_l, \lambda_u\right)$ is assigned for the threshold parameter $\lambda$. For $\mu_{0j}$ and $\alpha_k$, the normal priors  with mean 0 and variance $\sigma_{0j}^2$ and $\sigma_{\alpha_k}^2$ are assumed, respectively. These priors have been popular in the existing Bayesian studies.

\subsection{Bayesian Inference}
\label{sec:bayesian inference}
The model parameter space consists of $\left(\boldsymbol{\gamma}, \boldsymbol{R}, \boldsymbol{\mu}_0, \boldsymbol{\alpha}, \phi, \lambda \right) $, where $\boldsymbol{R} = (r_{ij})_{(n \times p)}$, and $\boldsymbol{\mu}_0$ and $\boldsymbol{\alpha}$ are the vectors consisting of $\mu_{0j}$'s and $\alpha_k$'s, respectively. The posterior distribution is
\begin{equation}
\begin{aligned}
& f\left(\boldsymbol{\gamma}, \boldsymbol{R}, \boldsymbol{\mu}_0, \boldsymbol{\alpha}, \phi, \lambda \mid \mathbf{Y}\right) \propto  \\
& \int f\left(\mathbf{Y} \mid \boldsymbol{\gamma}, \boldsymbol{R}, \boldsymbol{\mu}_0, \boldsymbol{\alpha}, \phi, \lambda \right) f\left(\boldsymbol{\gamma} \mid \sigma_\gamma^2\right) f\left(\sigma_\gamma^2\right) f(\lambda)  f\left(\boldsymbol{\mu}_0\right) f\left(\boldsymbol{\alpha}\right) f(\phi) f(\boldsymbol{R} \mid \boldsymbol{\pi}) f(\boldsymbol{\pi}) d \boldsymbol{\pi} \\
& = \left\{ \prod_{\left\{(i,j):r_{i j}=0\right\}} \operatorname{NB}\left(Y_{ij} \mid \mu_{ij}, \phi\right) \right\} \times \mathrm{N}\left(\boldsymbol{\gamma} \mid \mathbf{0}, \sigma_\gamma^2\left(\mathbf{L}+\varepsilon \mathbf{I}_{2p}\right)^{-1}\right) \times \operatorname{IG}\left(\sigma_{\boldsymbol{\gamma}} \mid a_\gamma, b_\gamma\right)  \\
& \quad \times  \operatorname{Unif}\left(\lambda \mid \lambda_l, \lambda_u\right) \times \mathrm{N}\left(\boldsymbol{\mu}_0 \mid \mathbf{0},  \operatorname{diag} \left(\sigma_{\mu_{01}}^2, \ldots, \sigma_{\mu_{0p}}^2 \right)\right)  \times \mathrm{N}\left(\boldsymbol{\alpha} \mid \mathbf{0},  \operatorname{diag} \left(\sigma_{\alpha_{1}}^2, \ldots, \sigma_{\alpha_{K}}^2 \right)\right)\\
& \quad \times  \operatorname{Ga}\left(\phi \mid a_\phi, b_\phi\right) \times \prod_{i=1}^{n} \prod_{j=1}^{p} \operatorname{Be-Bern}\left( r_{ij} \mid a_\pi, b_\pi\right),
\end{aligned}
\end{equation}
where $\operatorname{Be-Bern}$ denotes the Beta-Bernoulli distribution with the probability mass function  $\operatorname{Be-Bern}\left( r_{ij}\mid a_\pi, b_\pi\right) = \frac{\operatorname{Beta} \left( a_\pi + r_{ij}, b_\pi - r_{ij} + 1 \right) }{\operatorname{Beta} \left( a_\pi, b_\pi \right)}$.

The posterior sampling is conducted based on the MCMC algorithm. We first introduce the sampling variances $\tau_{\mu_0}^2, \tau_{\alpha}^2, \tau_{\phi}^2, \tau_{\gamma}^2,$ and $\tau_{\lambda}^2$
for $\mu_{0j}$'s, $\alpha_k$'s, $\phi$, $\boldsymbol{\gamma}$, and $\lambda$, respectively, and then conduct the following steps at each MCMC iteration, where the symbol ``$\boldsymbol{-}$'' in the condition position denotes ``the rest parameters''.

\begin{enumerate}
    \item[(a)] Sequentially update $r_{ij}$ for $ \left\{ \left(i,j\right): Y_{ij}=0 \right\}$ with the conditional distribution of $r_{ij}$ given by
$f\left(r_{ij} \mid \boldsymbol{-}\right) \propto \left(\frac{\phi}{\mu_{ij}+\phi}\right)^{\phi (1-r_{ij})} \times \operatorname{Be-Bern}\left( r_{ij}\mid a_\pi, b_\pi\right)$.
    \item[(b)] Sequentially sample $\mu_{0j}^{*} \sim  N( \mu_{0j}, \tau_{\mu_0}^2)$ for $ j = 1, \ldots, p $, and accept $\mu_{0j}^{*}$ with probability $\min \left\{1,
\frac{N\left(\mu_{0j} \mid \mu_{0j}^*, \tau_{\mu_0}^2\right)  \times N\left(\mu_{0j}^* \mid 0, \sigma_{0j}^2\right)  \times \prod_{\left\{i: r_{i j}=0\right\}} \text{NB} \left(Y_{ij} \mid \mu_{0j}^*, \boldsymbol{-} \right)}
{N\left(\mu^*_{0j} \mid \mu_{0j}, \tau_{\mu_0}^2\right) \times N\left(\mu_{0j} \mid 0, \sigma_{0j}^2\right) \times \prod_{\left\{i:r_{i j}=0\right\}} \text{NB} \left(Y_{ij} \mid \mu_{0j},  \boldsymbol{-} \right)}\right\}.
$

\item[(c)] Sequentially sample $\alpha_{k}^{*} \sim  N( \alpha_{k}, \tau_{\alpha}^2)$ for $ k = 1, \ldots, K$, and accept $\alpha_k^{*}$ with probability $\min \left\{1,
\frac{N\left(\alpha_{k} \mid \alpha_{k}^*, \tau_{\alpha}^2\right)  \times N\left(\alpha_{k}^* \mid 0, \sigma_{\alpha_k}^2\right)  \times \prod_{\left\{(i,j): r_{i j}=0\right\}} \text{NB} \left(Y_{ij} \mid \alpha_{k}^*, \boldsymbol{-} \right)}
{N\left(\alpha^*_{k} \mid \alpha_{k}, \tau_{\alpha}^2\right) \times N\left(\alpha_{k} \mid 0, \sigma_{\alpha_k}^2\right) \times \prod_{\left\{(i,j):r_{i j}=0\right\}} \text{NB} \left(Y_{ij} \mid \alpha_{k},  \boldsymbol{-} \right)}\right\}.
$
\item [(d)]  Sample $\phi^{*} $ from $ N( \phi, \tau_{\phi}^2)$ truncated at 0, and accept $\phi^{*}$ with probability $\min \left\{1,\right.$ $\left.
\frac{N_{+}\left(\phi\mid \phi^*, 0, \infty, \tau_\phi^2\right) \times \mathrm{Ga}\left(\phi^{*} ; a_{\phi}, b_{\phi}\right) \times \prod_{\left\{\left(i,j\right): r_{i j}=0\right\}} \text{NB} \left(Y_{ij} \mid \phi^*, \boldsymbol{-} \right)}
{N_{+}\left(\phi^*\mid \phi, 0, \infty, \tau_\phi^2\right) \times \mathrm{Ga}\left(\phi; a_{\phi}, b_{\phi}\right)  \times \prod_{\left\{\left(i,j\right):r_{i j}=0\right\}} \text{NB} \left(Y_{ij} \mid \phi, \boldsymbol{-} \right)}\right\}$.

\item [(e)] Sample $\boldsymbol{\gamma}^{*}$ from
$\mathrm{N}\left(\mu \left( \boldsymbol{\gamma}\right), \tau_{\boldsymbol{\gamma}}^2 \cdot \sigma_\gamma^2 \left(\mathbf{L}+\varepsilon \mathbf{I}_{2p} \right)^{-1}\right)$, where
\begin{equation}\label{sample_gamma}
\mu \left( \boldsymbol{\gamma}\right) = \sqrt{1 - \tau_{\boldsymbol{\gamma}}^2} \boldsymbol{\gamma} + \left( 1 - \sqrt{1 - \tau_{\boldsymbol{\gamma}}^2} \right)   \sigma_\gamma^2 \left(\mathbf{L}+\varepsilon \mathbf{I}_{2p} \right)^{-1} \nabla_{\boldsymbol{\gamma}} \log f( \mathbf{Y}\mid \boldsymbol{\gamma}, \boldsymbol{-}),
\end{equation}
with $\nabla_{\boldsymbol{\gamma}} \log f( \mathbf{Y}\mid \boldsymbol{\gamma}, \boldsymbol{-})$ being the first derivative of the log-likelihood function with respect to $\boldsymbol{\gamma}$. Then, accept $\boldsymbol{\gamma}^{*}$ with probability $\min \left\{1,
\frac{\prod_{\left\{(i,j): r_{i j}=0\right\}} \text{NB} \left(Y_{ij} \mid \boldsymbol{\gamma}^*, \boldsymbol{-} \right)}
{\prod_{\left\{(i,j):r_{i j}=0\right\}} \text{NB} \left(Y_{ij} \mid \boldsymbol{\gamma}, \boldsymbol{-} \right)}\right\}$.

\item[(f)] Update $\rho_{j}^{(d)} = \prod_{s=1}^{S} \left( \min_{ \left\{l: l \in V_{s} \right\} } \frac{1}{ \left|\boldsymbol{\gamma}^{(d)}_l \right|^{1/2} } \right)^{ \operatorname{I}(j \in V_s)} $ for $j = 1, \ldots, p$ and $d = 1, 2$, and update $\mathbf{t}_{\lambda,\boldsymbol{\rho}}(\boldsymbol{\gamma})$ accordingly, where $V_1, V_2, \ldots, V_S$ are the index sets of the $S$ disconnected sub-networks in $G(V,E)$.

 \item[(g)] Sample $\sigma_{\boldsymbol{\gamma}}^2$ from $\operatorname{IG}\left(\tilde{a}_{\boldsymbol{\gamma}}, \tilde{b}_{\boldsymbol{\gamma}} \right)$ with shape parameter $\tilde{a}_{\boldsymbol{\gamma} }= a_{\boldsymbol{\gamma}} + p $ and scale parameter $\tilde{b}_{\boldsymbol{\gamma} }= b_{\boldsymbol{\gamma}} + \frac{\boldsymbol{\gamma}^{\mathrm{T}} \left(\mathbf{L}+\varepsilon \mathbf{I}_{2p}\right) \boldsymbol{\gamma}}{2}$.

 \item [(h)] Sample $\lambda^*$ from
 $N \left(\lambda, \tau_\lambda^2\right)$ truncated at interval $[\lambda_l, \lambda_u]$, and accept $\lambda^*$ with probability $ \min \left\{1,
\frac{N_{+}\left(\lambda \mid \lambda^*, \lambda_{l}, \lambda_{u}, \tau_\lambda^2\right) \prod_{\left\{\left(i,j\right) : r_{i j}=0\right\}}\text{NB} \left(Y_{ij} \mid \lambda^*,  \boldsymbol{-} \right)}
{ N_{+}\left(\lambda^* \mid \lambda, \lambda_{l}, \lambda_{u}, \tau_\lambda^2\right)
 \prod_{\left\{ \left(i,j\right): r_{i j}=0\right\}}\text{NB} \left(Y_{ij} \mid \lambda,  \boldsymbol{-} \right)}\right\}.$
\end{enumerate}

Here, updating $r_{ij}$'s and $\sigma_{\boldsymbol{\gamma}}^2$ is achieved through the Gibbs sampler, while the Metropolis-Hasting (MH) algorithm is adopted for sampling $\mu_{0j}$'s, $\alpha_{k}$'s, $\phi$, and $\lambda$. For sampling $\boldsymbol{\gamma}$, we resort to the preconditioned Crank-Nicolson Langevin dynamics (pCNLD), which takes advantage of the gradient information of the target distribution to speed up convergence. Furthermore, with (\ref{sample_gamma}), pCNLD explicitly incorporates the network dependency into the sampling process. The details of the proposed pCNLD sampling are given in Supplementary Section S1 \citep{wu2025supp}. For $\rho_{j}^{(d)}$'s, since biological networks are often composed of multiple disconnected sub-networks, we propose adopting a set of group-wise weights for the $S$ sub-networks for more effectively utilizing network information. Specifically, a series of data-driven weights inversely proportional to the absolute effect sizes of the genes involved in the specific sub-networks are introduced, potentially facilitating the identification of SV genes with weak signals, which may be involved in the sub-networks consisting of SV genes with strong signals and small thresholds. The settings for the hyperparameters and sampling variances are provided in Supplementary Section S2 \citep{wu2025supp}.

For the spatial modeling function $\mathcal{K}(x_{id})$, as discussed above, we consider three most popular functions as recommended in the published studies. Specifically, for $x_{id}~(d=1,2)$ which has been transformed to have mean 0 and standard deviation 1, we consider one linear function $\mathcal{K}_1(x_{id}) = x_{id}$, two exponential functions $\mathcal{K}_2(x_{id})  =  \exp \left(-\frac{x_{id}}{2 \left(l_d^{(1)}\right)^2}\right)$ and $\mathcal{K}_3(x_{id})  =  \exp \left(-\frac{x_{id}}{2 \left(l_d^{(2)}\right)^2}\right)$, and two periodic functions $\mathcal{K}_4(x_{id})  =  \cos \left(\frac{2 \pi x_{id}}{l_d^{(1)}}\right)$ and $\mathcal{K}_5(x_{id})  =  \cos \left(\frac{2 \pi x_{id}}{l_d^{(2)}}\right)$, where $l_d^{(1)}$ and $l_d^{(2)}$ are the 40\% and 60\% quantiles of $|x_{1d}|,\cdots,|x_{nd}|$, respectively. To accommodate the fact that results may be sensitive to the choice of scale parameter, two values are considered for the exponential and periodic patterns. Then, we conduct the MCMC sampling for each $\mathcal{K}_s(\cdot) (s=1,\cdots,5)$ and obtain the estimated posterior expectation $
\hat{\boldsymbol{\beta}}^{{\mathcal{K}_s}} =\frac{\sum_{m=1}^M \boldsymbol{\gamma}^{(m)}  \boldsymbol{\mathbf{t}}_{\lambda^{(m)},\boldsymbol{\rho}^{(m)}}\left (\boldsymbol{\gamma}^{(m)}\right) }{\sum_{m=1}^M \boldsymbol{\mathbf{t}}_{\lambda^{(m)},\boldsymbol{\rho}^{(m)}}\left (\boldsymbol{\gamma}^{(m)}\right)},
\hat{\boldsymbol{\mu}}_{0}^{{\mathcal{K}_s}} = \frac{\sum_{m=1}^M \boldsymbol{\mu}_{0}^{(m)}}{M}, \hat{\boldsymbol{\alpha}}^{{\mathcal{K}_s}} = \frac{\sum_{m=1}^M \boldsymbol{\alpha}^{(m)}}{M}$, $\hat{\phi}^{{\mathcal{K}_s}} = \frac{\sum_{m=1}^M \phi^{(m)}}{M}$, and $\hat{\boldsymbol{R}}^{{\mathcal{K}_s}} = \frac{\sum_{m=1}^M \boldsymbol{R}^{(m)}}{M}$, where $\left\{ \boldsymbol{\gamma}^{(m)}, \lambda ^{(m)}, \boldsymbol{\rho}^{(m)}, \boldsymbol{\mu}_0^{(m)}, \boldsymbol{\alpha}^{(m)}, \phi^{(m)} , \boldsymbol{R}^{(m)}\right\}_{m=1}^{M}$ denotes the $M$ samples obtained after burn-in and thinning (we omit the dependence on $\mathcal{K}_s$ to simplify
notation).

To facilitate the combination of five models, instead of directly considering the values of $\hat{\boldsymbol{\beta}}^{{\mathcal{K}_s}}$ for SV gene identification, we further introduce
a posterior inclusion probability vector estimated as ${\mathrm{PIP}}^{{\mathcal{K}_s}}=\frac{1}{M} \sum_{m=1}^M \boldsymbol{\mathbf{t}}_{\lambda^{(m)},\boldsymbol{\rho}^{(m)}}\left (\boldsymbol{\gamma}^{(m)}\right)$ and consider the posterior model probability $f\left(\mathcal{M}_{\mathcal{K}_s} \mid\mathbf{Y}\right)$ introduced in \cite{quintana2011incorporating}, where $\mathcal{M}_{\mathcal{K}_s}$ denotes the model with $\mathcal{K}_s(\cdot)$. Specifically,
for each $\mathcal{M}_{\mathcal{K}_s}$, we calculate
$
f\left(\mathcal{M}_{\mathcal{K}_s} \mid\mathbf{Y}\right)=\frac{f\left(\mathbf{Y} \mid \mathcal{M}_{\mathcal{K}_s}\right)f\left(\mathcal{M}_{\mathcal{K}_s} \right)}{\sum_{s'=1}^5 f\left(\mathbf{Y} \mid \mathcal{M}_{\mathcal{K}_{s'}}\right)
f\left(\mathcal{M}_{\mathcal{K}_{s'}} \right)}
$. Here, $f\left(\mathcal{M}_{\mathcal{K}_s} \right)$ is the prior probability of model $\mathcal{M}_{\mathcal{K}_s}$, and we set a non-informative prior with $f\left(\mathcal{M}_{\mathcal{K}_s} \right)=1/5$. $f\left(\mathbf{Y} \mid \mathcal{M}_{\mathcal{K}_s}\right)$ is the likelihood function with the estimated parameters
$\hat{\boldsymbol{\beta}}^{{\mathcal{K}_s}},  \hat{\boldsymbol{\mu}}_0^{{\mathcal{K}_s}}, \hat{\boldsymbol{\alpha}}^{{\mathcal{K}_s}}, \hat{\phi}^{\mathcal{K}_s}$, and $\hat{\boldsymbol{R}}^{{\mathcal{K}_s}}$ under model $\mathcal{M}_{\mathcal{K}_s}$. We then utilize
$f\left(\mathcal{M}_{\mathcal{K}_s}\mid\mathbf{Y}\right)$'s as the model-specific weights to obtain the combined $\operatorname{PIP}= \sum\limits_{s=1}^5 f\left(\mathcal{M}_{\mathcal{K}_s}\mid\mathbf{Y}\right)\mathrm{PIP}^{{\mathcal{K}_s}}
$. Such an ensemble strategy allows us to comprehensively consider the effects across different kernel functions to improve identification accuracy. {Moreover, as gene $j$ with at least one non-zero $\beta^{(d)}_{j} \left(d = 1,2\right) $ is identified as SV, we introduce $\widetilde{\operatorname{PIP}}_j = \max \left(\operatorname{PIP}_j, \mathrm{PIP}_{j+p}\right)$ for $ j = 1,\ldots, p$.
Based on $\widetilde{\operatorname{PIP}}_j$'s, the Bayesian false discovery rate (BFDR) control strategy is adopted for controlling multiplicity. Specifically,
$\operatorname{BFDR}(c)=\frac{\sum_{j=1}^{p}\left(1-\widetilde{\operatorname{PIP}}_j\right) \mathrm{I}\left(1-\widetilde{\operatorname{PIP}}_j<c\right)}{\sum_{j=1}^{p} \mathrm{I}\left(1-\widetilde{\operatorname{PIP}}_j<c\right)}$} with $\operatorname{BFDR}(c)$ being the desired significance level. We set $\operatorname{BFDR}(c)$ as 0.05 in our numerical studies. Then, the SV gene set is defined as $\left\{ j : \widetilde{\operatorname{PIP}}_{j} \geq c \right\}$. To achieve improved stability and better false negative control, in numerical studies, we run five MCMC chains independently, and the genes identified in more than 80\% of the chains are finally identified as SV.

To improve computational efficiency, parallelization is implemented with the R package \textit{RcppParallel}. Specifically, the marginal sequential sampling is divided into parallel programming to reduce computer time. For $\boldsymbol{\gamma}$ with the dependency Laplacian matrix incorporated, benefiting from the sparse and block-wise nature of biological networks, the parallelism block-wise sampler is adopted to avoid sampling from a high-dimensional multivariate normal distribution, which further accelerates computation. More discussions on the computer time of the proposed algorithm are provided in Supplementary Section S3 \citep{wu2025supp}.

\section{Simulations}

\subsection{Basic Simulations}

Simulation studies are conducted under the following settings. (a) $n = 1,024$ spots located on a 32 by 32 square lattice, $p=5,000$ genes, and $K=6$ cell types. (b) The square lattice is partitioned into three regions as displayed in Figure S1, where the cellular compositions $\boldsymbol{w}_i$'s are independently sampled from Dirichlet distributions $\operatorname{Dirc}(1,1,1,1,1,1)$ (Region 1), $\operatorname{Dirc}(3,5,7,9,11,13)$ (Region 2), and $\operatorname{Dirc}(18,16,14,12,10,8)$ (Region 3). (c) Consider three types of spatial pattern $\mathcal{K}\left(x_{id}\right)$, including $x_{id}$ (Linear), $\exp\left(-\frac{x_{id}}{2}\right)$ (Exponential), and $\cos \left(2 \pi x_{id}\right)$ (Periodic). (d) The networks are block-wise and composed of 100 disconnected sub-networks with 50 nodes each.
For each sub-network, two types of network structure, namely \textit{Star} and \textit{Scale-free}, are considered to mimic the real-world transcription factor regulatory network and interaction network with scale-free properties, respectively. Illustrative examples of the sub-networks are presented in Figure S2(A) and S2(B). (e) All genes in the first ten sub-networks are SV, leading to a total of 500 SV genes. Both positive and negative signals are considered with various levels of magnitude. More detailed settings are provided in Supplementary Section S4 \citep{wu2025supp}. (f) The spatial transcriptomics count data is generated from model (\ref{eq:2}) with the dispersion parameter $\phi$ being 10. Two dropout rates settings, 0.1 and 0.5, are considered, representing low and high sparsity. The baseline parameter ${\mu}_{0j}$ and the cell-type-specific effect ${\alpha}_k$ are independently generated from $N(2,0.5^2)$ and $N(0,3.5^2)$, respectively. There are 12 scenarios (Table S1), comprehensively covering a wide spectrum with different patterns of spatial expressions, different structures of networks and the corresponding spatially variable signals, and different degrees of sparsity. 

In addition to the proposed approach, nine alternatives are considered. SpatialDE \citep{svensson2018spatialde} is a likelihood ratio test method based on Gaussian process regression. SPARK \citep{sun2020statistical} is a method built on a Poisson log-linear model with a Gaussian process incorporated. SPARK-X \citep{zhu2021spark} is a scalable non-parametric test constructed on a robust covariance test framework. {HRG \citep{wu2022highly} serves as a method to detect informative genes that exhibit regional expression patterns within the cell-cell similarity network.} MERINGUE \citep{miller2021characterizing} is a graph-based testing method relying on spatial cross-correlation analysis. CTSV and CTSV-g \citep{yu2022identification} are tests based on the ZINB model, with CTSV used for cell-type-specific SV gene identification and CTSV-g for global SV gene identification, respectively. {SINFONIA \citep{jiang2023sinfonia} is a graph-based method that ranks genes according to spatial autocorrelation measurements. nnSVG \citep{weber2023nnsvg} is a test based on nearest-neighbor Gaussian processes.} HEARTSVG \citep{yuan2024heartsvg} is a distribution-free method which employs the exclusion of non-SV genes to infer the presence of SV genes.  {Among these alternatives, SpatialDE, SPARK, CTSV, CTSV-g, and nnSVG are parametric, while SPARK-X, HRG, MERINGUE, SINFONIA, and HEARTSVG are non-parametric. In addition, HRG, MERINGUE, SINFONIA, and nnSVG adopt graph strategies. All the alternatives except HRG and SINFONIA conduct tests marginally and implement multiple testing control, whereas HRG and SINFONIA provide gene ranking.} The implementation details for the competing methods are given in Supplementary Section S5 \citep{wu2025supp}.

\begin{table}[htb]
\caption{Simulation results under the scenarios with a low dropout rate, where FDR (BFDR) is controlled to be $<$0.05. In each cell, mean (SD) is based on 50 replicates. \label{tab:simlow}}
\resizebox{\textwidth}{!}{
\begin{tabular}{lcccccc}
\hline
                    & Recall       & Precision    & F1           & Recall       & Precision    & F1           \\ \hline
Linear pattern      &              & \textit{Star} Network           &              &              & \textit{Scale-free} Network           &              \\
proposed            & 0.992(0.016) & 1.000(0.000) & 0.996(0.008) & 0.991(0.016) & 1.000(0.000) & 0.995(0.008) \\
spatialDE           & 0.408(0.172) & 0.990(0.030) & 0.554(0.194) & 0.549(0.183) & 0.990(0.045) & 0.685(0.179) \\
SPARK               & 0.907(0.111) & 0.977(0.010) & 0.937(0.065) & 0.911(0.079) & 0.991(0.005) & 0.947(0.046) \\
SPARK-X             & 0.887(0.259) & 0.154(0.193) & 0.181(0.021) & 0.772(0.143) & 0.485(0.370) & 0.502(0.287) \\
HRG              & 0.914(0.126) & 0.249(0.034) & 0.391(0.054) & 0.819(0.248)&	0.571(0.338) &	0.566(0.264)    \\
MERINGUE            & 0.562(0.272) & 0.944(0.027) & 0.658(0.241) & 0.642(0.256) & 0.958(0.021) & 0.732(0.219) \\
CTSV       & 0.118(0.039) & 0.642(0.249) & 0.186(0.054) & 0.162(0.076) & 0.729(0.193) & 0.249(0.100) \\
CTSV-g                & 0.978(0.057) & 0.810(0.104) & 0.880(0.063) & 0.978(0.044) & 0.936(0.013) & 0.956(0.023) \\
SINFONIA               &0.741(0.097)	&	0.686(0.111)&	0.712(0.105)      &0.766(0.065)	&	0.715(0.078)	&0.740(0.072)    \\
nnSVG                & 0.341(0.132) & 0.988(0.011) & 0.491(0.157) & 0.499(0.126)	&	0.987(0.012)	& 0.653(0.124)   \\
HEARTSVG            & 0.762(0.246) & 0.284(0.265) & 0.298(0.129) & 0.920(0.142) & 0.178(0.159) & 0.259(0.106) \\ \hline
Exponential pattern &              & \textit{Star} Network           &              &              & \textit{Scale-free} Network           &              \\
proposed            & 0.933(0.141) & 1.000(0.001) & 0.959(0.091) & 0.979(0.035) & 1.000(0.001) & 0.989(0.019) \\
spatialDE           & 0.036(0.032) & 0.989(0.058) & 0.084(0.053) & 0.278(0.186) & 0.986(0.061) & 0.424(0.232) \\
SPARK               & 0.229(0.103) & 0.988(0.013) & 0.360(0.139) & 0.622(0.133) & 0.993(0.005) & 0.756(0.113) \\
SPARK-X             & 0.860(0.296) & 0.098(0.004) & 0.177(0.011) & 0.702(0.262) & 0.473(0.369) & 0.442(0.276) \\
HRG & 0.711(0.245)	&	0.193(0.067)	&0.304(0.105) & 0.766(0.258)	&	0.506(0.345)&	0.524(0.298) \\
MERINGUE            & 0.153(0.127) & 0.949(0.042) & 0.248(0.174) & 0.479(0.252) & 0.957(0.023) & 0.592(0.254) \\
CTSV       & 0.012(0.022) & 0.116(0.150) & 0.043(0.033) & 0.025(0.034) & 0.366(0.315) & 0.058(0.045) \\
CTSV-g              & 0.899(0.173) & 0.918(0.022) & 0.897(0.121) & 0.957(0.078) & 0.934(0.015) & 0.944(0.041) \\
SINFONIA &0.421(0.062)	&	0.366(0.063)	&0.392(0.063) &0.647(0.071)	&	0.587(0.080)	& 0.615(0.076) \\
nnSVG &0.040(0.023)	&	0.991(0.026)&	0.076(0.042) & 0.233(0.117)	&	0.984(0.026)& 	0.362(0.165)  \\
HEARTSVG            & 0.282(0.169) & 0.839(0.278) & 0.361(0.195) & 0.634(0.184) & 0.623(0.313) & 0.536(0.188) \\ \hline
Periodic pattern    &              & \textit{Star} Network           &              &              & \textit{Scale-free} Network           &              \\
proposed            & 0.996(0.008) & 1.000(0.000) & 0.998(0.004) & 0.979(0.130) & 1.000(0.000) & 0.981(0.121) \\
spatialDE           & 0.087(0.027) & 0.991(0.053) & 0.158(0.049) & 0.174(0.097) & 0.990(0.056) & 0.286(0.137) \\
SPARK               & 0.215(0.088) & 0.987(0.012) & 0.345(0.117) & 0.620(0.130) & 0.992(0.006) & 0.754(0.111) \\
SPARK-X             & 0.887(0.268) & 0.156(0.195) & 0.178(0.023) & 0.817(0.238) & 0.492(0.366) & 0.501(0.299) \\
HRG &0.895(0.147)	&	0.244(0.040)	&0.383(0.063)& 0.921(0.140)	 &	0.621(0.313)& 	0.678(0.282)\\
MERINGUE            & 0.433(0.265) & 0.952(0.025) & 0.543(0.263) & 0.688(0.277) & 0.961(0.019) & 0.761(0.236) \\
CTSV      & 0.088(0.055) & 0.664(0.268) & 0.142(0.072) & 0.215(0.163) & 0.742(0.217) & 0.291(0.187) \\

CTSV-g                & 0.969(0.076) & 0.884(0.045) & 0.922(0.042) & 0.990(0.028) & 0.947(0.009) & 0.968(0.015) \\
SINFONIA &0.547(0,088)	&	0.483(0.090) &	0.513(0.089) &0.764(0.065)	&	0.713(0.078)& 	0.737(0.072)  \\
nnSVG &0.122(0.019)	&	0.990(0.015)	&0.216(0.030)&0.491(0.144)	&	0.988(0.011)& 	0.642(0.147) \\
HEARTSVG            & 0.230(0.135) & 0.980(0.043) & 0.355(0.165) & 0.598(0.184) & 0.976(0.084) & 0.720(0.159) \\ \hline
\end{tabular}
}
\end{table}

For evaluating identification performance, we adopt Recall = $\frac{\operatorname{TP}}{\operatorname{TP+FN}}$ , Precision =$\frac{\operatorname{TP}}{\operatorname{TP+FP}}$, and F1 score = $\frac{2 \cdot \text{Precision}\cdot \operatorname{Recall}}{\text{Precision+Recall}}, $ with TP, FP, and FN being the numbers of true positives, false positives, and false negatives, respectively. Under each simulation scenario, we simulate 50 replicates. Summary results under the scenarios with a low dropout rate are given in Table \ref{tab:simlow}. The rest of the results with a high dropout rate are provided in Supplementary Table S2 \citep{wu2025supp}. It is observed that the proposed approach achieves superior accuracy in identifying SV genes with higher F1 scores across all scenarios. Overall, CTSV-g achieves the second-best identification performance since it also accommodates dropouts and over-dispersion through the adoption of ZINB distribution. Improved performance over CTSV-g supports the validity of incorporating network information and accommodating cellular diversity. Furthermore, the majority of the alternatives exhibit good performance with the simple linear patterns while the proposed approach performs stably across different spatial expression patterns. Moreover, similar to those in some published studies, SPARK-X exhibits many false positives, probably due to its covariance test framework and neglection of cellular composition confounding, and SPARK achieves excellent FP control across all scenarios, mainly attributable to the robust Cauchy combination rule. {HRG also behaves poorly in FP control, which may be attributed to the potential error introduced by low-dimensional projection in the step of graph construction. The other two graph-based approaches, MERINGUE and nnSVG, are more conservative and tend to select fewer SV genes. As a result, they are poor at finding the true signals but show high precision values.} CTSV identifies the fewest global SV genes across all scenarios due to its concern with cell-type-specific SV genes and inability in accurately identifying genes with global spatial expression variations. Noticeably, under the scenarios with a scale-free network where the signals of the SV genes are larger, all approaches exhibit improved performance with higher F1 score values. However, the superiority of the proposed approach is again evidently observed. With a higher dropout rate, which is more common with practical ST data, all approaches have decayed performance, especially for SpatialDE which is not well-suited for sparse expression distribution. However, the superiority of the proposed approach becomes more prominent. 

\subsection{False Positive Rate Assessment Using SV-free Simulated Datasets}
{In addition to the basic simulations, we conduct another two SV-free simulation scenarios to examine false positive (FP) control performance. Specifically, we consider two different sparsity settings with dropout rates of 0.1 and 0.5. In each setting, we simulate $p = 5,000$ non-SV genes while keeping the other parameters the same as those in the above basic simulations.

In the SV-free simulations, for the proposed approach, we once again consider two networks, \textit{Star} and \textit{Scale-free}, to examine the FP control performance under different assigned network dependency structures. Comparison boxplots of the False Positive Rate (FPR), calculated as $= \frac{\operatorname{FP}}{\operatorname{TN}+\operatorname{FP}}$, based on 50 replicates, are shown in Supplementary Figure S3 \citep{wu2025supp}. We omit the results of SINFONIA since it requires a predefined number of target SV genes for identification.
We observe similar results to those in the basic simulation. When the dropout rate is lower, SPARK-X, HRG, and HEARTSVG perform poorly in false positive control, with much higher FPRs. In contrast, the remaining approaches all demonstrate satisfactory performance. When the dropout rate is higher, SPARK-X and HEARTSVG tend to select fewer genes in sparse data, resulting in improved false positive control. Meanwhile, the other approaches, except for HRG, can maintain their ability to control false positives.}

\subsection{Validation for Zero-Inflated Distribution} \label{sec:sim_zero}
{
To assess the necessity of adopting a zero-inflated distribution, we further compare our ZINB model with an NB model (which lacks zero inflation) under varying levels of data sparsity. Specifically, we focus on scenarios involving the star network and periodic spatial patterns, and we present the comparison results in Supplementary Figure S4 \citep{wu2025supp}. It is evident that the NB model consistently exhibits reduced performance in SV gene detection. The Recall and F1 metrics decay more significantly as the dropout rate increases. This suggests that zero-inflation modeling aids in recovering spatial signals that might otherwise be obscured by an excess of zeros, particularly under high-sparsity conditions.}

\subsection{Validation for the Network Assistance Strategy}
To better appreciate the operating characteristics of the proposed network-assisted strategy, in Figure S5, we take the scale-free network as an example and further examine the SV genes with large and small signals separately with two Recall indices (L:Recall and S:Recall). Here, besides the proposed approach, we also consider the corresponding approach without the assistance of the network (where $\mathbf{L}$ is a simple identity matrix). As shown in Figure S5, benefiting from network smoothness, the proposed approach has remarkably improved identification performance, especially for the SV genes with small signals. Such an improvement becomes more obvious for sparser data with a higher dropout rate. Moreover, selection stability is also improved.

\subsection{Model Misspecification}

We further evaluate performance of the proposed approach under scenarios where the data generation model is misspecified. In particular, we consider three types of model misspecification: (M1) the ZINB model considered in \cite{yu2022identification} where the logarithmic mean value is set as a mix of cell-type-specific expression levels; {(M2) the generalized linear spatial model where the spatial variability is introduced through the Gaussian covariance matrix as adopted in \cite{svensson2018spatialde} and \cite{sun2020statistical};} and (M3) the Poisson generalized model considered in \cite{sun2020statistical} with the parameters inferred from a real mouse olfactory bulb study. More detailed settings are provided in Supplementary Section S4 \citep{wu2025supp}, and summary results are reported in Figure S8.

It is evident that the proposed approach can either maintain its superiority or perform at least as well as the method that aligns with the data generation model. Specifically, under model (M1), which favors the CTSV method, the proposed approach achieves more accurate identification, with a median F1 score of 1. In contrast, CTSV shows the second-best performance. {Under model (M2), which involves covariance-based spatial variability, the proposed approach still attains the best F1 performance among all alternative methods, achieving a better balance between Recall and Precision. This enhanced performance once again validates the network assistance strategy.} Furthermore, even under model (M3), where there is no cellular composition confounding, the proposed approach can still accurately identify SV genes, with a median F1 score greater than 0.92. In this scenario, spatialDE and SPARK also perform well, as the expression pattern is inferred based on the analysis results of these two approaches.

\subsection{Examination on the Noise of Network and Cellular Composition}
\label{sec:sim_noise}
We continue to examine scenarios with noisy network and cellular composition information. In particular, the noisy network is likely to involve edges that connect both SV and non-SV genes. Two specific generation models are considered, including the ZINB model considered in Section 3.1 with {a} star network, {a} linear spatial pattern, and two degrees of sparsity, and the misspecified model (M1) considered in Section 3.5, where in each informative sub-network, about 20\% of the genes connected to the TF genes are uninformative. The summary results are reported in Supplementary Figure S9 \citep{wu2025supp}. It can be seen that the proposed approach can still achieve an effective balance between the prior network information and observation likelihood, leading to certain robustness. Moreover, to explore the impact of noisy cellular composition on identification performance, we consider the scale-free network with the linear pattern and high sparsity (dropout rate = 0.5), where the cell-type deconvolution estimates $\hat{\boldsymbol{w}}_i$ are sampled from $\operatorname{Dirc}(c \cdot \boldsymbol{w}_i)$ with $\boldsymbol{w}_i$ being the true value and concentration parameter $c = 100, 80,$ and 50. The lower value of $c$ indicates less accuracy of the cellular composition estimates. The summary results are provided in Figure S10. As expected, the noisy cellular compositions tend to result in decreased identification performance with increased variance. However, even with $c=50$, the identification performance is still satisfactory, and the superiority over the alternatives maintains.

\section{Data Analysis}
{
We applied the proposed method to three ST datasets generated using the Visium and Xenium In Situ platforms from 10x Genomics. These datasets include the Visium triple-negative breast cancer (TNBC), Visium primary liver cancer (PLC), and Xenium pancreatic cancer (PAC) datasets, which differ in spatial scale and resolution. Specifically, the Visium platform enables whole-transcriptome profiling at a spatial resolution of 55$\mu$m. In contrast, the Xenium platform offers analysis at sub-cellular resolution, targeting a pre-designed Human Multi-Tissue and Cancer Panel that comprises approximately 400 genes. In this section, we present the analysis of the Visium TNBC data. The analyses of the Visium PLC and Xenium PAC data are provided in Supplementary Sections S6 and S7 \citep{wu2025supp}, respectively.}

The TNBC dataset is obtained from a cohort study of TNBC tumors \citep{bassiouni2023spatial}, where a total of 28 tissue sections representing 14 primary TNBC tumors are subject to spatial transcriptomics using the 10x Genomics Visium platform. The raw expression data and the tissue information are available at the Gene Expression Omnibus (GEO) under record GSE210616. Here, we focus on the first tissue section from patient ID 1, which contains 1,109 spots and 36,601 genes. Following the published studies \citep{charitakis2023disparities,yan2024bayesian}, to improve efficiency and stability, we first remove the genes expressed in fewer than 5 spots and the spots with fewer than 500 expressed genes, resulting in a total of 1,108 spots and 17,781 genes, and then select the top 5,000 highly variable genes for downstream analysis. After preprocessing, the median value of expressed counts across spots is 7,036, and the proportion of zero expressions is about 68.89\%.

We consider the protein-protein interaction (PPI) network obtained from the STRING database \citep{szklarczyk2023string}. Integrating the PPI network has been widely adopted in recent breast cancer analysis, which provides a powerful complement for a mechanistic understanding of the genomic alterations related to breast cancer \citep{kim2021protein}. In particular, it has been demonstrated in published studies \citep{pranavathiyani2019integrated,chen2021establishing} that the oncogenes or prognostic genes associated with breast cancer are typically highly interconnected within distinct modules, underscoring the importance of considering the connections within networks in SV gene identification. Matching with this PPI network leads to 1,687 connected genes involved in 164 disconnected sub-networks, while the remaining 3,313 genes are singleton nodes.

For cell-type deconvolution, we adopt the Redeconve algorithm developed in \cite{zhou2023spatial}, which has been proven to have higher accuracy, robustness, and efficacy. More importantly, the superior performance of Redeconve on human breast cancer has been verified based on biological ground truths. To alleviate the impact of rare cell types and to improve interpretability, we follow \cite{yu2022identification} and remove the cell types whose 90 percentile of proportions across spots is less than 0.1, resulting in a total of seven major cell types. The pie charts of the distributions of different cell types across spots are shown in Figure \ref{fig:spatial}(A). It can be seen that most spots consist of cancer and normal epithelial cells, while some spots located at the middle and lower right parts of the tissue section are mostly CAFs.

\begin{figure}[htbp]
\centering
\includegraphics[width=5in]{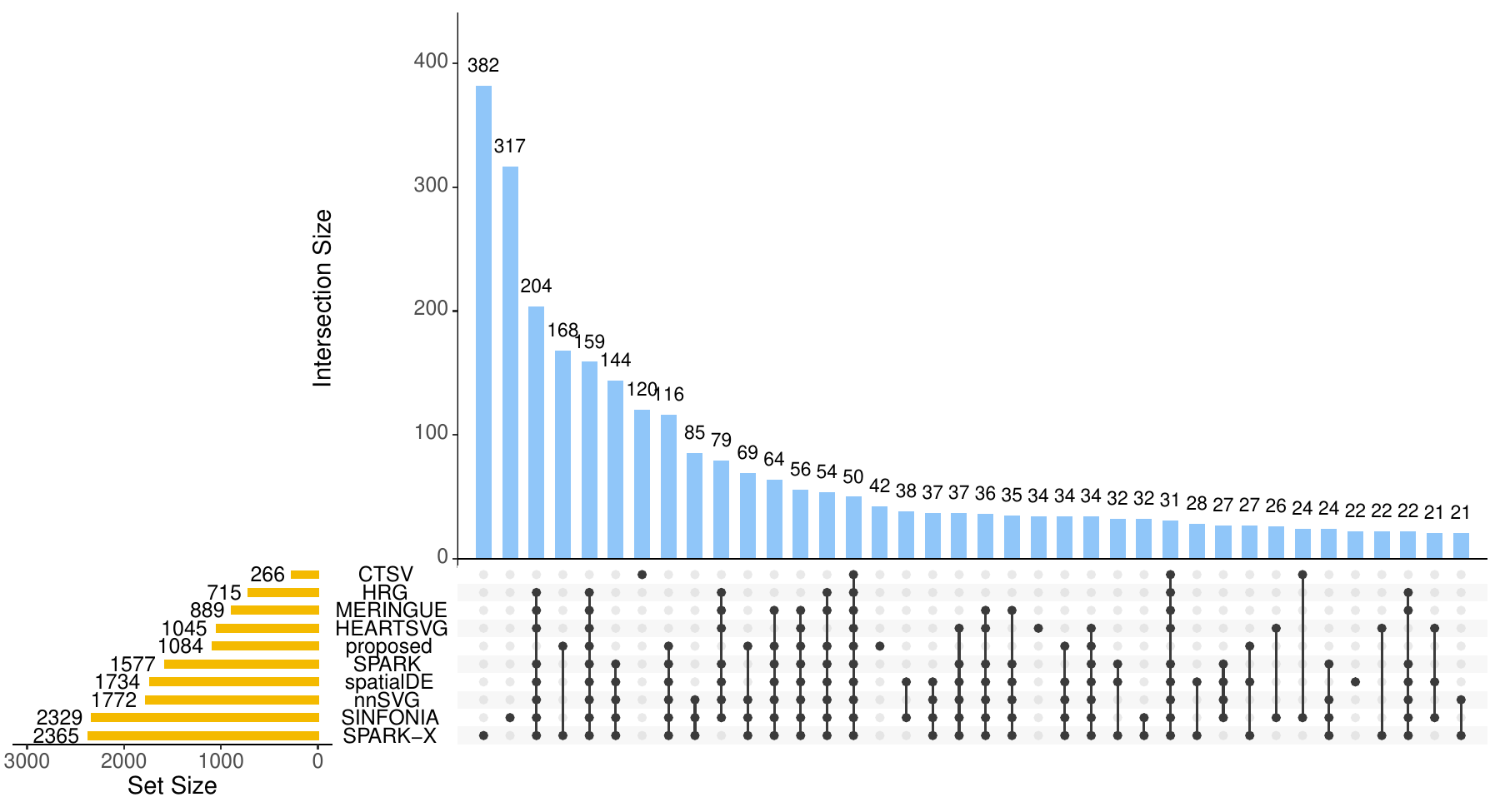}
\caption{Upset plot of the numbers of SV genes identified by different approaches and their overlaps for Visium TNBC data.}\label{fig:upset}
\end{figure}

\begin{figure}[htbp]
\centering
\includegraphics[width=4in]{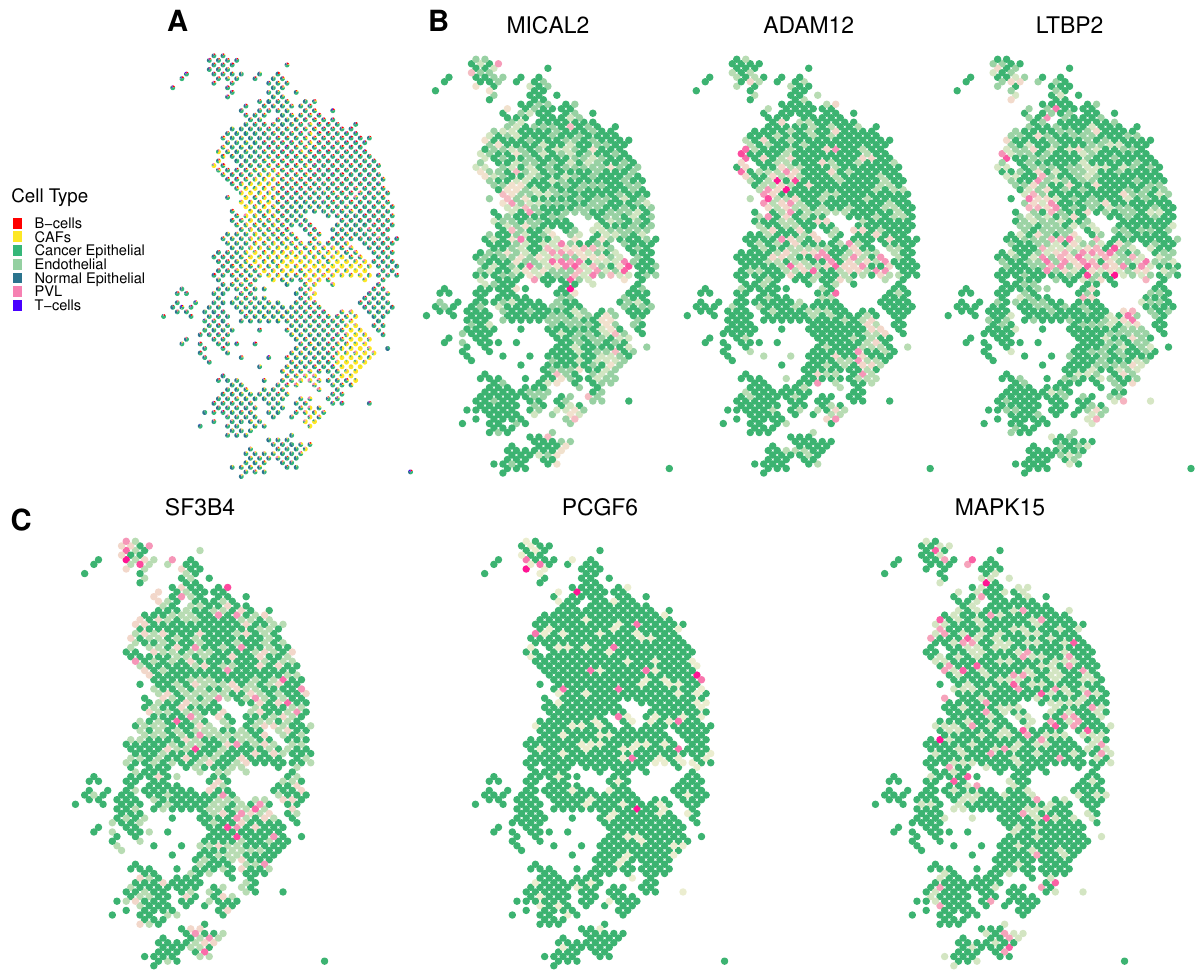}
\caption{(A) Pie charts of the distributions of cell types across spots for Visium TNBC data. {(B) Spatial expression patterns of three representative SV genes identified by all of SPARK-X, SINFONIA, nnSVG, spatialDE, SPARK, HEARTSVG, MERINGUE, and HRG. (C) Spatial expression patterns of three representative SV genes only identified by the proposed approach.}}\label{fig:spatial}
\end{figure}

\begin{figure}[htbp]
\centering
\includegraphics[width=4in]{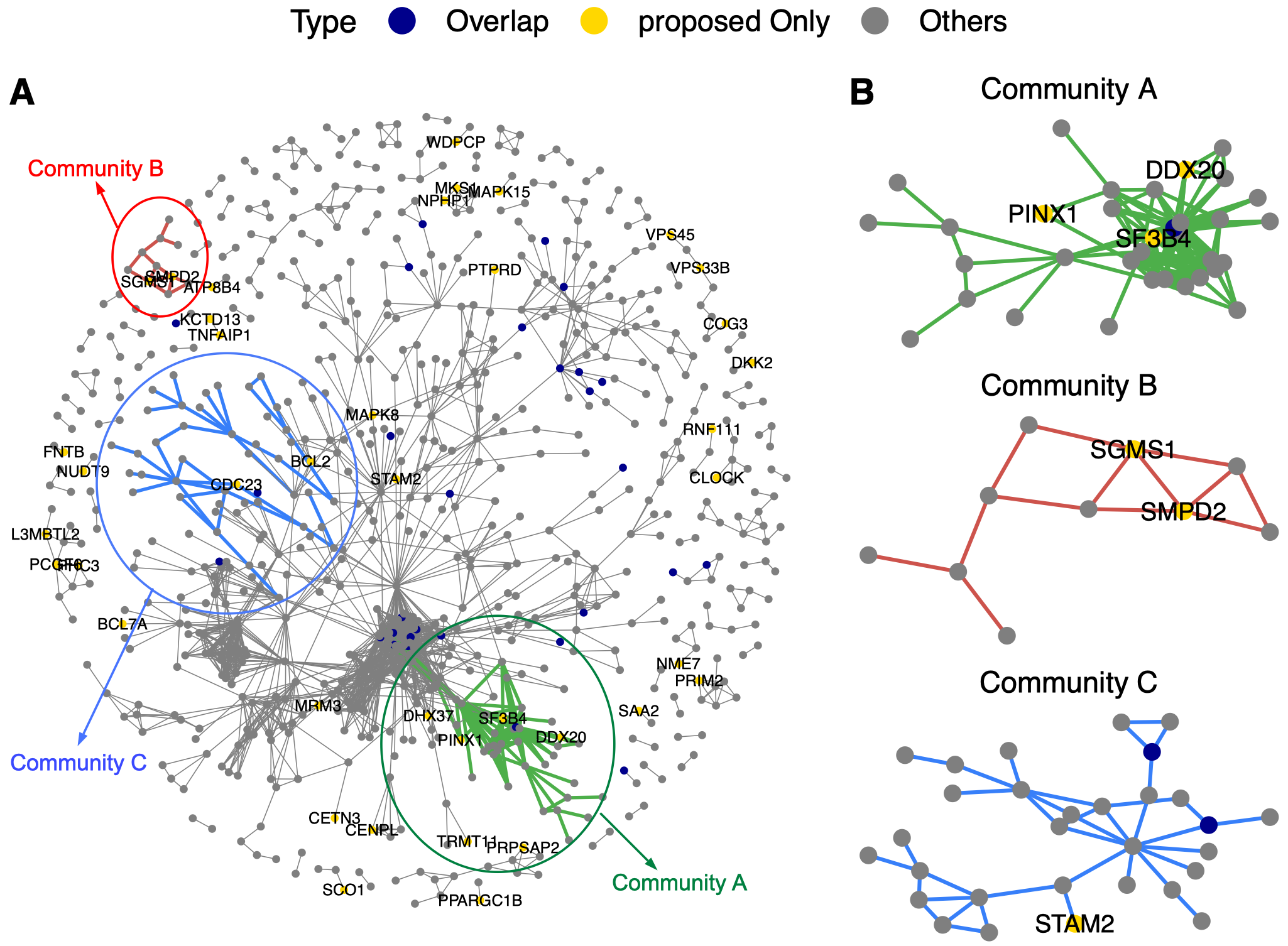}
\caption{(A) Network of 782 connected SV genes identified by the proposed approach for Visium TNBC data. (B) Network of three representative detected communities. }\label{fig:3module}
\end{figure}

The proposed approach identifies 1,084 SV genes. Analysis is also conducted using the alternatives. The upset plot, which provides the numbers of the SV genes identified by different approaches as well as their overlaps, is shown in Figure \ref{fig:upset}.  Specifically, SPARK-X identifies the largest number of SV genes (2,365), while CTSV detects the smallest (266). A total of 50 overlapping genes are identified by all 10 approaches. For SPARK-X, SINFONIA, nnSVG, spatialDE, SPARK, HEARTSVG, MERINGUE, and HRG, there are more overlapping SV genes (204), mainly due to the potential cell-type diversity confounding that is not well accommodated by these approaches. To provide a more intuitive illustration, among these 204 genes, we take three representative genes \textit{MICAL2}, \textit{ADAM12}, and \textit{LTBP2} and present their spatial expression patterns in Figure \ref{fig:spatial}(B). All three genes are observed to have significant cell type diversity.  Moreover, among 1,084 genes identified by the proposed approach, 782 are connected in the PPI network, of which the graph representation is shown in Figure \ref{fig:3module}(A). 47 genes (all connected) are only selected by the proposed approach. Among them, the expression patterns of three representative genes \textit{SF3B4}, \textit{PCGF6}, and \textit{MAPK15} are presented in Figure \ref{fig:spatial}(C), where clear spatial variability (not driven by cell type diversity) is observed.

To provide additional comparisons for the SV genes identified by different approaches, we obtain the lists of breast cancer driver genes from the Candidate Cancer Gene Database (CCGD), IntOGen, and DriverDB databases, which contain 579, 290, and 222 genes, respectively, and examine whether these genes are identified as SV. The inclusion rates per thousand are (20.30, 11.70, 23.06) for the proposed approach, (13.26, 7.50, 14.42) for spatialDE, (14.58, 7.61, 13.95) for SPARK, (19.45, 5.50, 14.80) for SPARK-X, {(8.39, 12.59, 15.38) for HRG}, (0.00, 0.00, 7.52) for CTSV, (12.37, 9.00, 13.50) for MERINGUE, {(13.31, 5.58, 13.31) for SINFONIA, (18.62, 6.21, 16.37) for nnSVG,} and (12.44, 8.61, 6.70) for HEARTSVG, respectively. This may support the higher power of the proposed approach to some extent.

We then conduct community detection with the 782 connected SV genes and show three representative communities in Figure \ref{fig:3module}(B), where the overlapping genes identified by all ten approaches (Overlap), genes only identified by the proposed approach (proposed Only), and the remaining genes (Others) are represented by different colors. It is observed that most overlapping genes are hubs with larger values of degrees. Some of the proposed Only genes are connected to the hub genes, for example, \textit{SF3B4} and \textit{PINX1} in community A, which may exhibit weak spatial patterns but are involved in the same functional organization or regulation mechanism with the hub ones. Some other genes seem to serve as bridge linkage genes with high betweenness centrality. Examples include \textit{SGMS1} in community B and \textit{STAM2} in community C, which expedites the formulation of the complete gene regulatory networks.

Then, we conduct Gene Ontology (GO) enrichment analysis. Specifically, for the aforementioned three communities, we list the corresponding top five significant GO terms in Table \ref{tab:go}, which have important biological implications for breast cancer.  For example, the genes involved in community A are mainly enriched in RNA splicing, and the relationships between RNA splicing and a variety of cancers have long been recognized and widely utilized in cancer diagnosis and therapy \citep{stanley2022dysregulation,yamauchi2022aberrant}. For community B, sphingolipid metabolism has been shown to be essential for breast cancer progression \citep{corsetto2023critical}, while the balance of ceramide metabolism has been confirmed as a critical step in breast cancer development and has been long adopted as a targeted pathway to induce apoptosis in breast cancer cell lines \citep{vethakanraj2015targeting}. In community C, cell cycle phase transition has been confirmed to play a critical role in breast cancer progression, as supported by previous studies \citep{kashyap2021oncogenic}.

\begin{figure}[htb]
\centering
\includegraphics[width=\textwidth]{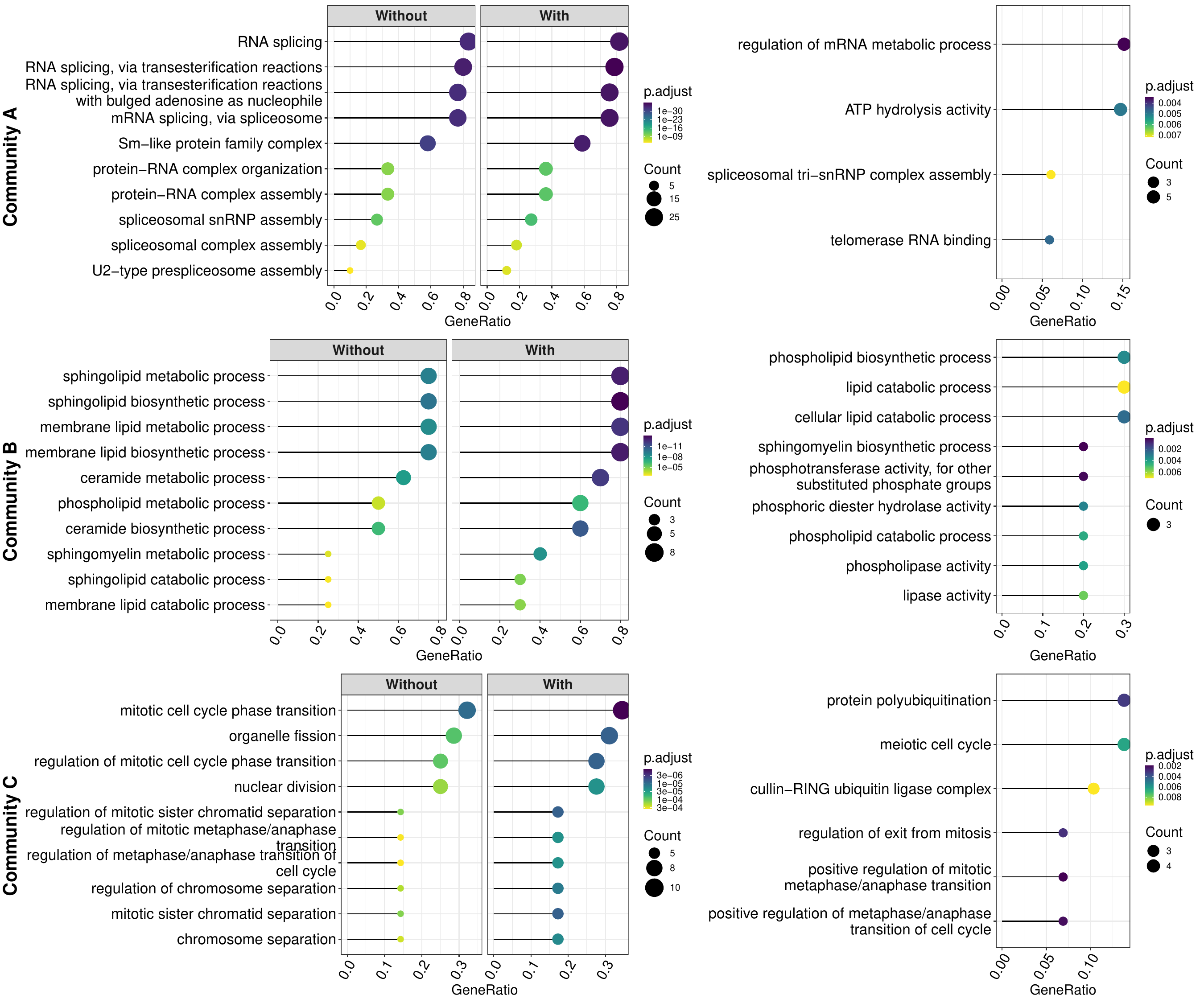}
\caption{Lollipop plots of GO enrichment analysis for the three communities detected in Visium TNBC data. \textit{Count} represents the number of the input genes that are annotated to the specific GO term, and \textit{GeneRatio} denotes the proportion of the \textit{Count} divided by the total number of input genes. Left: Comparison results of the analysis without and with the inclusion of proposed Only genes. Right: Newly detected significant GO terms with the proposed Only genes included.} \label{fig:GO}
\end{figure}

We further take a closer look at the proposed Only genes. Specifically, we conduct an GO enrichment analysis for each community again, but with the proposed Only genes eliminated. The comparison results are shown in Figure \ref{fig:GO} (Left), where the top ten significant GO terms with the proposed Only genes included are considered. It can be seen that when the proposed Only genes are included, some more significant GO terms are found. Moreover, the incorporation of the proposed Only genes contributes to some newly detected GO terms (shown in Figure \ref{fig:GO}, Right), such as regulation of mRNA metabolic process, which has been confirmed to be critical in targeted therapy, chemotherapy and immunotherapy in breast cancer \citep{xu2024metabolism}. These evidences suggest the importance of the proposed Only genes in the PPI network and their biological implications.

\begin{table}[htb]
\caption{Top five significant GO terms associated with the three representative communities. \label{tab:go}}
\centering
\begin{tabular}{ccl}
\hline
\multicolumn{1}{l}{ID} & \multicolumn{1}{l}{adjusted P value} & Description              \\ \hline
\multicolumn{3}{c}{\textit{Community A}}                                                                                                                      \\
GO:0000375            & 1.19$\times 10 ^{-37}$                             & RNA splicing, via transesterification reactions                                                                      \\
GO:0008380           & 4.22$\times 10 ^{-36}$                            & RNA splicing                                                                      \\
GO:0000377             & 4.22$\times 10 ^{-36}$                             & \begin{tabular}[c]{@{}l@{}}RNA splicing, via transesterification reactions  with \\ bulged adenosine as nucleophile\end{tabular} \\
GO:0000398            & 4.22$\times 10 ^{-36}$                             & mRNA splicing, via spliceosome                                               \\
GO:0000387            & 8.16$\times 10 ^{-15}$                             & spliceosomal snRNP assembly                                                 \\ \hline
\multicolumn{3}{c}{\textit{Community B}}                                                                                                                      \\
GO:0030148           & 1.18$\times 10 ^{-14}$                             &  sphingolipid biosynthetic process \\
GO:0046467            & 7.13$\times 10 ^{-14}$                             & membrane lipid biosynthetic process                                                     \\
GO:0006665             & 9.42$\times 10 ^{-14}$                            & sphingolipid metabolic process                              \\
GO:0006643           & 5.62$\times 10 ^{-13}$                             & membrane lipid metabolic process                                                                         \\
GO:0006672            & 9.39$\times 10 ^{-13}$                             & ceramide metabolic process       
               \\ \hline
\multicolumn{3}{c}{\textit{Community C}}                                                                                                                      \\
GO:0000152             & 3.16$\times 10 ^{-7}$                             & nuclear ubiquitin ligase complex                                                          \\
GO:0044772             & 1.30$\times 10 ^{-6}$                             & mitotic cell cycle phase transition                                                                   \\
GO:0051438            & 1.96$\times 10 ^{-6}$                           & regulation of ubiquitin-protein transferase activity                                   \\
GO:0010965         & 7.16$\times 10 ^{-6}$                           & regulation of mitotic sister chromatid separation                                              \\
GO:0048285          & 7.16$\times 10 ^{-6}$                             & organelle fission organization            
\\ \hline

\end{tabular}
\end{table}

\section{Discussion}

In this article, we have presented a novel Bayesian regularization approach for the joint identification of SV genes, with the network structure among genes accommodated. The proposed approach has two main advantages. First, attributing to the Bayesian framework, it can automatically incorporate the network structure through the Gaussian Graph Laplacian priors. Compared to most of the existing methods, the utilization of such network information provides more opportunities to search for more biologically sensible SV genes. Second, to tackle the confounding effects of cell type mixtures within spots, cell-type-specific parameters have been introduced to model the variations induced by diverse latent cellular compositions, which are not well accommodated by most existing methods. The extensive simulation studies have revealed that the proposed approach can achieve better performance. The biological implications of the findings from the application to the Visium TNBC, Visium PLC, and Xenium PAC datasets have further supported the utility of the proposed approach.

{The proposed approach has used a zero-inflated model to handle the excess zeros often seen in ST data, proving its necessity and effectiveness in our simulations. However, we recognize prior research \citep{zhao2022modeling} indicating that modeling zero inflation is not always essential. Inspired by this, a valuable future direction is to systematically explore alternative zero-inflation handling strategies, balancing modeling accuracy and computational efficiency. We have identified SV genes by checking if their corresponding spatial coordinate-specific regression coefficients are nonzero. Though our approach uses regularization to induce sparsity, we acknowledge the value of inference-based alternatives relying on statistical testing, which we will explore in the future. We have considered preset spatial patterns and introduced a parametric model. It would be interesting to expand the proposed framework to accommodate unknown spatial patterns using nonparametric methods. { Moreover, Su et al. (2025) noted that mean-value-based strategies may suffer from a lack of rotation invariance, wherein rotations alter spatial coordinates and consequently influence effect estimation and SV gene identification. We have carried out an exploratory analysis in Supplementary Section S8 \citep{wu2025supp} to evaluate the robustness of our method against spatial coordinate rotations. The results indicate that incorporating prior network information can partially mitigate the sensitivity of our approach to rotation. However, inconsistent SV gene detection persists across different rotation scales. Further investigation integrating rotation-invariant strategies appears promising, such as constructing predictors based on relative distances as suggested in \cite{su2025rotation}.}

In our real-data analysis, we have employed Redeconve \citep{zhou2023spatial} to estimate cell-type proportions before detecting SV genes. This is a widely-used strategy in existing studies \citep{zhu2021spark, yu2022identification,cable2022cell} and has demonstrated satisfactory performance with biological significance. Although Redeconve relies on expression data, the additional network information we have incorporated can mitigate this dependency to some extent. Furthermore, our simulation studies (Section \ref{sec:sim_noise}) indicate that the proposed method is quite robust to noise in cell-type proportion estimates. Exploring alternative cell-type estimation approaches is a potential avenue for future work. In addition to cellular composition information, we utilized PPI information for network construction. Other resources, such as the Kyoto Encyclopedia of Genes and Genomes (KEGG) and High-quality INTeractomes (HINT) databases, could also be used. Additionally, as an alternative, the network could be directly estimated from the analyzed data. A literature search and GO enrichment analysis have demonstrated the significant implications of the findings. A more definitive confirmation from functional validations may be needed.}

\begin{acks}[Acknowledgments]
The authors would like to thank the anonymous referees, an Associate
Editor, and the Editor for their constructive comments which led to a significant improvement of this article. Mingcong Wu and Yang Li are joint first authors, and Mengyun Wu is the corresponding author.
\end{acks}
\begin{funding}
This research was supported by the National Natural Science Foundation of China (12071273); Shanghai Rising-Star Program (22QA1403500); Shanghai Science and Technology Development Funds (23JC1402100); Shanghai Research Center for Data Science and Decision Technology; Public Health $\&$ Disease Control and Prevention, the MOE Project of Key Research Institute of Humanities and Social Sciences (22JJD910001); National Institutes of Health (CA204120 and CA121974); and National Science Foundation (2209685).
\end{funding}

\begin{supplement}
\stitle{Supplement.pdf}
\sdescription{ Supplement to ``Joint identification of spatially variable genes via a network-assisted Bayesian regularization approach'', including the details of the MCMC algorithm and additional settings and results of simulation studies and real data analysis.}
\end{supplement}
\begin{supplement}
\stitle{SV-network.zip}
\sdescription{The \textsf{R} package \texttt{SV-network} that implements the proposed approach are
provided as a supplement and can also be found at \href{https://github.com/mengyunwu2020/SV-network}{https://github.com/mengyunwu2020/SV-network}.}
\end{supplement}


\bibliographystyle{imsart-nameyear} 
\bibliography{main}       

\begin{thebibliography}{54}

\bibitem[\protect\citeauthoryear{Barrio-Hernandez
  et~al.}{2023}]{barrio2023network}
\begin{barticle}[author]
\bauthor{\bsnm{Barrio-Hernandez},~\bfnm{Inigo}\binits{I.}},
  \bauthor{\bsnm{Schwartzentruber},~\bfnm{Jeremy}\binits{J.}},
  \bauthor{\bsnm{Shrivastava},~\bfnm{Anjali}\binits{A.}},
  \bauthor{\bsnm{Del-Toro},~\bfnm{Noemi}\binits{N.}},
  \bauthor{\bsnm{Gonzalez},~\bfnm{Asier}\binits{A.}},
  \bauthor{\bsnm{Zhang},~\bfnm{Qian}\binits{Q.}},
  \bauthor{\bsnm{Mountjoy},~\bfnm{Edward}\binits{E.}},
  \bauthor{\bsnm{Suveges},~\bfnm{Daniel}\binits{D.}},
  \bauthor{\bsnm{Ochoa},~\bfnm{David}\binits{D.}},
  \bauthor{\bsnm{Ghoussaini},~\bfnm{Maya}\binits{M.}} \betal{et~al.}
(\byear{2023}).
\btitle{Network expansion of genetic associations defines a pleiotropy map of
  human cell biology}.
\bjournal{Nature Genetics}
\bvolume{55}
\bpages{389--398}.
\end{barticle}
\endbibitem

\bibitem[\protect\citeauthoryear{Bassiouni et~al.}{2023}]{bassiouni2023spatial}
\begin{barticle}[author]
\bauthor{\bsnm{Bassiouni},~\bfnm{Rania}\binits{R.}},
  \bauthor{\bsnm{Idowu},~\bfnm{Michael~O}\binits{M.~O.}},
  \bauthor{\bsnm{Gibbs},~\bfnm{Lee~D}\binits{L.~D.}},
  \bauthor{\bsnm{Robila},~\bfnm{Valentina}\binits{V.}},
  \bauthor{\bsnm{Grizzard},~\bfnm{Pamela~J}\binits{P.~J.}},
  \bauthor{\bsnm{Webb},~\bfnm{Michelle~G}\binits{M.~G.}},
  \bauthor{\bsnm{Song},~\bfnm{Jiarong}\binits{J.}},
  \bauthor{\bsnm{Noriega},~\bfnm{Ashley}\binits{A.}},
  \bauthor{\bsnm{Craig},~\bfnm{David~W}\binits{D.~W.}} \AND
  \bauthor{\bsnm{Carpten},~\bfnm{John~D}\binits{J.~D.}}
(\byear{2023}).
\btitle{Spatial transcriptomic analysis of a diverse patient cohort reveals a
  conserved architecture in triple-negative breast cancer}.
\bjournal{Cancer Research}
\bvolume{83}
\bpages{34--48}.
\end{barticle}
\endbibitem

\bibitem[\protect\citeauthoryear{BinTayyash et~al.}{2021}]{bintayyash2021non}
\begin{barticle}[author]
\bauthor{\bsnm{BinTayyash},~\bfnm{Nuha}\binits{N.}},
  \bauthor{\bsnm{Georgaka},~\bfnm{Sokratia}\binits{S.}},
  \bauthor{\bsnm{John},~\bfnm{ST}\binits{S.}},
  \bauthor{\bsnm{Ahmed},~\bfnm{Sumon}\binits{S.}},
  \bauthor{\bsnm{Boukouvalas},~\bfnm{Alexis}\binits{A.}},
  \bauthor{\bsnm{Hensman},~\bfnm{James}\binits{J.}} \AND
  \bauthor{\bsnm{Rattray},~\bfnm{Magnus}\binits{M.}}
(\byear{2021}).
\btitle{Non-parametric modelling of temporal and spatial counts data from
  RNA-seq experiments}.
\bjournal{Bioinformatics}
\bvolume{37}
\bpages{3788--3795}.
\end{barticle}
\endbibitem

\bibitem[\protect\citeauthoryear{Cable et~al.}{2022a}]{cable2022robust}
\begin{barticle}[author]
\bauthor{\bsnm{Cable},~\bfnm{Dylan~M}\binits{D.~M.}},
  \bauthor{\bsnm{Murray},~\bfnm{Evan}\binits{E.}},
  \bauthor{\bsnm{Zou},~\bfnm{Luli~S}\binits{L.~S.}},
  \bauthor{\bsnm{Goeva},~\bfnm{Aleksandrina}\binits{A.}},
  \bauthor{\bsnm{Macosko},~\bfnm{Evan~Z}\binits{E.~Z.}},
  \bauthor{\bsnm{Chen},~\bfnm{Fei}\binits{F.}} \AND
  \bauthor{\bsnm{Irizarry},~\bfnm{Rafael~A}\binits{R.~A.}}
(\byear{2022}a).
\btitle{Robust decomposition of cell type mixtures in spatial transcriptomics}.
\bjournal{Nature Biotechnology}
\bvolume{40}
\bpages{517--526}.
\end{barticle}
\endbibitem

\bibitem[\protect\citeauthoryear{Cable et~al.}{2022b}]{cable2022cell}
\begin{barticle}[author]
\bauthor{\bsnm{Cable},~\bfnm{Dylan~M}\binits{D.~M.}},
  \bauthor{\bsnm{Murray},~\bfnm{Evan}\binits{E.}},
  \bauthor{\bsnm{Shanmugam},~\bfnm{Vignesh}\binits{V.}},
  \bauthor{\bsnm{Zhang},~\bfnm{Simon}\binits{S.}},
  \bauthor{\bsnm{Zou},~\bfnm{Luli~S}\binits{L.~S.}},
  \bauthor{\bsnm{Diao},~\bfnm{Michael}\binits{M.}},
  \bauthor{\bsnm{Chen},~\bfnm{Haiqi}\binits{H.}},
  \bauthor{\bsnm{Macosko},~\bfnm{Evan~Z}\binits{E.~Z.}},
  \bauthor{\bsnm{Irizarry},~\bfnm{Rafael~A}\binits{R.~A.}} \AND
  \bauthor{\bsnm{Chen},~\bfnm{Fei}\binits{F.}}
(\byear{2022}b).
\btitle{Cell type-specific inference of differential expression in spatial
  transcriptomics}.
\bjournal{Nature Methods}
\bvolume{19}
\bpages{1076--1087}.
\end{barticle}
\endbibitem

\bibitem[\protect\citeauthoryear{Cai, Kang and Yu}{2020}]{cai2020bayesian}
\begin{barticle}[author]
\bauthor{\bsnm{Cai},~\bfnm{Qingpo}\binits{Q.}},
  \bauthor{\bsnm{Kang},~\bfnm{Jian}\binits{J.}} \AND
  \bauthor{\bsnm{Yu},~\bfnm{Tianwei}\binits{T.}}
(\byear{2020}).
\btitle{{Bayesian network marker selection via the thresholded graph Laplacian
  Gaussian prior}}.
\bjournal{Bayesian Analysis}
\bvolume{15}
\bpages{79}.
\end{barticle}
\endbibitem

\bibitem[\protect\citeauthoryear{Chakraborty and
  Lozano}{2019}]{chakraborty2019graph}
\begin{barticle}[author]
\bauthor{\bsnm{Chakraborty},~\bfnm{Sounak}\binits{S.}} \AND
  \bauthor{\bsnm{Lozano},~\bfnm{Aurelie~C}\binits{A.~C.}}
(\byear{2019}).
\btitle{{A graph Laplacian prior for Bayesian variable selection and
  grouping}}.
\bjournal{Computational Statistics \& Data Analysis}
\bvolume{136}
\bpages{72--91}.
\end{barticle}
\endbibitem

\bibitem[\protect\citeauthoryear{Charitakis
  et~al.}{2023}]{charitakis2023disparities}
\begin{barticle}[author]
\bauthor{\bsnm{Charitakis},~\bfnm{Natalie}\binits{N.}},
  \bauthor{\bsnm{Salim},~\bfnm{Agus}\binits{A.}},
  \bauthor{\bsnm{Piers},~\bfnm{Adam~T}\binits{A.~T.}},
  \bauthor{\bsnm{Watt},~\bfnm{Kevin~I}\binits{K.~I.}},
  \bauthor{\bsnm{Porrello},~\bfnm{Enzo~R}\binits{E.~R.}},
  \bauthor{\bsnm{Elliott},~\bfnm{David~A}\binits{D.~A.}} \AND
  \bauthor{\bsnm{Ramialison},~\bfnm{Mirana}\binits{M.}}
(\byear{2023}).
\btitle{Disparities in spatially variable gene calling highlight the need for
  benchmarking spatial transcriptomics methods}.
\bjournal{Genome Biology}
\bvolume{24}
\bpages{209}.
\end{barticle}
\endbibitem

\bibitem[\protect\citeauthoryear{Chaudhary and
  Kim}{2021}]{chaudhary2021insight}
\begin{barticle}[author]
\bauthor{\bsnm{Chaudhary},~\bfnm{Preeti~Kumari}\binits{P.~K.}} \AND
  \bauthor{\bsnm{Kim},~\bfnm{Soochong}\binits{S.}}
(\byear{2021}).
\btitle{An insight into GPCR and G-proteins as cancer drivers}.
\bjournal{Cells}
\bvolume{10}
\bpages{3288}.
\end{barticle}
\endbibitem

\bibitem[\protect\citeauthoryear{Chen, Verbeek and
  Wolstencroft}{2021}]{chen2021establishing}
\begin{barticle}[author]
\bauthor{\bsnm{Chen},~\bfnm{Yi}\binits{Y.}},
  \bauthor{\bsnm{Verbeek},~\bfnm{Fons~J}\binits{F.~J.}} \AND
  \bauthor{\bsnm{Wolstencroft},~\bfnm{Katherine}\binits{K.}}
(\byear{2021}).
\btitle{Establishing a consensus for the hallmarks of cancer based on gene
  ontology and pathway annotations}.
\bjournal{BMC Bioinformatics}
\bvolume{22}
\bpages{1--20}.
\end{barticle}
\endbibitem

\bibitem[\protect\citeauthoryear{Chen et~al.}{2021}]{chen2021cell}
\begin{barticle}[author]
\bauthor{\bsnm{Chen},~\bfnm{Xin}\binits{X.}},
  \bauthor{\bsnm{Zeh},~\bfnm{Herbert~J}\binits{H.~J.}},
  \bauthor{\bsnm{Kang},~\bfnm{Rui}\binits{R.}},
  \bauthor{\bsnm{Kroemer},~\bfnm{Guido}\binits{G.}} \AND
  \bauthor{\bsnm{Tang},~\bfnm{Daolin}\binits{D.}}
(\byear{2021}).
\btitle{Cell death in pancreatic cancer: from pathogenesis to therapy}.
\bjournal{Nature Reviews Gastroenterology \& Hepatology}
\bvolume{18}
\bpages{804--823}.
\end{barticle}
\endbibitem

\bibitem[\protect\citeauthoryear{Corsetto et~al.}{2023}]{corsetto2023critical}
\begin{barticle}[author]
\bauthor{\bsnm{Corsetto},~\bfnm{Paola~Antonia}\binits{P.~A.}},
  \bauthor{\bsnm{Zava},~\bfnm{Stefania}\binits{S.}},
  \bauthor{\bsnm{Rizzo},~\bfnm{Angela~Maria}\binits{A.~M.}} \AND
  \bauthor{\bsnm{Colombo},~\bfnm{Irma}\binits{I.}}
(\byear{2023}).
\btitle{The critical impact of sphingolipid metabolism in breast cancer
  progression and drug response}.
\bjournal{International Journal of Molecular Sciences}
\bvolume{24}
\bpages{2107}.
\end{barticle}
\endbibitem

\bibitem[\protect\citeauthoryear{Elyanow et~al.}{2020}]{elyanow2020netnmf}
\begin{barticle}[author]
\bauthor{\bsnm{Elyanow},~\bfnm{Rebecca}\binits{R.}},
  \bauthor{\bsnm{Dumitrascu},~\bfnm{Bianca}\binits{B.}},
  \bauthor{\bsnm{Engelhardt},~\bfnm{Barbara~E}\binits{B.~E.}} \AND
  \bauthor{\bsnm{Raphael},~\bfnm{Benjamin~J}\binits{B.~J.}}
(\byear{2020}).
\btitle{net{NMF}-sc: leveraging gene--gene interactions for imputation and
  dimensionality reduction in single-cell expression analysis}.
\bjournal{Genome Research}
\bvolume{30}
\bpages{195--204}.
\end{barticle}
\endbibitem

\bibitem[\protect\citeauthoryear{Jiang et~al.}{2023}]{jiang2023sinfonia}
\begin{barticle}[author]
\bauthor{\bsnm{Jiang},~\bfnm{Rui}\binits{R.}},
  \bauthor{\bsnm{Li},~\bfnm{Zhen}\binits{Z.}},
  \bauthor{\bsnm{Jia},~\bfnm{Yuhang}\binits{Y.}},
  \bauthor{\bsnm{Li},~\bfnm{Siyu}\binits{S.}} \AND
  \bauthor{\bsnm{Chen},~\bfnm{Shengquan}\binits{S.}}
(\byear{2023}).
\btitle{SINFONIA: scalable identification of spatially variable genes for
  deciphering spatial domains}.
\bjournal{Cells}
\bvolume{12}
\bpages{604}.
\end{barticle}
\endbibitem

\bibitem[\protect\citeauthoryear{Kashyap et~al.}{2021}]{kashyap2021oncogenic}
\begin{barticle}[author]
\bauthor{\bsnm{Kashyap},~\bfnm{Dharambir}\binits{D.}},
  \bauthor{\bsnm{Garg},~\bfnm{Vivek~Kumar}\binits{V.~K.}},
  \bauthor{\bsnm{Sandberg},~\bfnm{Elise~N}\binits{E.~N.}},
  \bauthor{\bsnm{Goel},~\bfnm{Neelam}\binits{N.}} \AND
  \bauthor{\bsnm{Bishayee},~\bfnm{Anupam}\binits{A.}}
(\byear{2021}).
\btitle{Oncogenic and tumor suppressive components of the cell cycle in breast
  cancer progression and prognosis}.
\bjournal{Pharmaceutics}
\bvolume{13}
\bpages{569}.
\end{barticle}
\endbibitem

\bibitem[\protect\citeauthoryear{Kim et~al.}{2021}]{kim2021protein}
\begin{barticle}[author]
\bauthor{\bsnm{Kim},~\bfnm{Minkyu}\binits{M.}},
  \bauthor{\bsnm{Park},~\bfnm{Jisoo}\binits{J.}},
  \bauthor{\bsnm{Bouhaddou},~\bfnm{Mehdi}\binits{M.}},
  \bauthor{\bsnm{Kim},~\bfnm{Kyumin}\binits{K.}},
  \bauthor{\bsnm{Rojc},~\bfnm{Ajda}\binits{A.}},
  \bauthor{\bsnm{Modak},~\bfnm{Maya}\binits{M.}},
  \bauthor{\bsnm{Soucheray},~\bfnm{Margaret}\binits{M.}},
  \bauthor{\bsnm{McGregor},~\bfnm{Michael~J}\binits{M.~J.}},
  \bauthor{\bsnm{O’Leary},~\bfnm{Patrick}\binits{P.}},
  \bauthor{\bsnm{Wolf},~\bfnm{Denise}\binits{D.}} \betal{et~al.}
(\byear{2021}).
\btitle{A protein interaction landscape of breast cancer}.
\bjournal{Science}
\bvolume{374}
\bpages{eabf3066}.
\end{barticle}
\endbibitem

\bibitem[\protect\citeauthoryear{Li and Li}{2010}]{li2010variable}
\begin{barticle}[author]
\bauthor{\bsnm{Li},~\bfnm{Caiyan}\binits{C.}} \AND
  \bauthor{\bsnm{Li},~\bfnm{Hongzhe}\binits{H.}}
(\byear{2010}).
\btitle{Variable selection and regression analysis for graph-structured
  covariates with an application to genomics}.
\bjournal{The Annals of Applied Statistics}
\bvolume{4}
\bpages{1498}.
\end{barticle}
\endbibitem

\bibitem[\protect\citeauthoryear{Li et~al.}{2021}]{li2021bayesian}
\begin{barticle}[author]
\bauthor{\bsnm{Li},~\bfnm{Qiwei}\binits{Q.}},
  \bauthor{\bsnm{Zhang},~\bfnm{Minzhe}\binits{M.}},
  \bauthor{\bsnm{Xie},~\bfnm{Yang}\binits{Y.}} \AND
  \bauthor{\bsnm{Xiao},~\bfnm{Guanghua}\binits{G.}}
(\byear{2021}).
\btitle{{Bayesian modeling of spatial molecular profiling data via Gaussian
  process}}.
\bjournal{Bioinformatics}
\bvolume{37}
\bpages{4129--4136}.
\end{barticle}
\endbibitem

\bibitem[\protect\citeauthoryear{Li et~al.}{2022}]{li2022g}
\begin{barticle}[author]
\bauthor{\bsnm{Li},~\bfnm{Nan}\binits{N.}},
  \bauthor{\bsnm{Shan},~\bfnm{Shan}\binits{S.}},
  \bauthor{\bsnm{Li},~\bfnm{Xiuqin}\binits{X.}},
  \bauthor{\bsnm{Chen},~\bfnm{Tingting}\binits{T.}},
  \bauthor{\bsnm{Qi},~\bfnm{Meng}\binits{M.}},
  \bauthor{\bsnm{Zhang},~\bfnm{Shengnan}\binits{S.}},
  \bauthor{\bsnm{Wang},~\bfnm{Ziying}\binits{Z.}},
  \bauthor{\bsnm{Zhang},~\bfnm{Lingling}\binits{L.}},
  \bauthor{\bsnm{Wei},~\bfnm{Wei}\binits{W.}} \AND
  \bauthor{\bsnm{Sun},~\bfnm{Wuyi}\binits{W.}}
(\byear{2022}).
\btitle{G Protein-coupled receptor kinase 2 as novel therapeutic target in
  fibrotic diseases}.
\bjournal{Frontiers in Immunology}
\bvolume{12}
\bpages{822345}.
\end{barticle}
\endbibitem

\bibitem[\protect\citeauthoryear{Lin et~al.}{2021}]{lin2021trpm2}
\begin{barticle}[author]
\bauthor{\bsnm{Lin},~\bfnm{Rui}\binits{R.}},
  \bauthor{\bsnm{Bao},~\bfnm{Xunxia}\binits{X.}},
  \bauthor{\bsnm{Wang},~\bfnm{Hui}\binits{H.}},
  \bauthor{\bsnm{Zhu},~\bfnm{Sibo}\binits{S.}},
  \bauthor{\bsnm{Liu},~\bfnm{Zhongyan}\binits{Z.}},
  \bauthor{\bsnm{Chen},~\bfnm{Quanning}\binits{Q.}},
  \bauthor{\bsnm{Ai},~\bfnm{Kaixing}\binits{K.}} \AND
  \bauthor{\bsnm{Shi},~\bfnm{Baomin}\binits{B.}}
(\byear{2021}).
\btitle{TRPM2 promotes pancreatic cancer by PKC/MAPK pathway}.
\bjournal{Cell death \& disease}
\bvolume{12}
\bpages{585}.
\end{barticle}
\endbibitem

\bibitem[\protect\citeauthoryear{Liu et~al.}{2016}]{liu2016Gprotein}
\begin{barticle}[author]
\bauthor{\bsnm{Liu},~\bfnm{Ying}\binits{Y.}},
  \bauthor{\bsnm{An},~\bfnm{Su}\binits{S.}},
  \bauthor{\bsnm{Ward},~\bfnm{Richard}\binits{R.}},
  \bauthor{\bsnm{Yang},~\bfnm{Yang}\binits{Y.}},
  \bauthor{\bsnm{Guo},~\bfnm{Xiao-Xi}\binits{X.-X.}},
  \bauthor{\bsnm{Li},~\bfnm{Wei}\binits{W.}} \AND
  \bauthor{\bsnm{Xu},~\bfnm{Tian-Rui}\binits{T.-R.}}
(\byear{2016}).
\btitle{G protein-coupled receptors as promising cancer targets}.
\bjournal{Cancer Letters}
\bvolume{376}
\bpages{226--239}.
\end{barticle}
\endbibitem

\bibitem[\protect\citeauthoryear{Liu et~al.}{2023}]{liu2023sonar}
\begin{barticle}[author]
\bauthor{\bsnm{Liu},~\bfnm{Zhiyuan}\binits{Z.}},
  \bauthor{\bsnm{Wu},~\bfnm{Dafei}\binits{D.}},
  \bauthor{\bsnm{Zhai},~\bfnm{Weiwei}\binits{W.}} \AND
  \bauthor{\bsnm{Ma},~\bfnm{Liang}\binits{L.}}
(\byear{2023}).
\btitle{SONAR enables cell type deconvolution with spatially weighted
  Poisson-Gamma model for spatial transcriptomics}.
\bjournal{Nature Communications}
\bvolume{14}
\bpages{4727}.
\end{barticle}
\endbibitem

\bibitem[\protect\citeauthoryear{Miller
  et~al.}{2021}]{miller2021characterizing}
\begin{barticle}[author]
\bauthor{\bsnm{Miller},~\bfnm{Brendan~F}\binits{B.~F.}},
  \bauthor{\bsnm{Bambah-Mukku},~\bfnm{Dhananjay}\binits{D.}},
  \bauthor{\bsnm{Dulac},~\bfnm{Catherine}\binits{C.}},
  \bauthor{\bsnm{Zhuang},~\bfnm{Xiaowei}\binits{X.}} \AND
  \bauthor{\bsnm{Fan},~\bfnm{Jean}\binits{J.}}
(\byear{2021}).
\btitle{Characterizing spatial gene expression heterogeneity in spatially
  resolved single-cell transcriptomic data with nonuniform cellular densities}.
\bjournal{Genome Research}
\bvolume{31}
\bpages{1843--1855}.
\end{barticle}
\endbibitem

\bibitem[\protect\citeauthoryear{Pranavathiyani
  et~al.}{2019}]{pranavathiyani2019integrated}
\begin{barticle}[author]
\bauthor{\bsnm{Pranavathiyani},~\bfnm{G}\binits{G.}},
  \bauthor{\bsnm{Thanmalagan},~\bfnm{Raja~Rajeswary}\binits{R.~R.}},
  \bauthor{\bsnm{Devi},~\bfnm{Naorem~Leimarembi}\binits{N.~L.}} \AND
  \bauthor{\bsnm{Venkatesan},~\bfnm{Amouda}\binits{A.}}
(\byear{2019}).
\btitle{Integrated transcriptome interactome study of oncogenes and tumor
  suppressor genes in breast cancer}.
\bjournal{Genes \& Diseases}
\bvolume{6}
\bpages{78--87}.
\end{barticle}
\endbibitem

\bibitem[\protect\citeauthoryear{Qin, Ma and Wu}{2023}]{qin2023two}
\begin{barticle}[author]
\bauthor{\bsnm{Qin},~\bfnm{Xing}\binits{X.}},
  \bauthor{\bsnm{Ma},~\bfnm{Shuangge}\binits{S.}} \AND
  \bauthor{\bsnm{Wu},~\bfnm{Mengyun}\binits{M.}}
(\byear{2023}).
\btitle{{Two-level Bayesian interaction analysis for survival data
  incorporating pathway information}}.
\bjournal{Biometrics}
\bvolume{79}
\bpages{1761--1774}.
\end{barticle}
\endbibitem

\bibitem[\protect\citeauthoryear{Quintana
  et~al.}{2011}]{quintana2011incorporating}
\begin{barticle}[author]
\bauthor{\bsnm{Quintana},~\bfnm{Melanie~A}\binits{M.~A.}},
  \bauthor{\bsnm{Berstein},~\bfnm{Jonine~L}\binits{J.~L.}},
  \bauthor{\bsnm{Thomas},~\bfnm{Duncan~C}\binits{D.~C.}} \AND
  \bauthor{\bsnm{Conti},~\bfnm{David~V}\binits{D.~V.}}
(\byear{2011}).
\btitle{Incorporating model uncertainty in detecting rare variants: the
  Bayesian risk index}.
\bjournal{Genetic Epidemiology}
\bvolume{35}
\bpages{638--649}.
\end{barticle}
\endbibitem

\bibitem[\protect\citeauthoryear{Rao et~al.}{2021}]{rao2021exploring}
\begin{barticle}[author]
\bauthor{\bsnm{Rao},~\bfnm{Anjali}\binits{A.}},
  \bauthor{\bsnm{Barkley},~\bfnm{Dalia}\binits{D.}},
  \bauthor{\bsnm{Fran{\c{c}}a},~\bfnm{Gustavo~S}\binits{G.~S.}} \AND
  \bauthor{\bsnm{Yanai},~\bfnm{Itai}\binits{I.}}
(\byear{2021}).
\btitle{Exploring tissue architecture using spatial transcriptomics}.
\bjournal{Nature}
\bvolume{596}
\bpages{211--220}.
\end{barticle}
\endbibitem

\bibitem[\protect\citeauthoryear{Seal, Bitler and Ghosh}{2023}]{seal2023smash}
\begin{barticle}[author]
\bauthor{\bsnm{Seal},~\bfnm{Souvik}\binits{S.}},
  \bauthor{\bsnm{Bitler},~\bfnm{Benjamin~G}\binits{B.~G.}} \AND
  \bauthor{\bsnm{Ghosh},~\bfnm{Debashis}\binits{D.}}
(\byear{2023}).
\btitle{SMASH: Scalable Method for Analyzing Spatial Heterogeneity of genes in
  spatial transcriptomics data}.
\bjournal{PLoS Genetics}
\bvolume{19}
\bpages{e1010983}.
\end{barticle}
\endbibitem

\bibitem[\protect\citeauthoryear{Shang, Wu and
  Zhou}{2025}]{shang2025statistical}
\begin{barticle}[author]
\bauthor{\bsnm{Shang},~\bfnm{Lulu}\binits{L.}},
  \bauthor{\bsnm{Wu},~\bfnm{Peijun}\binits{P.}} \AND
  \bauthor{\bsnm{Zhou},~\bfnm{Xiang}\binits{X.}}
(\byear{2025}).
\btitle{Statistical identification of cell type-specific spatially variable
  genes in spatial transcriptomics}.
\bjournal{Nature Communications}
\bvolume{16}
\bpages{1059}.
\end{barticle}
\endbibitem

\bibitem[\protect\citeauthoryear{Song et~al.}{2021}]{song2021bayesian}
\begin{barticle}[author]
\bauthor{\bsnm{Song},~\bfnm{Yanyi}\binits{Y.}},
  \bauthor{\bsnm{Zhou},~\bfnm{Xiang}\binits{X.}},
  \bauthor{\bsnm{Kang},~\bfnm{Jian}\binits{J.}},
  \bauthor{\bsnm{Aung},~\bfnm{Max~T}\binits{M.~T.}},
  \bauthor{\bsnm{Zhang},~\bfnm{Min}\binits{M.}},
  \bauthor{\bsnm{Zhao},~\bfnm{Wei}\binits{W.}},
  \bauthor{\bsnm{Needham},~\bfnm{Belinda~L}\binits{B.~L.}},
  \bauthor{\bsnm{Kardia},~\bfnm{Sharon~LR}\binits{S.~L.}},
  \bauthor{\bsnm{Liu},~\bfnm{Yongmei}\binits{Y.}},
  \bauthor{\bsnm{Meeker},~\bfnm{John~D}\binits{J.~D.}} \betal{et~al.}
(\byear{2021}).
\btitle{Bayesian sparse mediation analysis with targeted penalization of
  natural indirect effects}.
\bjournal{Journal of the Royal Statistical Society Series C: Applied
  Statistics}
\bvolume{70}
\bpages{1391--1412}.
\end{barticle}
\endbibitem

\bibitem[\protect\citeauthoryear{St{\aa}hl
  et~al.}{2016}]{staahl2016visualization}
\begin{barticle}[author]
\bauthor{\bsnm{St{\aa}hl},~\bfnm{Patrik~L}\binits{P.~L.}},
  \bauthor{\bsnm{Salm{\'e}n},~\bfnm{Fredrik}\binits{F.}},
  \bauthor{\bsnm{Vickovic},~\bfnm{Sanja}\binits{S.}},
  \bauthor{\bsnm{Lundmark},~\bfnm{Anna}\binits{A.}},
  \bauthor{\bsnm{Navarro},~\bfnm{Jos{\'e}~Fern{\'a}ndez}\binits{J.~F.}},
  \bauthor{\bsnm{Magnusson},~\bfnm{Jens}\binits{J.}},
  \bauthor{\bsnm{Giacomello},~\bfnm{Stefania}\binits{S.}},
  \bauthor{\bsnm{Asp},~\bfnm{Michaela}\binits{M.}},
  \bauthor{\bsnm{Westholm},~\bfnm{Jakub~O}\binits{J.~O.}},
  \bauthor{\bsnm{Huss},~\bfnm{Mikael}\binits{M.}} \betal{et~al.}
(\byear{2016}).
\btitle{Visualization and analysis of gene expression in tissue sections by
  spatial transcriptomics}.
\bjournal{Science}
\bvolume{353}
\bpages{78--82}.
\end{barticle}
\endbibitem

\bibitem[\protect\citeauthoryear{Stanley and
  Abdel-Wahab}{2022}]{stanley2022dysregulation}
\begin{barticle}[author]
\bauthor{\bsnm{Stanley},~\bfnm{Robert~F}\binits{R.~F.}} \AND
  \bauthor{\bsnm{Abdel-Wahab},~\bfnm{Omar}\binits{O.}}
(\byear{2022}).
\btitle{Dysregulation and therapeutic targeting of RNA splicing in cancer}.
\bjournal{Nature Cancer}
\bvolume{3}
\bpages{536--546}.
\end{barticle}
\endbibitem

\bibitem[\protect\citeauthoryear{Su and Cui}{2025}]{su2025rotation}
\begin{barticle}[author]
\bauthor{\bsnm{Su},~\bfnm{Haohao}\binits{H.}} \AND
  \bauthor{\bsnm{Cui},~\bfnm{Yuehua}\binits{Y.}}
(\byear{2025}).
\btitle{Rotation-invariance is essential for accurate detection of spatially
  variable genes in spatial transcriptomics}.
\bjournal{Nature Communications}
\bvolume{16}
\bpages{7122}.
\end{barticle}
\endbibitem

\bibitem[\protect\citeauthoryear{Sun, Zhu and Zhou}{2020}]{sun2020statistical}
\begin{barticle}[author]
\bauthor{\bsnm{Sun},~\bfnm{Shiquan}\binits{S.}},
  \bauthor{\bsnm{Zhu},~\bfnm{Jiaqiang}\binits{J.}} \AND
  \bauthor{\bsnm{Zhou},~\bfnm{Xiang}\binits{X.}}
(\byear{2020}).
\btitle{Statistical analysis of spatial expression patterns for spatially
  resolved transcriptomic studies}.
\bjournal{Nature Methods}
\bvolume{17}
\bpages{193--200}.
\end{barticle}
\endbibitem

\bibitem[\protect\citeauthoryear{Svensson, Teichmann and
  Stegle}{2018}]{svensson2018spatialde}
\begin{barticle}[author]
\bauthor{\bsnm{Svensson},~\bfnm{Valentine}\binits{V.}},
  \bauthor{\bsnm{Teichmann},~\bfnm{Sarah~A}\binits{S.~A.}} \AND
  \bauthor{\bsnm{Stegle},~\bfnm{Oliver}\binits{O.}}
(\byear{2018}).
\btitle{Spatial{DE}: identification of spatially variable genes}.
\bjournal{Nature Methods}
\bvolume{15}
\bpages{343--346}.
\end{barticle}
\endbibitem

\bibitem[\protect\citeauthoryear{Szklarczyk
  et~al.}{2023}]{szklarczyk2023string}
\begin{barticle}[author]
\bauthor{\bsnm{Szklarczyk},~\bfnm{Damian}\binits{D.}},
  \bauthor{\bsnm{Kirsch},~\bfnm{Rebecca}\binits{R.}},
  \bauthor{\bsnm{Koutrouli},~\bfnm{Mikaela}\binits{M.}},
  \bauthor{\bsnm{Nastou},~\bfnm{Katerina}\binits{K.}},
  \bauthor{\bsnm{Mehryary},~\bfnm{Farrokh}\binits{F.}},
  \bauthor{\bsnm{Hachilif},~\bfnm{Radja}\binits{R.}},
  \bauthor{\bsnm{Gable},~\bfnm{Annika~L}\binits{A.~L.}},
  \bauthor{\bsnm{Fang},~\bfnm{Tao}\binits{T.}},
  \bauthor{\bsnm{Doncheva},~\bfnm{Nadezhda~T}\binits{N.~T.}},
  \bauthor{\bsnm{Pyysalo},~\bfnm{Sampo}\binits{S.}} \betal{et~al.}
(\byear{2023}).
\btitle{The {STRING} database in 2023: protein--protein association networks
  and functional enrichment analyses for any sequenced genome of interest}.
\bjournal{Nucleic Acids Research}
\bvolume{51}
\bpages{D638--D646}.
\end{barticle}
\endbibitem

\bibitem[\protect\citeauthoryear{Vethakanraj
  et~al.}{2015}]{vethakanraj2015targeting}
\begin{barticle}[author]
\bauthor{\bsnm{Vethakanraj},~\bfnm{Helen~Shiphrah}\binits{H.~S.}},
  \bauthor{\bsnm{Babu},~\bfnm{Thabraz~Ahmed}\binits{T.~A.}},
  \bauthor{\bsnm{Sudarsanan},~\bfnm{Ganesh~Babu}\binits{G.~B.}},
  \bauthor{\bsnm{Duraisamy},~\bfnm{Prabhu~Kumar}\binits{P.~K.}} \AND
  \bauthor{\bsnm{Kumar},~\bfnm{Sekar~Ashok}\binits{S.~A.}}
(\byear{2015}).
\btitle{Targeting ceramide metabolic pathway induces apoptosis in human breast
  cancer cell lines}.
\bjournal{Biochemical and Biophysical Research Communications}
\bvolume{464}
\bpages{833--839}.
\end{barticle}
\endbibitem

\bibitem[\protect\citeauthoryear{Weber et~al.}{2023}]{weber2023nnsvg}
\begin{barticle}[author]
\bauthor{\bsnm{Weber},~\bfnm{Lukas~M}\binits{L.~M.}},
  \bauthor{\bsnm{Saha},~\bfnm{Arkajyoti}\binits{A.}},
  \bauthor{\bsnm{Datta},~\bfnm{Abhirup}\binits{A.}},
  \bauthor{\bsnm{Hansen},~\bfnm{Kasper~D}\binits{K.~D.}} \AND
  \bauthor{\bsnm{Hicks},~\bfnm{Stephanie~C}\binits{S.~C.}}
(\byear{2023}).
\btitle{nnSVG for the scalable identification of spatially variable genes using
  nearest-neighbor Gaussian processes}.
\bjournal{Nature Communications}
\bvolume{14}
\bpages{4059}.
\end{barticle}
\endbibitem

\bibitem[\protect\citeauthoryear{Wu, Guo and Kang}{2024}]{wu2024bayesian}
\begin{barticle}[author]
\bauthor{\bsnm{Wu},~\bfnm{Ben}\binits{B.}},
  \bauthor{\bsnm{Guo},~\bfnm{Ying}\binits{Y.}} \AND
  \bauthor{\bsnm{Kang},~\bfnm{Jian}\binits{J.}}
(\byear{2024}).
\btitle{{Bayesian spatial blind source separation via the thresholded Gaussian
  process}}.
\bjournal{Journal of the American Statistical Association}
\bvolume{119}
\bpages{422--433}.
\end{barticle}
\endbibitem

\bibitem[\protect\citeauthoryear{Wu et~al.}{2021}]{wu2021comprehensive}
\begin{barticle}[author]
\bauthor{\bsnm{Wu},~\bfnm{Rui}\binits{R.}},
  \bauthor{\bsnm{Guo},~\bfnm{Wenbo}\binits{W.}},
  \bauthor{\bsnm{Qiu},~\bfnm{Xinyao}\binits{X.}},
  \bauthor{\bsnm{Wang},~\bfnm{Shicheng}\binits{S.}},
  \bauthor{\bsnm{Sui},~\bfnm{Chengjun}\binits{C.}},
  \bauthor{\bsnm{Lian},~\bfnm{Qiuyu}\binits{Q.}},
  \bauthor{\bsnm{Wu},~\bfnm{Jianmin}\binits{J.}},
  \bauthor{\bsnm{Shan},~\bfnm{Yiran}\binits{Y.}},
  \bauthor{\bsnm{Yang},~\bfnm{Zhao}\binits{Z.}},
  \bauthor{\bsnm{Yang},~\bfnm{Shuai}\binits{S.}} \betal{et~al.}
(\byear{2021}).
\btitle{Comprehensive analysis of spatial architecture in primary liver
  cancer}.
\bjournal{Science Advances}
\bvolume{7}
\bpages{eabg3750}.
\end{barticle}
\endbibitem

\bibitem[\protect\citeauthoryear{Wu et~al.}{2022}]{wu2022highly}
\begin{barticle}[author]
\bauthor{\bsnm{Wu},~\bfnm{Yanhong}\binits{Y.}},
  \bauthor{\bsnm{Hu},~\bfnm{Qifan}\binits{Q.}},
  \bauthor{\bsnm{Wang},~\bfnm{Shicheng}\binits{S.}},
  \bauthor{\bsnm{Liu},~\bfnm{Changyi}\binits{C.}},
  \bauthor{\bsnm{Shan},~\bfnm{Yiran}\binits{Y.}},
  \bauthor{\bsnm{Guo},~\bfnm{Wenbo}\binits{W.}},
  \bauthor{\bsnm{Jiang},~\bfnm{Rui}\binits{R.}},
  \bauthor{\bsnm{Wang},~\bfnm{Xiaowo}\binits{X.}} \AND
  \bauthor{\bsnm{Gu},~\bfnm{Jin}\binits{J.}}
(\byear{2022}).
\btitle{Highly Regional Genes: graph-based gene selection for single-cell
  RNA-seq data}.
\bjournal{Journal of Genetics and Genomics}
\bvolume{49}
\bpages{891--899}.
\end{barticle}
\endbibitem

\bibitem[\protect\citeauthoryear{Wu et~al.}{2025}]{wu2025supp}
\begin{bmisc}[author]
\bauthor{\bsnm{Wu},~\bfnm{Mingcong}\binits{M.}},
  \bauthor{\bsnm{Li},~\bfnm{Yang}\binits{Y.}},
  \bauthor{\bsnm{Ma},~\bfnm{Shuangge}\binits{S.}} \AND
  \bauthor{\bsnm{Wu},~\bfnm{Mengyun}\binits{M.}}
(\byear{2025}).
\btitle{Supplement to ``Joint identification of spatially variable genes via a
  network-assisted Bayesian regularization approach''}.
\bdoi{10.1214/[provided by typesetter]}
\end{bmisc}
\endbibitem

\bibitem[\protect\citeauthoryear{Xu et~al.}{2024}]{xu2024metabolism}
\begin{barticle}[author]
\bauthor{\bsnm{Xu},~\bfnm{Shiliang}\binits{S.}},
  \bauthor{\bsnm{Wang},~\bfnm{Lingxia}\binits{L.}},
  \bauthor{\bsnm{Zhao},~\bfnm{Yuexin}\binits{Y.}},
  \bauthor{\bsnm{Mo},~\bfnm{Tong}\binits{T.}},
  \bauthor{\bsnm{Wang},~\bfnm{Bo}\binits{B.}},
  \bauthor{\bsnm{Lin},~\bfnm{Jun}\binits{J.}} \AND
  \bauthor{\bsnm{Yang},~\bfnm{Huan}\binits{H.}}
(\byear{2024}).
\btitle{Metabolism-regulating non-coding RNAs in breast cancer: roles,
  mechanisms and clinical applications}.
\bjournal{Journal of Biomedical Science}
\bvolume{31}
\bpages{25}.
\end{barticle}
\endbibitem

\bibitem[\protect\citeauthoryear{Yamauchi, Nishimura and
  Yoshimi}{2022}]{yamauchi2022aberrant}
\begin{barticle}[author]
\bauthor{\bsnm{Yamauchi},~\bfnm{Hirofumi}\binits{H.}},
  \bauthor{\bsnm{Nishimura},~\bfnm{Kazuki}\binits{K.}} \AND
  \bauthor{\bsnm{Yoshimi},~\bfnm{Akihide}\binits{A.}}
(\byear{2022}).
\btitle{Aberrant {RNA} splicing and therapeutic opportunities in cancers}.
\bjournal{Cancer Science}
\bvolume{113}
\bpages{373--381}.
\end{barticle}
\endbibitem

\bibitem[\protect\citeauthoryear{Yan and Luo}{2024}]{yan2024bayesian}
\begin{barticle}[author]
\bauthor{\bsnm{Yan},~\bfnm{Yinqiao}\binits{Y.}} \AND
  \bauthor{\bsnm{Luo},~\bfnm{Xiangyu}\binits{X.}}
(\byear{2024}).
\btitle{Bayesian integrative region segmentation in spatially resolved
  transcriptomic studies}.
\bjournal{Journal of the American Statistical Association}
\bvolume{119}
\bpages{1709-1721}.
\end{barticle}
\endbibitem

\bibitem[\protect\citeauthoryear{Yang et~al.}{2023}]{yang2023metabolic}
\begin{barticle}[author]
\bauthor{\bsnm{Yang},~\bfnm{Flora}\binits{F.}},
  \bauthor{\bsnm{Hilakivi-Clarke},~\bfnm{Leena}\binits{L.}},
  \bauthor{\bsnm{Shaha},~\bfnm{Aurpita}\binits{A.}},
  \bauthor{\bsnm{Wang},~\bfnm{Yuanguo}\binits{Y.}},
  \bauthor{\bsnm{Wang},~\bfnm{Xianghu}\binits{X.}},
  \bauthor{\bsnm{Deng},~\bfnm{Yibin}\binits{Y.}},
  \bauthor{\bsnm{Lai},~\bfnm{Jinping}\binits{J.}} \AND
  \bauthor{\bsnm{Kang},~\bfnm{Ningling}\binits{N.}}
(\byear{2023}).
\btitle{Metabolic reprogramming and its clinical implication for liver cancer}.
\bjournal{Hepatology}
\bvolume{78}
\bpages{1602--1624}.
\end{barticle}
\endbibitem

\bibitem[\protect\citeauthoryear{Yu and Li}{2024}]{yu2024spvc}
\begin{barticle}[author]
\bauthor{\bsnm{Yu},~\bfnm{Shan}\binits{S.}} \AND
  \bauthor{\bsnm{Li},~\bfnm{Wei~Vivian}\binits{W.~V.}}
(\byear{2024}).
\btitle{spVC for the detection and interpretation of spatial gene expression
  variation}.
\bjournal{Genome Biology}
\bvolume{25}
\bpages{103}.
\end{barticle}
\endbibitem

\bibitem[\protect\citeauthoryear{Yu and Luo}{2022}]{yu2022identification}
\begin{barticle}[author]
\bauthor{\bsnm{Yu},~\bfnm{Jinge}\binits{J.}} \AND
  \bauthor{\bsnm{Luo},~\bfnm{Xiangyu}\binits{X.}}
(\byear{2022}).
\btitle{Identification of cell-type-specific spatially variable genes
  accounting for excess zeros}.
\bjournal{Bioinformatics}
\bvolume{38}
\bpages{4135--4144}.
\end{barticle}
\endbibitem

\bibitem[\protect\citeauthoryear{Yuan et~al.}{2024}]{yuan2024heartsvg}
\begin{barticle}[author]
\bauthor{\bsnm{Yuan},~\bfnm{Xin}\binits{X.}},
  \bauthor{\bsnm{Ma},~\bfnm{Yanran}\binits{Y.}},
  \bauthor{\bsnm{Gao},~\bfnm{Ruitian}\binits{R.}},
  \bauthor{\bsnm{Cui},~\bfnm{Shuya}\binits{S.}},
  \bauthor{\bsnm{Wang},~\bfnm{Yifan}\binits{Y.}},
  \bauthor{\bsnm{Fa},~\bfnm{Botao}\binits{B.}},
  \bauthor{\bsnm{Ma},~\bfnm{Shiyang}\binits{S.}},
  \bauthor{\bsnm{Wei},~\bfnm{Ting}\binits{T.}},
  \bauthor{\bsnm{Ma},~\bfnm{Shuangge}\binits{S.}} \AND
  \bauthor{\bsnm{Yu},~\bfnm{Zhangsheng}\binits{Z.}}
(\byear{2024}).
\btitle{{HEARTSVG}: a fast and accurate method for identifying spatially
  variable genes in large-scale spatial transcriptomics}.
\bjournal{Nature Communications}
\bvolume{15}
\bpages{5700}.
\end{barticle}
\endbibitem

\bibitem[\protect\citeauthoryear{Zhang
  et~al.}{2022}]{zhang2022isoliquiritigenin}
\begin{barticle}[author]
\bauthor{\bsnm{Zhang},~\bfnm{Zhu}\binits{Z.}},
  \bauthor{\bsnm{Chen},~\bfnm{{Wenqing}}\binits{W.}},
  \bauthor{\bsnm{Zhang},~\bfnm{{Shiqing}}\binits{S.}},
  \bauthor{\bsnm{Bai},~\bfnm{{Jingxuan}}\binits{J.}},
  \bauthor{\bsnm{Liu},~\bfnm{Bin}\binits{B.}}, \bauthor{\bsnm{Yung},~\bfnm{{Ken
  Kin Lam}}\binits{K.}} \AND \bauthor{\bsnm{Ko},~\bfnm{Joshua~{Ka
  Shun}}\binits{J.~K.}}
(\byear{2022}).
\btitle{Isoliquiritigenin inhibits pancreatic cancer progression through
  blockade of p38 MAPK-regulated autophagy}.
\bjournal{Phytomedicine}
\bvolume{106}
\bpages{154406}.
\end{barticle}
\endbibitem

\bibitem[\protect\citeauthoryear{Zhao et~al.}{2022}]{zhao2022modeling}
\begin{barticle}[author]
\bauthor{\bsnm{Zhao},~\bfnm{Peiyao}\binits{P.}},
  \bauthor{\bsnm{Zhu},~\bfnm{Jiaqiang}\binits{J.}},
  \bauthor{\bsnm{Ma},~\bfnm{Ying}\binits{Y.}} \AND
  \bauthor{\bsnm{Zhou},~\bfnm{Xiang}\binits{X.}}
(\byear{2022}).
\btitle{Modeling zero inflation is not necessary for spatial transcriptomics}.
\bjournal{Genome Biology}
\bvolume{23}
\bpages{118}.
\end{barticle}
\endbibitem

\bibitem[\protect\citeauthoryear{Zhou et~al.}{2023}]{zhou2023spatial}
\begin{barticle}[author]
\bauthor{\bsnm{Zhou},~\bfnm{Zixiang}\binits{Z.}},
  \bauthor{\bsnm{Zhong},~\bfnm{Yunshan}\binits{Y.}},
  \bauthor{\bsnm{Zhang},~\bfnm{Zemin}\binits{Z.}} \AND
  \bauthor{\bsnm{Ren},~\bfnm{Xianwen}\binits{X.}}
(\byear{2023}).
\btitle{Spatial transcriptomics deconvolution at single-cell resolution using
  Redeconve}.
\bjournal{Nature Communications}
\bvolume{14}
\bpages{7930}.
\end{barticle}
\endbibitem

\bibitem[\protect\citeauthoryear{Zhu, Sun and Zhou}{2021}]{zhu2021spark}
\begin{barticle}[author]
\bauthor{\bsnm{Zhu},~\bfnm{Jiaqiang}\binits{J.}},
  \bauthor{\bsnm{Sun},~\bfnm{Shiquan}\binits{S.}} \AND
  \bauthor{\bsnm{Zhou},~\bfnm{Xiang}\binits{X.}}
(\byear{2021}).
\btitle{{SPARK-X}: non-parametric modeling enables scalable and robust
  detection of spatial expression patterns for large spatial transcriptomic
  studies}.
\bjournal{Genome Biology}
\bvolume{22}
\bpages{184}.
\end{barticle}
\endbibitem

\bibitem[\protect\citeauthoryear{Zuo, Xia and Chen}{2024}]{zuo2024dissecting}
\begin{barticle}[author]
\bauthor{\bsnm{Zuo},~\bfnm{Chunman}\binits{C.}},
  \bauthor{\bsnm{Xia},~\bfnm{Junjie}\binits{J.}} \AND
  \bauthor{\bsnm{Chen},~\bfnm{Luonan}\binits{L.}}
(\byear{2024}).
\btitle{Dissecting tumor microenvironment from spatially resolved
  transcriptomics data by heterogeneous graph learning}.
\bjournal{Nature Communications}
\bvolume{15}
\bpages{5057}.
\end{barticle}
\endbibitem

\end{thebibliography}


\newpage

  \begin{center}
    {\LARGE\bf Supplement to ``Joint identification of spatially variable genes via a network-assisted Bayesian regularization approach'' }
\end{center}
  \medskip
  
  \renewcommand{\thefigure}{S\arabic{figure}}
\renewcommand{\thetable}{S\arabic{table}}
\renewcommand{\theequation}{S\arabic{equation}}
\renewcommand{\thesection}{S\arabic{section}}

\setcounter{section}{0}
\setcounter{figure}{0}
\numberwithin{equation}{section}
\numberwithin{figure}{section}

\section{Derivation of step (e) in posterior sampling}

The preconditioned Crank-Nicolson Langevin dynamics (pCNLD) sampler is adopted for sampling of $\boldsymbol{\gamma}$. The proposed sample of $\boldsymbol{\gamma}^*$ is generated as follows.
\begin{equation*}
  \boldsymbol{\gamma}^*  =   \sqrt{1 - \tau_\gamma^2} \boldsymbol{\gamma} + \left( 1 - \sqrt{1 - \tau_\gamma^2} \right)  \sigma_\gamma^2 \left(\mathbf{L}+\varepsilon \mathbf{I}_{2p} \right)^{-1} \nabla_{\boldsymbol{\gamma}} \log f( \mathbf{Y}\mid \boldsymbol{\gamma}, \boldsymbol{-}) +   \tau_\gamma N \left\{ \boldsymbol{0}, \sigma_\gamma^2 \left(\mathbf{L}+\varepsilon \mathbf{I}_{2p} \right)^{-1}\right\}
\end{equation*}
with
\begin{equation*}
     \nabla_{\gamma^{(d)}_{j}} \log f( \mathbf{Y}\mid \boldsymbol{\gamma}, \boldsymbol{-}) =   \sum_{ \left \{i: r_{ij} = 0 \right\} } \frac{\phi \left( Y_{ij} - c_i \mu_{ij} \right)}{\mu_{ij} \left( c_i \mu_{ij} + \phi \right)} \, \mu_{ij} \,  \mathcal{K}\left(x_{id}\right) \, \nabla_{\gamma^{(d)}_{j}} \left\{ \gamma^{(d)}_{j} \operatorname{I}\left(\left|\gamma^{(d)}_{j}\right| >\lambda \, \rho^{(d)}_j\right) \right\}.
\end{equation*}

For the derivative of the hard-thresholding function, a smooth approximation is introduced that $\operatorname{I}\left(\left|\gamma_j^{(d)}\right|> \lambda \rho^{(d)}_j\right)  \simeq \frac{1}{2}\left\{1+\frac{2}{\pi} \arctan \left(\frac{\left(\gamma_j^{(d)}\right)^2-\left( \lambda  \rho^{(d)}_j\right)^2}{\epsilon}\right)\right\}$ for $ \epsilon \rightarrow 0$, leading to
\begin{equation*}
    \nabla_{\gamma^{(d)}_{j}} \left\{ \gamma^{(d)}_{j} \operatorname{I}\left(\left|\gamma^{(d)}_{j}\right| > \lambda  \rho^{(d)}_j \right) \right\} \simeq  \operatorname{I}\left(\left|\gamma^{(d)}_{j}\right| > \lambda  \rho^{(d)}_j\right) + \gamma^{(d)}_{j}  \frac{2 \gamma^{(d)}_{j}/ \epsilon}{\pi\left\{1+\left(\left(\gamma^{(d)}_{j}\right)^2-\left( \lambda  \rho^{(d)}_j \right)^2\right)^2 / \epsilon^2\right\}},
\end{equation*}
where $\epsilon$ is set as $10^{-4}$ throughout the article.

\section{Detailed settings for the hyperparameter set}

Regarding the hyperparameter set, we assume $a_{\pi} = b_{\pi}= 1$ for a weakly informative prior on $\pi_{j}$. We set $a_\phi = 10$ and $b_\phi=0.1$ for $\text{Ga}(a_\phi, b_\phi)$, and $a_\gamma = 3.5$ and $b_\gamma = 0.5$ for $\operatorname{IG}\left(a_\gamma, b_\gamma\right)$. $\sigma_{0j}^2$'s and $\sigma_{\alpha_k}^2$'s are set as $3^2$ for all $j$ and $k$ to yield vague priors for the baseline expression levels and cell-type-specific effects. We set $\lambda_l=0$ and $\lambda_u$ as the 90\% quantile of $\left|\tilde{\boldsymbol{\beta}}\right|$ to help against false positive detection as suggested in \cite{song2021bayesian}. The sampling variances $\tau_{\mu_0}^2, \tau_{\alpha}^2, \tau_{\phi}^2, \tau_{\gamma}^2,$ and $\tau_{\lambda}^2$ are all adaptively chosen by tuning acceptance rates to 30\% in simulation studies and 15\% in real data analysis. We perform 2,000 iterations with the first 1000 discarded as burn-in and thinning by 10 in the simulations, and 3,000 iterations with the first 2000 burn-in and thinning by 10 in the data analysis.

{
\section{Computation efficiency}
We assess the computational efficiency of our proposed method using both simulated and real datasets. For simulated datasets, with a fixed gene size of 5,000 and 2000 MCMC iterations, the average computer times for spot sizes of 1,000, 5,000, 10,000, and 50,000 are 0.37, 1.53, 2.82, and 15.84 hours, respectively. In real data analysis, the computer times for a full 3,000 iteration MCMC run (which is sufficient to ensure convergence) are 1.2, 2.7, and 5.3 hours for the Visium TNBC (1,108 spots with 5,000 genes), Visium PLC (3,181 spots with 5,000 genes), and Xenium PAC (164,274 single cells with 474 genes) datasets, respectively. All analyses are implemented on a MacBook Pro computer with Intel Core i5, 4 cores, and 16 GB RAM, using 8 threads for paralleling. These results demonstrate that our proposed approach is computationally feasible even for large-scale cell-resolution ST data (mostly with a few genes). 
}

\clearpage
\section{Detailed simulation settings and additional simulation results}

\subsection{Graphical representation for cellular compositions and network structures in the basic simulations}
\quad

\begin{figure}[!h]
\centering
\includegraphics[width=0.5\textwidth]{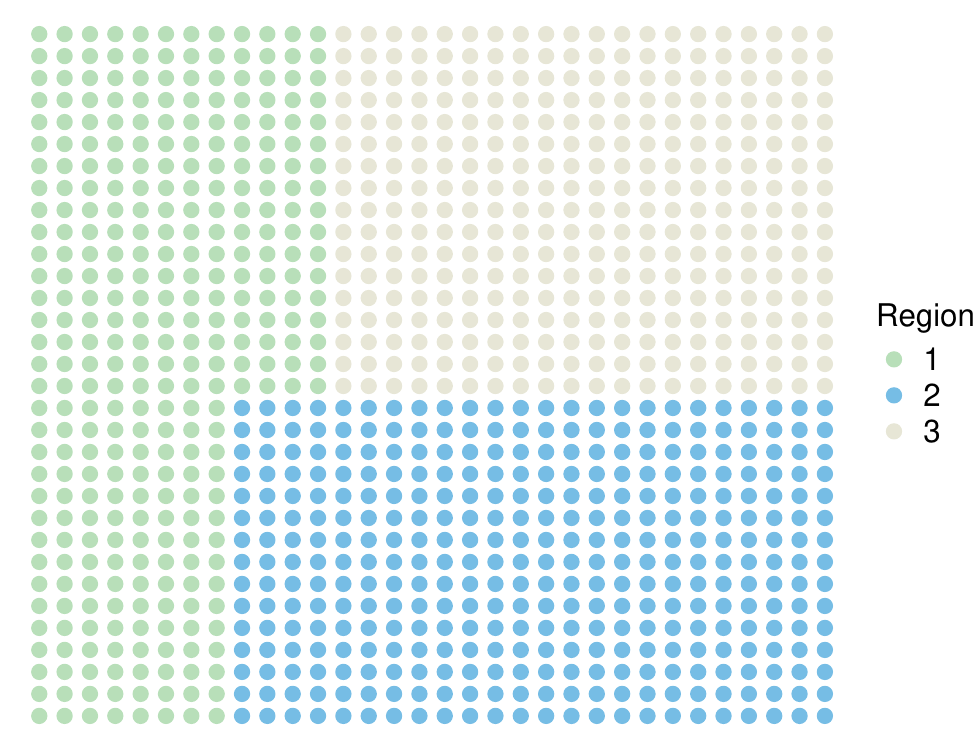}
\caption{Partitioned spot regions considered in the basic simulations introduced in Section 3.1 of the main text. Regions are denoted by different colors, where the cellular compositions $\boldsymbol{w}_i$'s are independently sampled from Dirichlet distributions Dirc(1,1,1,1,1,1)(Region 1), Dirc(3,5,7,9,11,13) (Region 2), and Dirc(18,16,14,12,10,8) (Region 3). \label{fig:sim_region}} 
\end{figure}

\begin{figure}[htb]
\centering
\includegraphics[width=\textwidth]{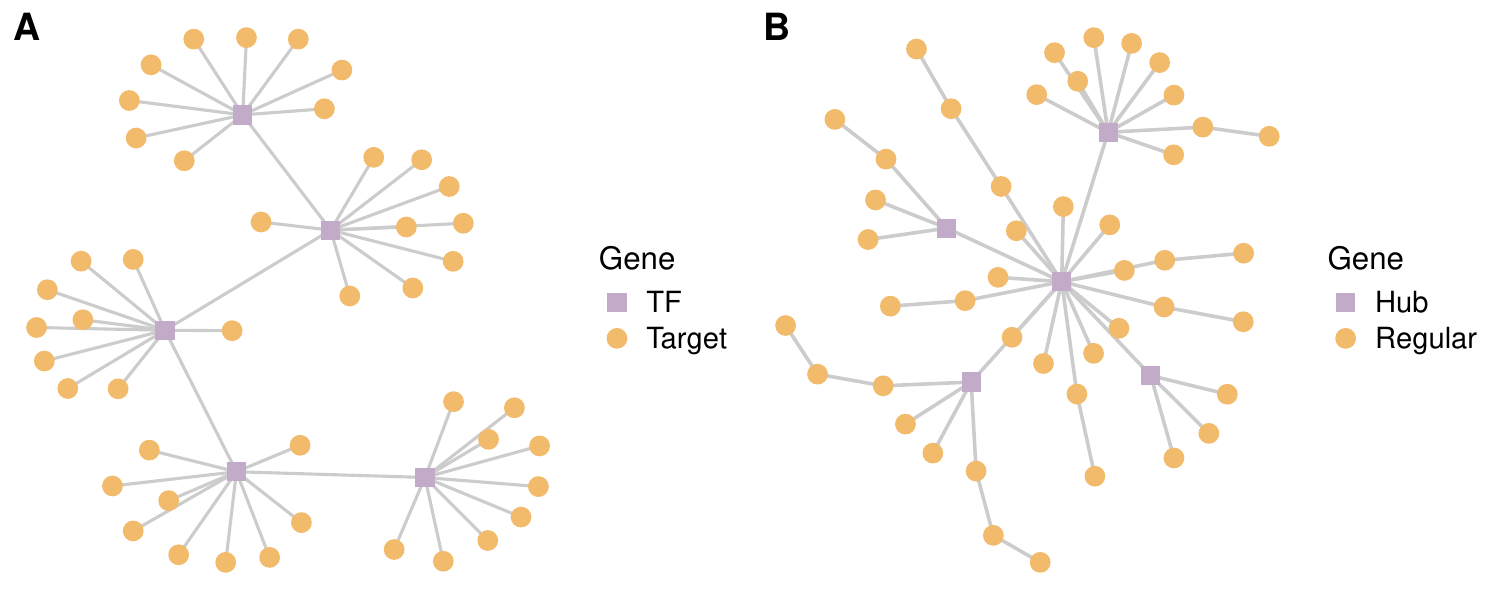}
\caption{
Illustrative examples of the (A) \textit{Star} network and (B) \textit{Scale-free} network. \label{fig:sim_network}}
\end{figure}

\subsection{Detailed settings for spatial effects in the basic simulations}  


{
The spatial effects associated with the two specified network structures are set as follows.
}

\begin{enumerate}
    \item \textit{Star} network:
    
    { Denote $\beta_{\operatorname{TF}}$ and $\beta_{\operatorname{target}}$ as the spatial effect parameters for the transcription factor (TF) gene and connected target gene, respectively.}
     For the non-informative sub-networks, $\beta^{(1)}_{\operatorname{TF}} = \beta^{(2)}_{\operatorname{TF}} = \beta^{(1)}_{\operatorname{target}} = \beta^{(2)}_{\operatorname{target}} = 0$. For each informative sub-network, the transcription factor (TF) gene and its connected  target genes have the same sign of spatial effect (either all positive or all negative), while the magnitudes satisfy that $ \frac{\left|\beta_{\operatorname{TF}}^{(d)}\right| }{\sqrt{n_{\operatorname{target}}}} > \left| \beta_{\operatorname{target}}^{(d)} \right|$ with $n_{\operatorname{target}}$ being the number of the target genes regulated by TF gene. {The specified values of $\beta_{\operatorname{TF}}$ and $\beta_{\operatorname{target}}$ for informative sub-networks  under different spatial patterns are detailed in Table \ref{tab:sim_beta}.}

    \item \textit{Scale-free} network:
    
    { Denote $\beta_{\operatorname{Hub}}$ and $\beta_{\operatorname{Regular}}$ as the spatial effect parameters for the hub gene and connected regular gene, respectively. } For the non-informative sub-networks, $\beta^{(1)}_{\operatorname{Hub}} = \beta^{(2)}_{\operatorname{Hub}} = \beta^{(1)}_{\operatorname{Regular}} = \beta^{(2)}_{\operatorname{Regular}} = 0$. For each informative sub-network, the spatial effects of the hub gene and its connected regular genes are randomly assigned as positive or negative. In addition, the magnitudes of $\beta^{(d)}_{\operatorname{Hub}} (d = 1, 2)$ are simulated from $\operatorname{Unif}(\operatorname{Hub}_a, \operatorname{Hub}_b)$, while the magnitudes of  $\beta^{(d)}_{\operatorname{Regular}} (d = 1, 2)$ are randomly generated from $\operatorname{Unif}(\operatorname{Regular}_a, \operatorname{Regular}_b)$ or $\operatorname{Unif}(2\operatorname{Regular}_a, 2\operatorname{Regular}_b)$ with the probability generated from a $\operatorname{Bern}(0.5)$ distribution,  representing large and small signals, respectively.
 { The specified values of these parameters for informative sub-networks under different spatial patterns are detailed in Table \ref{tab:sim_beta}.  }
    
\end{enumerate}

\begin{table}[htb]
\centering
\caption{
 Detailed values of spatial effects for informative sub-networks in the basic simulations. \label{tab:sim_beta}}
\renewcommand{\arraystretch}{1.5} 
\begin{tabular}{c|c|cc}
\hline
Network                          & Spatial pattern     & \multicolumn{2}{c}{Signal strength}  \\ \hline
\multirow{3}{*}{\textit{Star}}       & Linear      & \multicolumn{1}{c|}{\multirow{3}{*}{$\beta^{(1)}_{\operatorname{TF}}  = 1, \beta^{(2)}_{\operatorname{TF}}  =- 1$}} &  $ \beta^{(1)}_{\operatorname{target}} =0.15,   \beta^{(2)}_{\operatorname{target}} = -0.25 $ \\
                            & Exponential & \multicolumn{1}{c|}{}                  &  $\beta^{(1)}_{\operatorname{target}} =0.4,   \beta^{(2)}_{\operatorname{target}} = -0.5 $ \\
                            & Periodic    & \multicolumn{1}{c|}{}                  &  $\beta^{(1)}_{\operatorname{target}} =0.2,   \beta^{(2)}_{\operatorname{target}} = -0.3 $ \\ \hline
\multirow{3}{*}{\textit{Scale-free}} & Linear      & \multicolumn{1}{c|}{\multirow{3}{*}{ $\text{Hub}_a = 1.0, \text{Hub}_b = 1.2$ }} &  $\text{Regular}_a = 0.1,  \text{Regular}_b = 0.2 $ \\
                            & Exponential & \multicolumn{1}{c|}{}                  & $\text{Regular}_a = 0.3,  \text{Regular}_b = 0.4$  \\
                            & Periodic    & \multicolumn{1}{c|}{}                  &$\text{Regular}_a = 0.2,  \text{Regular}_b = 0.3$\\ \hline
\end{tabular}
\end{table}

\clearpage
\subsection{Additional results for the basic simulations}
\quad

\begin{table}[htb]
\caption{ Simulation results under the scenarios with a high dropout rate, where FDR (BFDR) is controlled to be $<$0.05. In each cell, mean (SD) is based on 50 replicates. \label{tab:simhigh}}
\resizebox{\textwidth}{!}{
\begin{tabular}{lcccccc}
\hline
                    & Recall       & Precision    & F1           & Recall       & Precision    & F1           \\ \hline
Linear  pattern     &              & \textit{Star} Network           &              &              & \textit{Scale-free} Network           &              \\
proposed            & 0.975(0.081) & 1.000(0.000) & 0.985(0.053) & 0.965(0.106) & 1.000(0.000) & 0.978(0.078) \\
spatialDE           & 0.079(0.036) & 0.999(0.004) & 0.168(0.028) & 0.093(0.039) & 1.000(0.000) & 0.174(0.060) \\
SPARK               & 0.169(0.065) & 0.985(0.013) & 0.284(0.094) & 0.312(0.135) & 0.983(0.016) & 0.457(0.163) \\
SPARK-X             & 0.735(0.361) & 0.220(0.276) & 0.179(0.025) & 0.465(0.158) & 0.847(0.226) & 0.567(0.171) \\
HRG  & 0.644(0.069)		& 0.188(0.019)	&0.275(0.030) &  0.500(0.258)	&	0.475(0.296) &	0.349(0.100) \\
MERINGUE            & 0.133(0.044) & 0.967(0.026) & 0.230(0.069) & 0.238(0.111) & 0.974(0.022) & 0.369(0.145) \\
CTSV       & 0.085(0.051) & 0.551(0.293) & 0.128(0.059) & 0.103(0.060) & 0.571(0.267) & 0.147(0.073) \\

CTSV-g                & 0.921(0.131) & 0.883(0.039) & 0.894(0.074) & 0.932(0.100) & 0.939(0.015) & 0.932(0.055) \\
SINFONIA & 0.336(0.064)	&	0.308(0.054)	&0.322(0.059) &  0.415(0.081)	&	0.383(0.068)	& 0.398(0.074) \\
nnSVG & 0.090(0.028)	&	0.990(0.013) &	0.168(0.045) & 0.111(0.025)	&	0.986(0.017)	& 0.199(0.041) \\
HEARTSVG            & 0.205(0.079) & 0.674(0.261) & 0.284(0.075) & 0.534(0.178) & 0.616(0.258) & 0.494(0.114) \\ \hline
Exponential pattern &              & \textit{Star} Network          &              &              & \textit{Scale-free} Network           &              \\
proposed            & 0.873(0.215) & 1.000(0.000) & 0.913(0.170) & 0.937(0.152) & 1.000(0.001) & 0.958(0.128) \\
spatialDE           & 0.000(0.000) & -(-)         & -(-)         & 0.000(0.000) & -(-)         & -(-)         \\
SPARK               & 0.001(0.002) & 0.897(0.285) & 0.007(0.003) & 0.009(0.009) & 0.965(0.098) & 0.024(0.015) \\
SPARK-X             & 0.698(0.394) & 0.131(0.157) & 0.165(0.041) & 0.211(0.181) & 0.820(0.263) & 0.319(0.224) \\
HRG &    0.430(0.317)	&	0.105(0.015)	 & 0.139(0.052)          & 0.301(0.291) &  0.157(0.072)	& 0.142(0.044) \\
MERINGUE            & 0.004(0.005) & 0.900(0.251) & 0.013(0.011) & 0.038(0.035) & 0.899(0.173) & 0.082(0.061) \\
CTSV      & 0.018(0.029) & 0.103(0.129) & 0.044(0.032) & 0.023(0.032) & 0.250(0.035) & 0.043(0.036) \\
CTSV-g                & 0.733(0.278) & 0.929(0.018) & 0.783(0.215) & 0.878(0.154) & 0.937(0.013) & 0.898(0.095) \\
SINFONIA & 0.190(0.035)	&	0.171(0.027)	& 0.180(0.031) & 0.281(0.065) & 0.253(0.052)	& 0.266(0.058) \\
nnSVG &  0.000(0.000) & -(-) & -(-)   & 0.000(0.000) & -(-) & -(-) \\
HEARTSVG            & 0.029(0.020) & 0.890(0.186) & 0.057(0.036) & 0.193(0.112) & 0.933(0.094) & 0.304(0.154) \\ \hline
Periodic  pattern   &              & \textit{Star} Network           &              &              & \textit{Scale-free} Network           &              \\
proposed            & 0.983(0.099) & 1.000(0.000) & 0.988(0.076) & 0.977(0.134) & 1.000(0.000) & 0.980(0.127) \\
spatialDE           & 0.000(0.000) & -(-)         & -(-)         & 0.001(0.002) & 1.000(0.000) & 0.006(0.004) \\
SPARK               & 0.029(0.026) & 0.954(0.177) & 0.070(0.043) & 0.050(0.037) & 0.975(0.038) & 0.112(0.056) \\
SPARK-X             & 0.723(0.370) & 0.213(0.267) & 0.173(0.033) & 0.437(0.222) & 0.845(0.236) & 0.531(0.246) \\
HRG & 0.436(0.292)	&	0.272(0.159)	 & 0.230(0.041) &  0.595(0.265) &		0.482(0.354)	&0.374(0.128) \\
MERINGUE            & 0.092(0.033) & 0.965(0.035) & 0.166(0.057) & 0.272(0.149) & 0.972(0.024) & 0.401(0.189) \\
CTSV      & 0.061(0.062) & 0.442(0.315) & 0.094(0.074) & 0.098(0.092) & 0.565(0.302) & 0.152(0.114) \\
CTSV-g               & 0.890(0.166) & 0.912(0.024) & 0.889(0.106) & 0.959(0.080) & 0.945(0.010) & 0.950(0.044) \\
SINFONIA & 0.259(0.042)	&		0.234(0.033) &		0.246(0.037) & 0.420(0.089) &		0.380(0.073)	&0.398(0.080) \\
nnSVG & 0.044(0.029)		&	0.995(0.014)		& 0.096(0.045) & 0.064(0.038)	&	0.968(0.148)	& 0.134(0.057) \\ 

HEARTSVG            & 0.075(0.017) & 0.988(0.021) & 0.139(0.029) & 0.159(0.055) & 0.994(0.010) & 0.270(0.083) \\ \hline
\end{tabular}
}
\end{table}

\clearpage
{
\subsection{Results for the SV-free simulated datasets} \quad

\begin{figure}[h]

\centering
\includegraphics[width=0.8\textwidth]{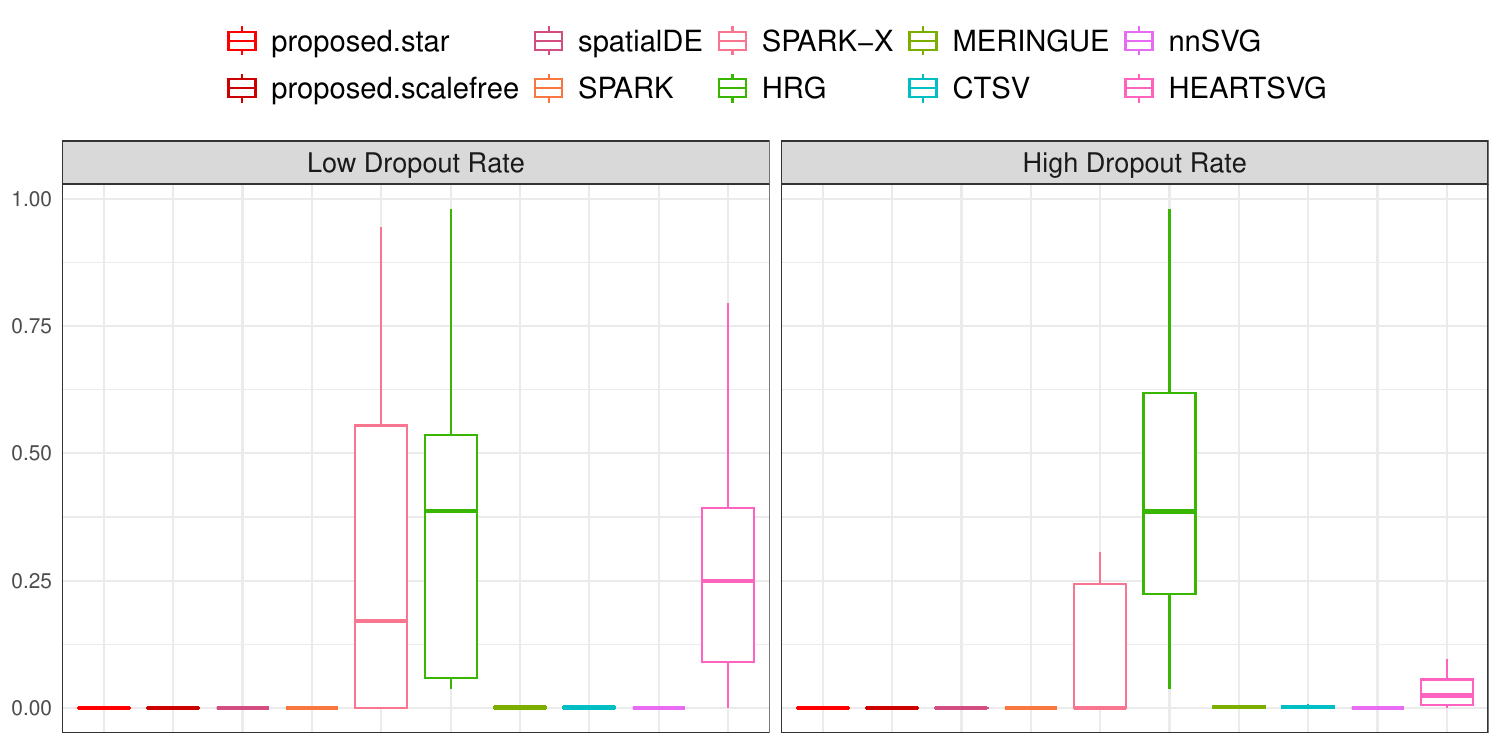}
\caption{ Comparison boxplots of False Positive Rate (FPR) based on 50 replicates under SV-free simulation scenarios, where FDR(BFDR) is controlled to be $<$0.05.
\label{sim_nullcase}}
\end{figure}
}

\clearpage
{
\subsection{Results for the validation of zero-inflation distribution} \quad
\begin{figure}[htb]
\centering
\includegraphics[width=0.8\textwidth]{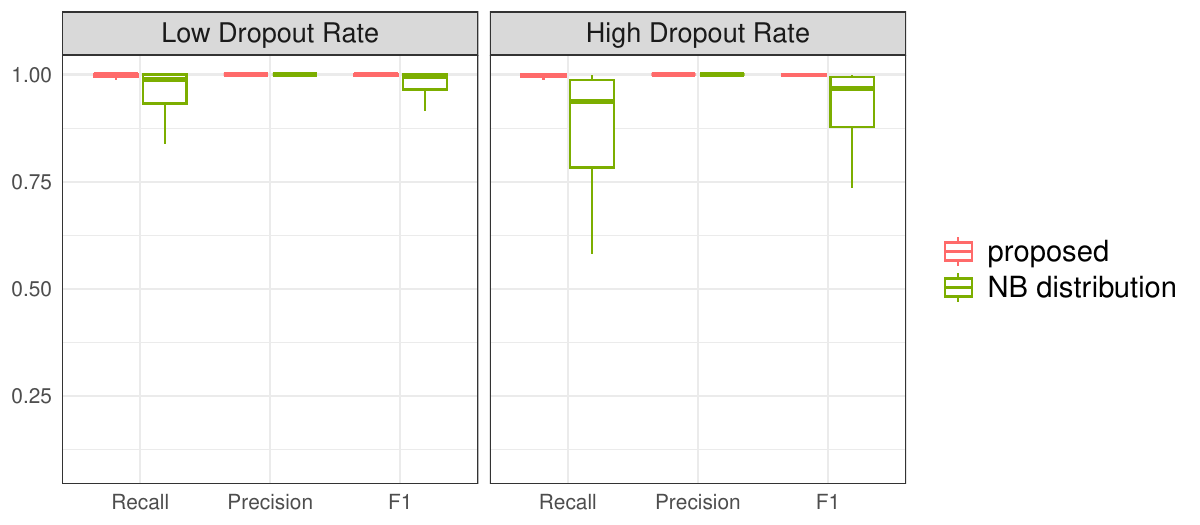}
\caption{ Comparison boxplots of Recall, Precision, and F1 based on 50 replicates under scenarios with star network and periodic pattern, where FDR (BFDR) is controlled to be $<$0.05. \label{fig:sim_NB}}
\end{figure}
}

\clearpage
\subsection{Results for the validation of network assistance strategy} \quad
\begin{figure}[htb]
\centering
\includegraphics[width=\textwidth]{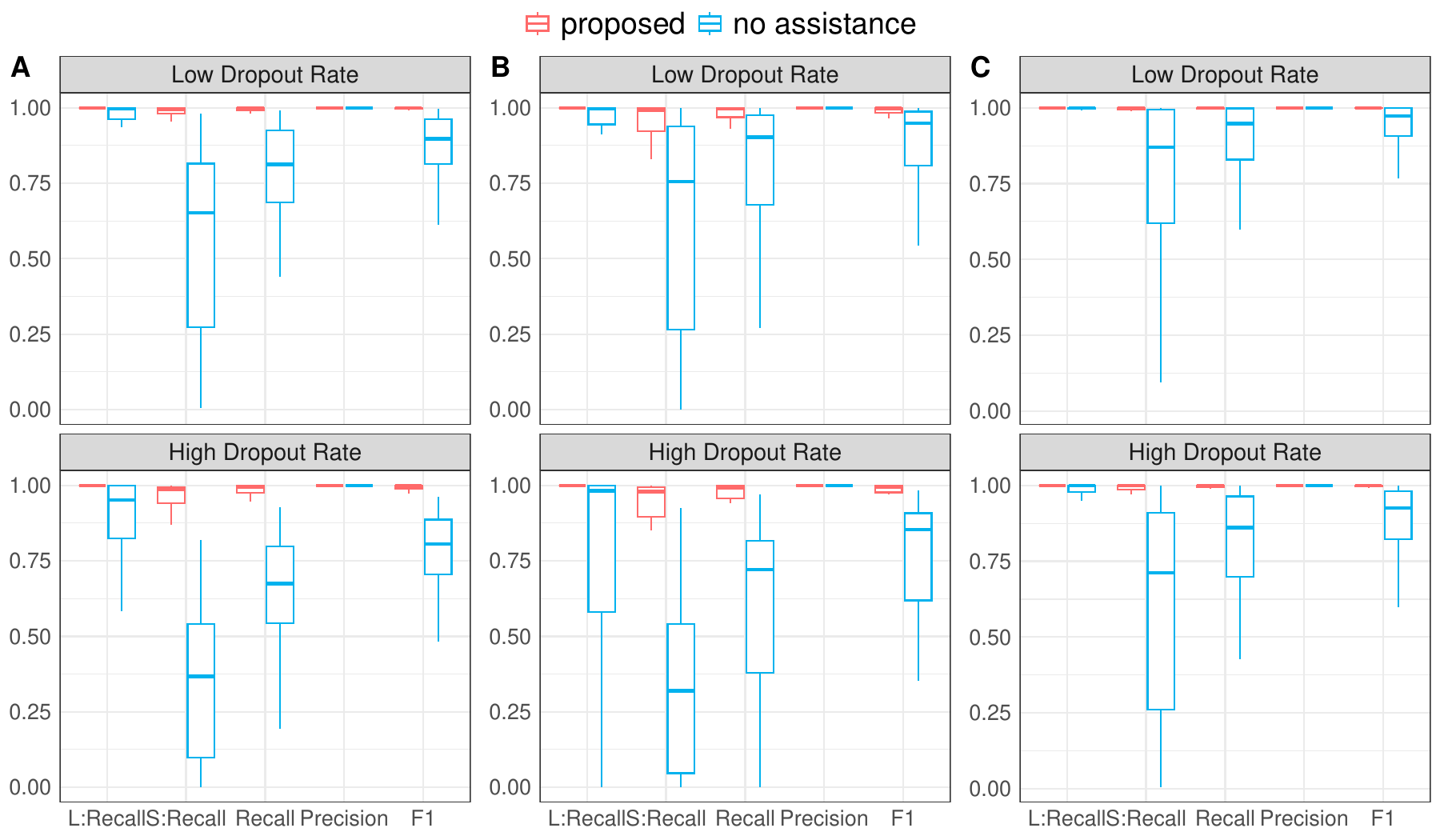}
\caption{Comparison boxplots of Recall for large (L:Recall) and small (S:Recall) signals, overall Recall, Precision, and F1 based on 50 replicates under scenarios with scale-free network and (A) linear pattern, (B) exponential pattern, and (C) periodic pattern, where FDR (BFDR) is controlled to be $<$0.05. \label{fig:sim_null}}
\end{figure}

\clearpage
\subsection{Detailed settings for model misspecified simulations and results} 

We consider three types of model misspecification scenarios as follows.

\begin{figure}[htb]
\centering
\includegraphics[width=0.9\textwidth]{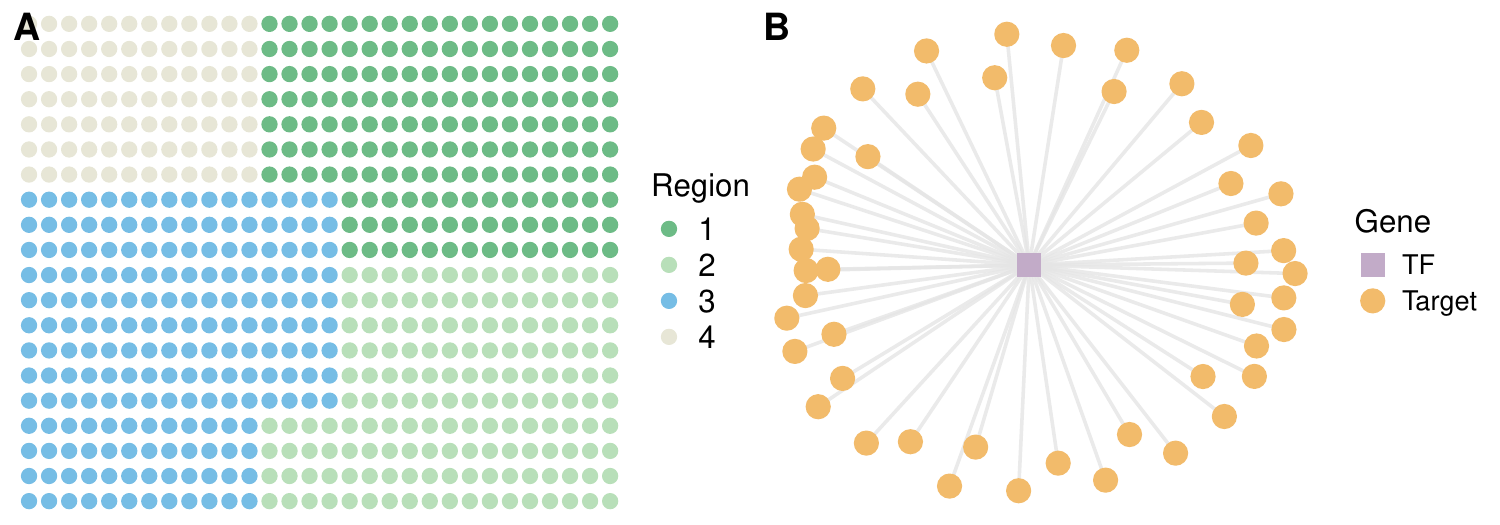}
\caption{(A) Partitioned spot regions considered in the model misspecification scenario (M1). Regions are denoted by different colors, where the cellular compositions $\boldsymbol{w}_i$'s are independently sampled from Dirichlet distributions Dirc(1,1,1,1,1,1) (Region 1), Dirc(1,3,5,7,9,11) (Region 2), Dirc(16,14,12,10,8,6) (Region 3), and Dirc(1,4,4,4,4,1) (Region 4). (B) Illustrative example of the simple regulatory network considered in model misspecification scenarios.}
\label{fig:sim_mis_region_net}
\end{figure}

\begin{figure}[htb]
\centering
\includegraphics[width=0.4\textwidth]{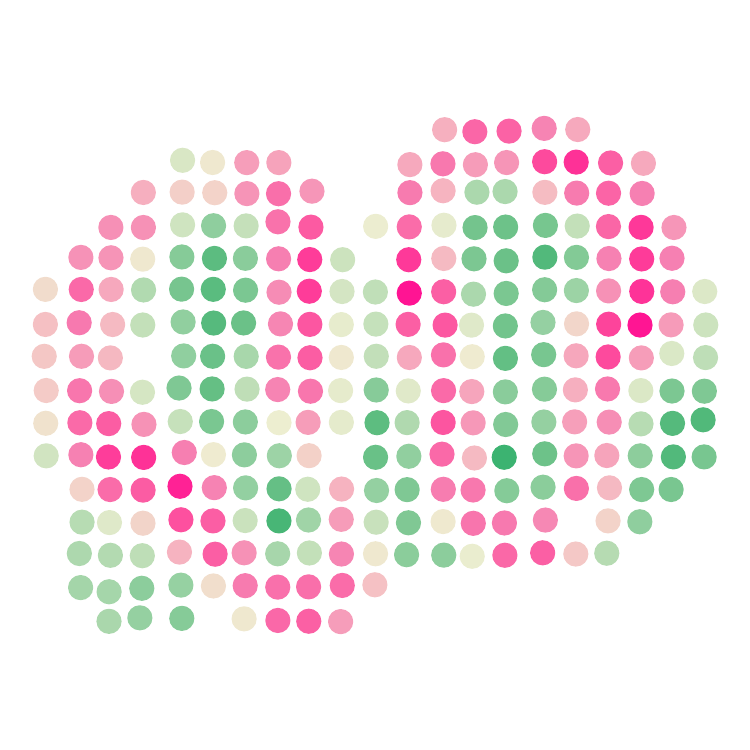}
\caption{Summarized spatial expression pattern inferred from the data analysis of mouse olfactory bulb study. \label{fig:sim_spark_mob}}
\end{figure}

\begin{enumerate}
    \item[(M1)] ZINB model considered in \cite{yu2022identification}. Specifically,  the raw count data $Y_{ij}$ is generated from $\mathrm{NB}\left(Y_{ij}\mid\mu_{ij},\phi\right)^{(1-r_{ij})} \delta_0(Y_{ij})^{r_{ij}},$ where $\log \mu_{i j}=\sum_{k=1}^K \mu^{(k)}_{ij} w_{i k}$ with $\mu^{(k)}_{ij}=\eta_{jk}+\beta^{(1)}_{jk} \mathcal{K}\left(x_{i 1}\right)+\beta^{(2)}_{jk} \mathcal{K}\left(x_{i 2}\right)$. Following \cite{yu2022identification}, consider $n=600$ spots located on a $30 \times  20 $ grid and partitioned into four regions as shown in Figure \ref{fig:sim_mis_region_net}(A). Consider $K=6$ underlying cell types where the cellular composition of spots in regions 1, 2, 3, and 4 are independently sampled from $\operatorname{Dir}(1,1,1,1,1,1)$, $\operatorname{Dir}(1,3$, $5,7,9,11), \operatorname{Dir}(16,14,12,10,8,6)$, and $\operatorname{Dir}(1,4,4,4,4,1)$, respectively. Consider $p=5,000$ genes, involving in a block-wise network composed of 100 disconnected sub-networks with 50 nodes each. Here, each sub-network includes one TF gene and 49 connected target genes, as shown in Figure \ref{fig:sim_mis_region_net}(B). All genes in the first seven sub-networks are SV. Specifically, the $ \left( 50 \cdot (k-1) + 1 \right)$th to $\left( 50 \cdot (k+1) \right)$th genes are set as cell-type-$k$ SV for $k = 1, \ldots, K$, resulting in a total of 350 SV genes. The spatial effect function is fixed as a linear pattern with $\beta^{(1)}_{jk} = 1.8 $ and $\beta^{(2)}_{jk} = 0.8$ for SV genes and $\beta^{(1)}_{jk}$ = $\beta^{(2)}_{jk}$ = 0 for non-SV genes. For $\eta_{jk}$, first simulate $\eta_{j1}$ from $\mathrm{N}\left(2,0.2^2\right)$ for $j=1, \ldots, p$, and then simulate 150 differentially expressed genes for each cell type $k (2 \leq k \leq K)$ from $\mathrm{N}\left(\theta_k, \xi_k^2\right)$ independently, where $\left(\theta_2, \xi_2\right)=(3,0.2),\left(\theta_3, \xi_3\right)=$ $(2,0.2),\left(\theta_4, \xi_4\right)=(4,0.2),\left(\theta_5, \xi_5\right)=(3,0.2),$ and $\left(\theta_6, \xi_6\right)=(4,0.2)$.  In addition, $\phi=100$ and $r_{ij}$'s are simulated from a Bernoulli distribution $\operatorname{Bern}\left(0.6\right)$.

  \item[(M2)] { 
  The generalized linear spatial model where the spatial variability is introduced through the Gaussian covariance matrix as adopted in \cite{svensson2018spatialde} and \cite{sun2020statistical}.  Specifically, for gene $j$, the raw count data is generated through a zero-inflated Poisson distribution $ y_j\left(x_i\right) \sim \operatorname{Poi}\left(\lambda_j\left(x_i\right)\right)^{(1 - r_{ij})} \delta_0(y_j\left(x_i\right))^{r_{ij}}$ with $\left(\log \left(\lambda_j\left(x_1\right)\right), \ldots, \log \left(\lambda_j \left(x_n\right)\right)\right)^{\mathrm{T}} \sim \mathrm{N}\left(\mu_j \mathbf{1}, \tau_j^2 \Sigma+\sigma_j^2 \mathbf{I}\right)$. Consider $n = 400$ spots located on a $20 \times 20 $ grid and $p  = 5, 000$ genes. The network is set as the same as (M1) (as shown in Figure \ref{fig:sim_mis_region_net}(B)). The genes in the first ten sub-networks are SV, resulting in a number of 500 SV genes. Following \cite{sun2020statistical}, the Gaussian covariance matrix $\Sigma$ is set as $(\Sigma_{ii'})_{n \times n } = \left( \exp \left(-\frac{\left\|x_i-x_{i'}\right\|^2}{2 l^2}\right)\right)_{n \times n }$ with length scale $l$ set as 1.  For SV gene, $\tau^2_j$ (accounts for the expression variance attributable to spatial effects) is set as 0.7 while $\sigma_j^2$ is set as 0.3. For non-SV gene, $\tau^2_j = 0$ and $\sigma^2_j = 1$. No cellular composition variations are considered in this setting. ${r}_{ij}$'s are simulated from a Bernoulli distribution $\operatorname{Bern}\left(0.5\right)$.

} 
  
    \item[(M3)] { 
    Consider 260 spots collected in the mouse olfactory bulb study \citep{staahl2016visualization} and generate simulated data based on Poisson generalized linear model adopted in \cite{sun2020statistical} with the corresponding parameters inferred from the data analysis results conducted by spatialDE and SPARK. } Specifically, following the setting of \cite{sun2020statistical}, the raw count data is generated as $Y_{ij} \sim \operatorname{Poi}( N_{i} \cdot \lambda_{ij})$, where $\log \lambda_{ij} = \operatorname{Intercept}_{ij} + \epsilon_{ij} + \sum_{k=1}^{K}\alpha_k w_{ik}$ with $\operatorname{Intercept}_{ij}$ and $\epsilon_{ij}$ being the intercept and residual error term, respectively. 
    Consider $p= 5,000$ genes, the network is set as the same as (M1) and (M2) (Figure \ref{fig:sim_mis_region_net}(B)) with all genes in the first ten sub-networks set as SV.  For non-SV genes, $\operatorname{Intercept}_{ij}$ is set to be $-10.2$ across all spots, which corresponds to the median of the intercept estimates in the mouse olfactory data analysis. For SV genes, the spots are categorized into two groups, including the group with low expression levels (green) and the group with high expression levels (pink), according to the spatial patterns of the identified SV genes in the mouse olfactory bulb data, as illustrated in Figure \ref{fig:sim_spark_mob}. In particular, $\operatorname{Intercept}_{ij}$'s for the low group are set as $-10.2$ while $\operatorname{Intercept}_{ij}$'s for the high group are set as $-8.8$ and $-17.8$ for the target and TF genes, respectively, according to the estimates for mean values. $\epsilon_{ij}$'s are independently simulated from $\mathrm{N}(0, 0.2^2)$, which is approximately the first quantile of the non-spatial variance estimates. For cellular composition variations, consider $K=6$ underlying cell types, where $w_{ik}$'s of spots from high and low groups are independently sampled from $\operatorname{Dirc}(3,5,7,9,11,13)$ and $\operatorname{Dirc}(18,16,14,12,10,8)$, respectively. $\alpha_k$ is set as 0 for $k = 1, \ldots, K$, so that the cellular composition variations do not contribute to the expression levels. $N_i$'s are obtained based on the real counts.




\end{enumerate}

\begin{figure}[htb]
\centering
\includegraphics[width=5in]{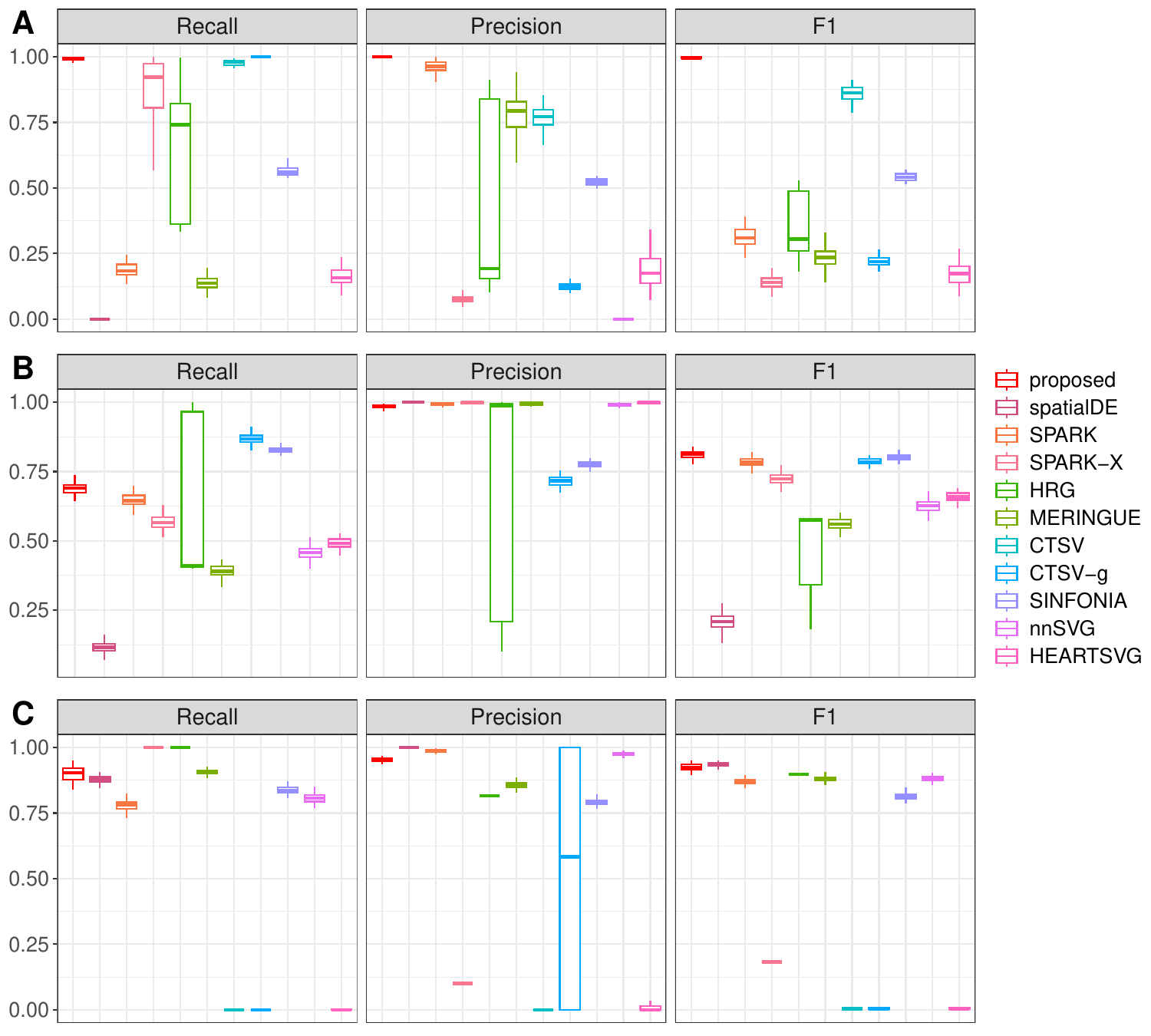}
\caption{
 Comparison boxplots of Recall, Precision, and F1 based on 50 replicates in model misspecification scenarios (A) M1, (B) M2, and (C) M3, where FDR (BFDR) is controlled to be $<$0.05.
\label{fig:sim_misspe}}
\end{figure}

\clearpage
\subsection{Results for the examination on the noise of network and cellular composition} \quad

\begin{figure}[htb]
\centering
\includegraphics[width=5in]{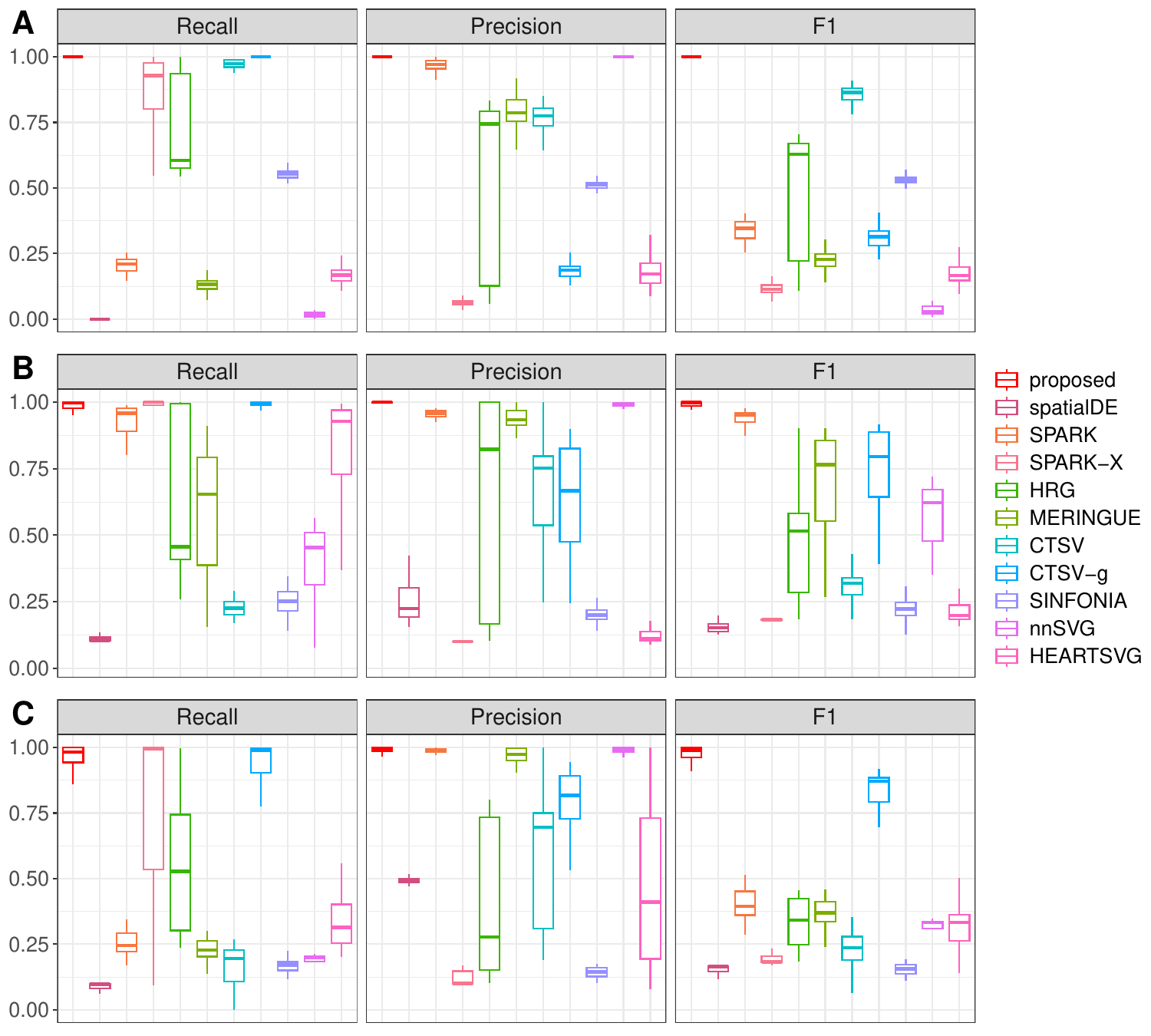}
\caption{
 Comparison boxplots of Recall, Precision, and F1 based on 50 replicates, under (A) ZINB model considered in \cite{yu2022identification}; (B) ZINB model considered in Section 3.1 of the main text with star network, linear spatial pattern, and low dropout rate = 0.1; (C) ZINB model considered in Section 3.1 of the main text with star network, linear spatial pattern, and high dropout rate = 0.5, where in each informative sub-network, about 20\% of the genes connected to the TF genes are set as uninformative and FDR (BFDR) is controlled to be $<$0.05.
\label{fig:sim_wrong_net}}
\end{figure}

\begin{figure}[htb]
\centering
\includegraphics[width=5in]{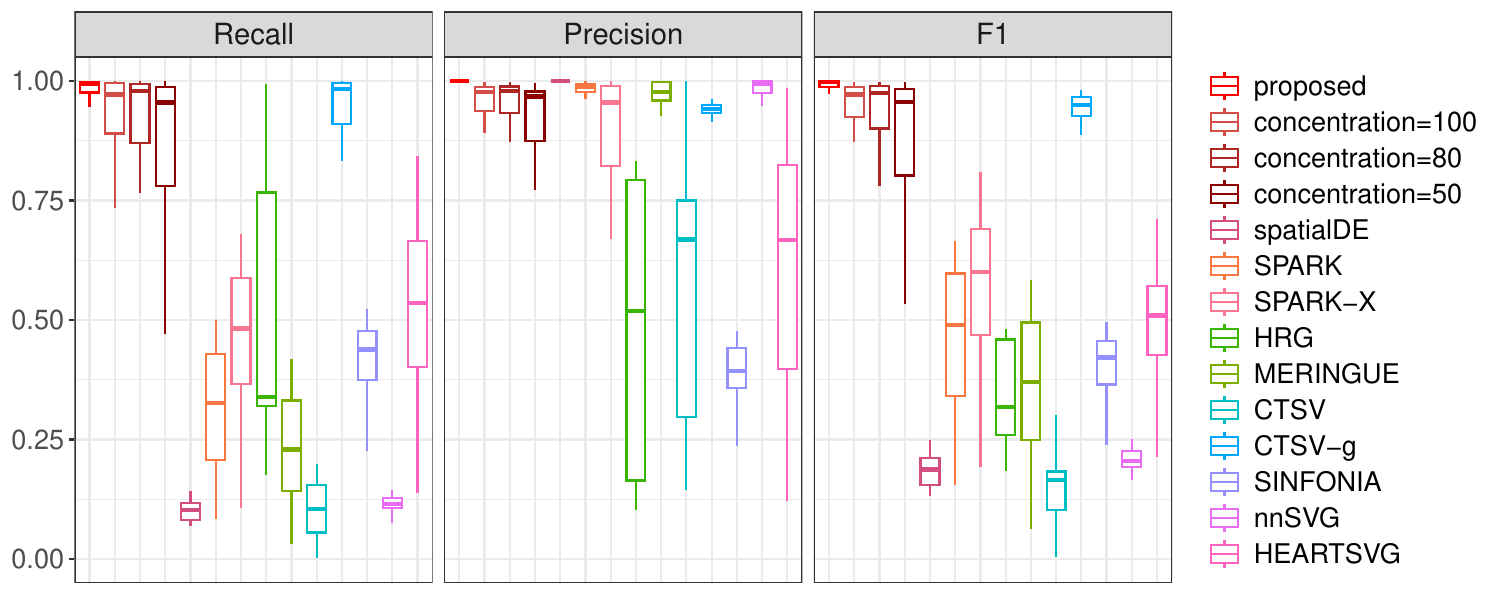}
\caption{ 
Comparison boxplots of Recall, Precision, and F1 with the proposed approach with true cellular compositions (proposed) and different degrees of cellular estimate errors (concentration = 100, 80, and 50) based on 50 replicates, under ZINB model considered in Section 3.1 of the main text with scale-free network, linear pattern, and high dropout rate = 0.5, where
FDR (BFDR) is controlled to be $<$0.05. \label{fig:sim_errorW}}
\end{figure}

\clearpage
\section{Implementation details of the competing methods} 

\begin{enumerate}
    \item SpatialDE: The source code for SpatialDE is publicly available at
    \url{https://github.com/Teichlab/SpatialDE}. Following the instructions, we filter the practically unobserved genes with total observations less than 3. The linear regression is conducted on the raw count matrix to account for the potential bias caused by library size or sequencing depth. The main function for SV gene detection is \texttt{SpatialDE.run} with the arguments being the coordinate information $\left(x_{i1}, x_{i2}\right)$ and the sample residual expressions corrected by the linear regression. To account for multiple testing, the genes with \textit{qval} less than 0.05 are identified as SV.

    \item SPARK: The source code for SPARK is publicly available at
    \url{https://github.com/xzhoulab/SPARK}. Following the instructions, we filter genes that are expressed in less than 10\% spots and spots with the total observations less than 10. The statistical model under the null hypothesis is first fit by employing the function \texttt{spark.vc}, and the function \texttt{spark.test} is subsequently employed for SV gene detection. Then, the genes with \textit{adjusted\_pvalue} less than 0.05 are identified as SV.

    \item SPARK-X: The source code for SPARK-X is publicly available at
    \url{https://github.com/xzhoulab/SPARK}. The SV gene detection is implemented with the main function \texttt{sparkx} where the argument \textit{option}  is set as ``mixture" for multiple kernels testing. To account for the impact of latent cellular compositions, as conducted in the original paper, each spot is assigned its major cell type, with the argument \textsl{X\_in} being the matrix of binary indicators. Then, the genes with \textsl{adjustedPval} less than 0.05 are identified as SV.

    \item { HRG: The source code for HRG is publicly available at \url{https://github.com/JulieBaker1/HighlyRegionalGenes}. We first run PCA by employing the function \texttt{runPCA} and use the first 10 principal components for subsequent analysis. The identification of highly regional genes is implemented with the main function \texttt{FindRegionalGenes}, where the gene number is automatically chosen by finding knee point by employing function \texttt{HRG\_elbowplot} as recommended.}

    \item  MERINGUE: The source code for MERINGUE is publicly available at  \url{https://github.com/JEFworks-Lab/MERINGUE}. The non-expressed genes and spots are filtered following the tutorial. Normalization is conducted for the raw count matrix by employing the function \texttt{normalizeCounts}. For SV gene detection, the neighborhood relationships are first constructed using the function \texttt{getSpatialNeighbors} with the argument \textit{filterDist} set as the default value 2.5. Then, the SV genes are identified with the function \texttt{filterSpatialPatterns} with the arguments \textit{minPercentCells} set as 0.05 to restrict that the SV genes are driven by more than 5\% of spots. The adjusted significance threshold is set as 0.05 through setting the arguments \textit{adjustPv} as TRUE and \textit{alpha} as 0.05.

    \item CTSV and CTSV-g: The source code for CTSV is publicly available at
    \url{https://github.com/jingeyu/CTSV}. The SV gene detection is conducted through the main function \texttt{ctsv}. Specifically, the cell-type-specific SV genes are identified with the argument $W$ being the $n \times K $ matrix composed of $w_i$'s ($i = 1, \ldots, n$). The final SV gene set is the union of all cell-type-specific SV genes. We also conduct the global SV gene detection without accommodation for the cellular composition in the simulation studies (CTSV-g). Specifically, the argument \textit{$W$} is set as $(1, \ldots, 1)^{\mathrm{T}}_{(n)}$. For both CTSV and CTSV-g, the SV genes are identified through function \texttt{svGene} with the significance threshold \textit{thre.alpha} set as 0.05.

\item { SINFONIA: The source code for SINFONIA is publicly available at \url{https://github.com/BioX-NKU/SINFONIA}. The SV gene identification is performed using the main function \texttt{sinfonia.spatially\_variable\_genes}, with the raw count matrix first normalized and log-transformed. 

\item  nnSVG: The source code for nnSVG is publicly available at \url{https://github.com/lmweber/nnSVG}. 
Following the recommended procedure, genes with at least 3 counts in at least 0.5\% of spatial locations are log-transformed prior to downstream analysis. SV gene detection is performed using the main function \texttt{nnSVG}, where the genes with adjusted P values \textit{padj} less than 0.05 are identified as SV.}

    \item HEARTSVG: The source code for HEARTSVG is publicly available at \url{https://github.com/cz0316/HEARTSVG}. The detection of SV genes is conducted through the main function \texttt{heartsvg} with the raw count matrix first scaled as recommended. The Holm method is further conducted for multiple testing control. The genes with adjusted P values \textit{p\_adj} less than 0.05 are identified as SV.

\end{enumerate}

\clearpage
{
\section{ Data analysis on  Visium PLC dataset}

We apply the proposed approach to the primary liver cancer (PLC) dataset \citep{wu2021comprehensive} sequenced by the 10x Genomics Visium platform. The original PLC data could be obtained from http://lifeome.net/supp/livercancer-st/data.htm. We first perform the quality control, filtering out spots with fewer than 500 expressed genes and genes with non-zero expressions in less than 5 spots. A total of 3,181 spots and 18,661 genes are retained. The top 5,000 highly variable genes are selected for downstream analysis. After preprocessing, the median value of expressed counts across spots is 7,287 with the zero expression proportion being about 71.31\%.

The PPI network is considered for network dependency information incorporation where 1,948 connected genes are involved in 160 disconnected sub-networks after matching with STRING database. The remaining 3,052 genes are treated as singleton nodes. The cellular proportion estimates across spots for six major cell types are obtained based on Redeconve algorithm.

Analysis is conducted by the proposed approach along with the nine alternatives. The proposed approach identifies 1,575 SV genes. An upset plot illustrating the number and overlap of SV genes identified by each method is shown in Figure \ref{fig:upset_PLC}. Among them, SPARK-X detects the most SV genes (3,054), including 365 uniquely identified genes. MERINGUE identifies the fewest SV genes (744). Notably, 362 SV genes are consistently selected by all ten methods. There is an overlapping shared gene set (108) by SPARK-X, SPARK, spatialDE, SINFONIA, HRG,  HEARTSVG, nnSVG, and MERINGUE. The spatial expression patterns of two representative genes \textit{ZNF627} and \textit{HSD17B6} from this shared gene set are shown in Figure \ref{fig:pie_PLC}(B), with noticeable cell type diversity as compared to Figure \ref{fig:pie_PLC}(A).

\begin{figure}[htb]
\centering
\includegraphics[width=0.8\textwidth]{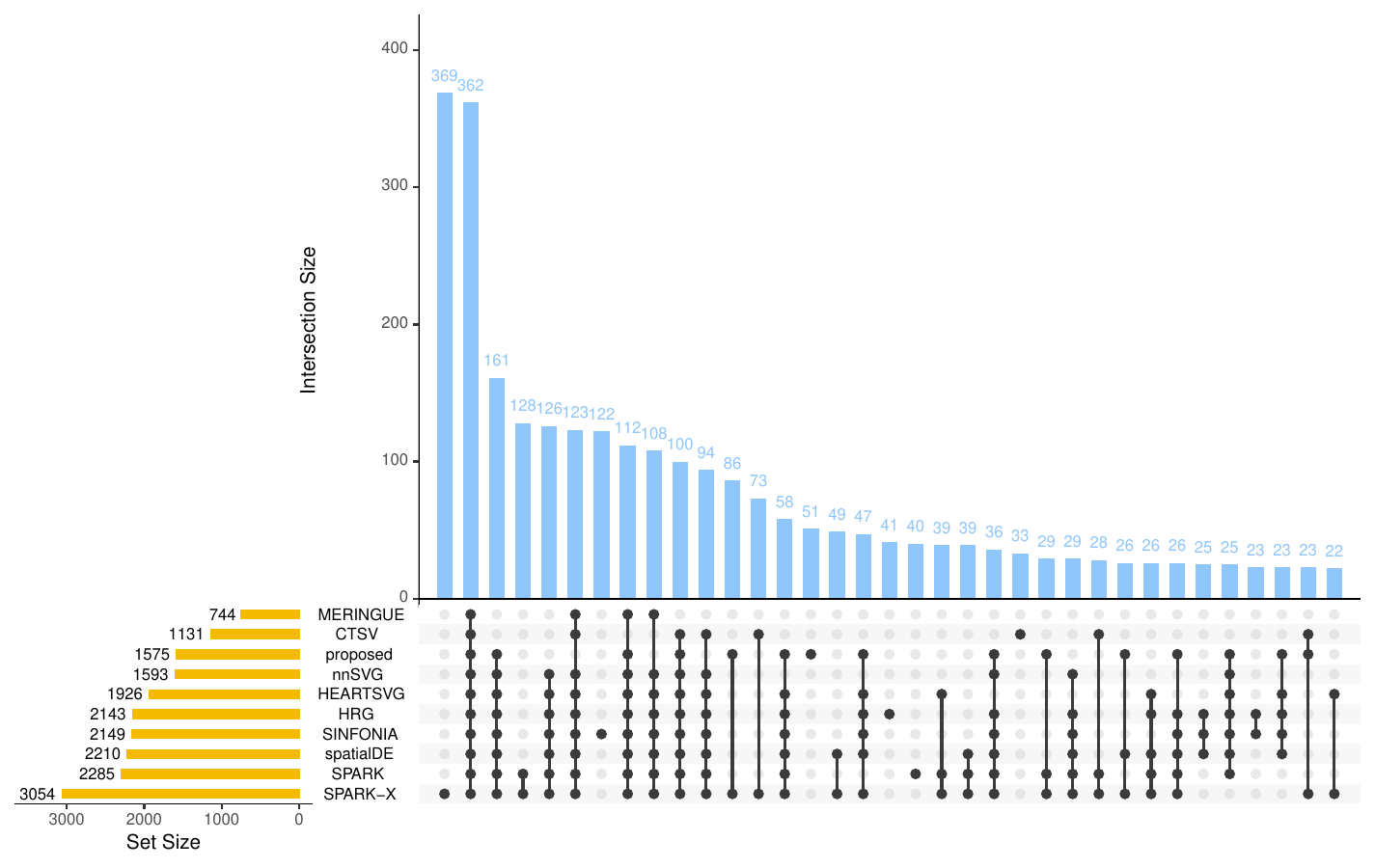}
\caption{Upset plot of the numbers of SV genes identified by different approaches and their overlaps for Visium PLC dataset.
\label{fig:upset_PLC}}
\end{figure}

\begin{figure}[htb]
\centering
\includegraphics[width=0.7\textwidth]{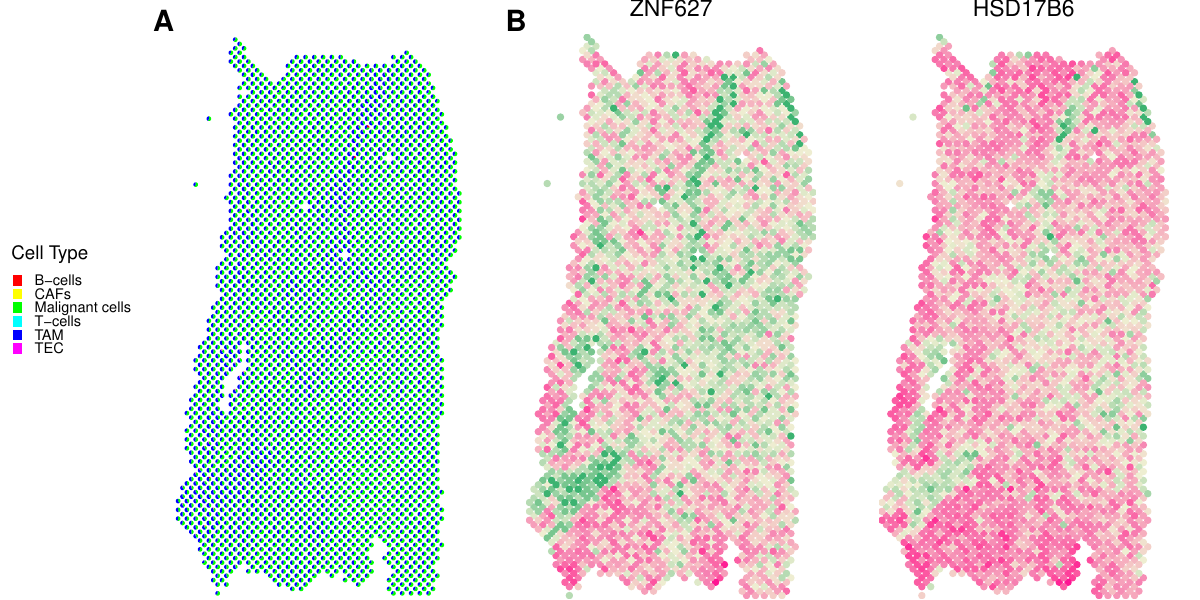}
\caption{(A) Pie charts of the cell type distributions across spots for Visium PLC dataset. (B) Spatial expression patterns of two representative SV genes identified by all of SPARK-X, SPARK, spatialDE, SINFONIA, HRG,  HEARTSVG, nnSVG, and MERINGUE.
\label{fig:pie_PLC}}
\end{figure}

Figure \ref{fig:PLC_oneside} further presents four representative SV genes identified by the proposed approach. It is observed that certain genes illustrate noticeable spatial expression patterns that vary predominantly along one spatial axis while remaining nearly constant along the other. This observation highlights a key advantage of our proposed approach to capture axis-specific spatial variation with finer biological insights and interpretability.

\begin{figure}[htb]
\centering
\includegraphics[width=\textwidth]{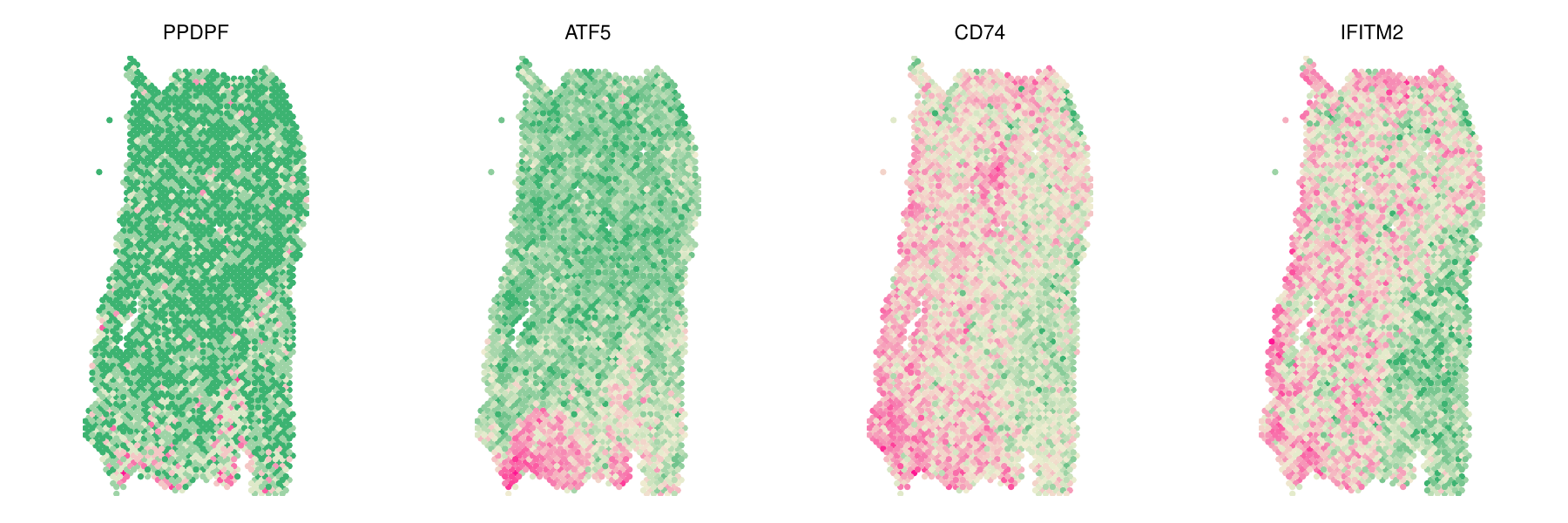}
\caption{Spatial expression patterns of four representative SV genes identified by the proposed approach with noticeable expression variations along a specific coordinate for Visium PLC dataset.
\label{fig:PLC_oneside}}
\end{figure}

A total of 1,035 connected SV genes are identified, and their network structure is presented in Figure \ref{fig:network_PLC}(A). We further perform community detection for these genes and visualize two representative communities in Figure \ref{fig:network_PLC}(B). Compared to the Overlap hub genes consistently identified by all methods, the proposed Only genes are often connected to these hubs or serve as bridging genes, suggesting they may help complete or coordinate the same biological functions. 
GO enrichment analysis is conducted for these two communities with the top five associated significant GO terms listed in Table \ref{tab:PLC_GO}. These enriched GO terms are found biologically significant. Specifically, The genes in community A are significantly involved in the carbohydrate catabolic process (GO:0016052), a key component of cancer metabolic reprogramming. In liver cancer, enhanced carbohydrate breakdown supports tumor growth and survival, as highlighted by recent findings \citep{yang2023metabolic}, highlighting its relevance to cancer progression. The G protein-coupled receptor signaling pathway enriched in community B has been implicated to be permissive for tumor formation and growth \citep{liu2016Gprotein}, and is considered as one of the most useful drug targets against many solid cancers \citep{chaudhary2021insight,li2022g}.

\begin{figure}[htb]
\centering
\includegraphics[width=3.5in]{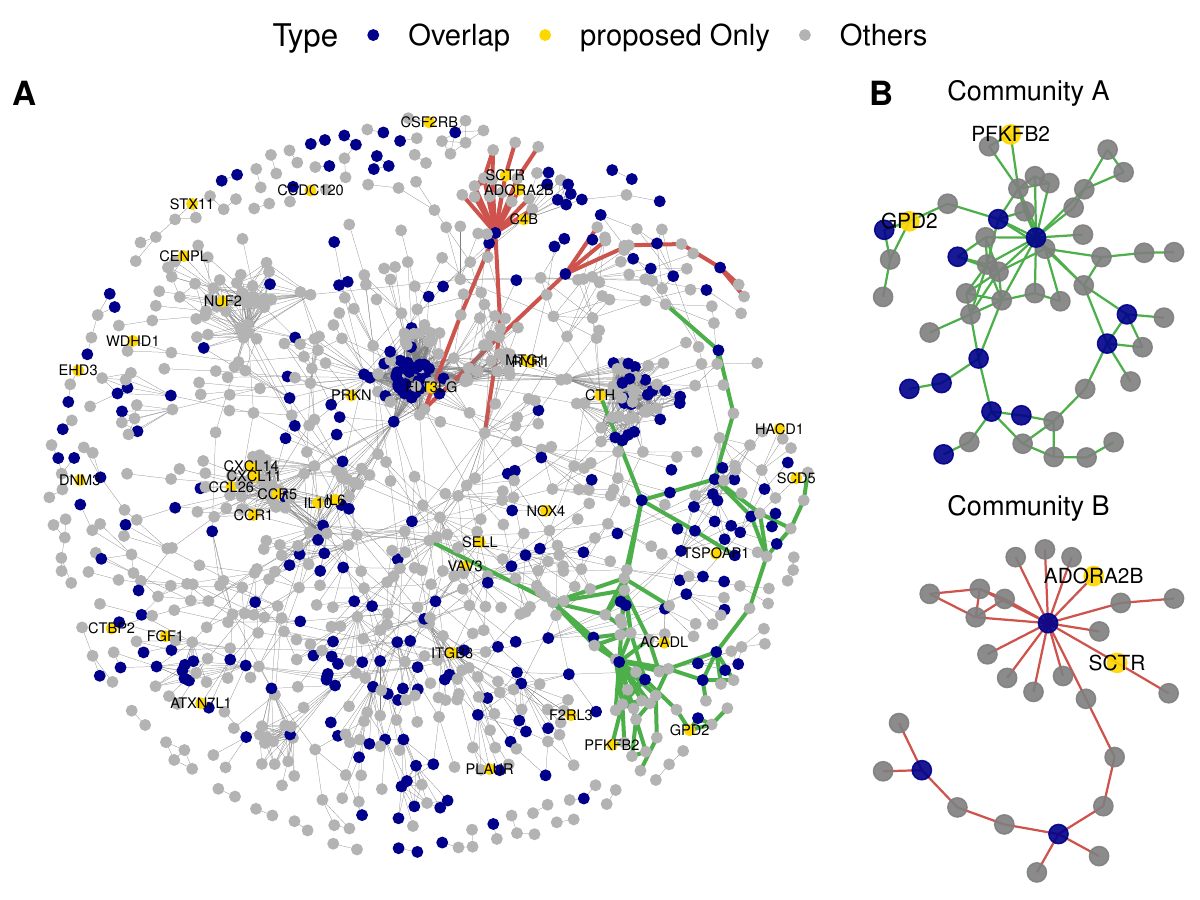}
\caption{ (A) Network of 1,035 connected SV genes identified by the proposed approach for Visium PLC dataset. (B) Network of two representative detected communities.
\label{fig:network_PLC}}
\end{figure}

\begin{table}[htb]
\caption{Top five significant GO terms associated with the two representative communities detected in Visium PLC dataset.\label{tab:PLC_GO}}
\centering
\begin{tabular}{cll}
\hline
\multicolumn{1}{l}{ID} & \multicolumn{1}{l}{adjusted P value} & Description              \\ \hline
\multicolumn{3}{c}{\textit{Community A}}                                                                                                                      \\
GO:0016052            & 4.76$\times 10 ^{-28}$                             & carbohydrate catabolic process                                                                      \\
GO:0006090           & 3.06$\times 10 ^{-23}$                            & pyruvate metabolic process                                                                     \\
GO:0044282            & 2.52$\times 10 ^{-22}$                             & small molecule catabolic process \\
GO:0005996            & 9.12$\times 10 ^{-21}$                             & monosaccharide metabolic process                                               \\
GO:0019318           & 4.18$\times 10 ^{-20}$                             & hexose metabolic process                                                 \\ \hline
\multicolumn{3}{c}{\textit{Community B}} \\
GO:0007188           & 2.80$\times 10 ^{-22}$                             &  adenylate cyclase-modulating G protein-coupled receptor signaling pathway\\
GO:0007189            & 1.89$\times 10 ^{-17}$                             &adenylate cyclase-activating G protein-coupled receptor signaling pathway                                               \\
GO:0042277             & 4.92$\times 10 ^{-9}$                            & peptide binding          \\
GO:0033218           & 2.12$\times 10 ^{-8}$                             & amide binding                                                      \\
GO:0008528            & 3.37$\times 10^{-8}$                             & G protein-coupled peptide receptor activity       
               \\ \hline
\end{tabular}
\end{table}

For further validation for the genes uniquely identified by the proposed approach, comparative GO enrichment analysis results with and without the inclusion of proposed Only genes are provided in Figure \ref{fig:GO_PLC}. The results indicate that incorporating these unique genes leads to the detection of much more significant GO terms. Additionally, several novel GO terms emerge only when the proposed Only genes are included, as illustrated in Figure \ref{fig:GO_PLC} (Right).

\begin{figure}[htb]
\centering
\includegraphics[width=0.9\textwidth]{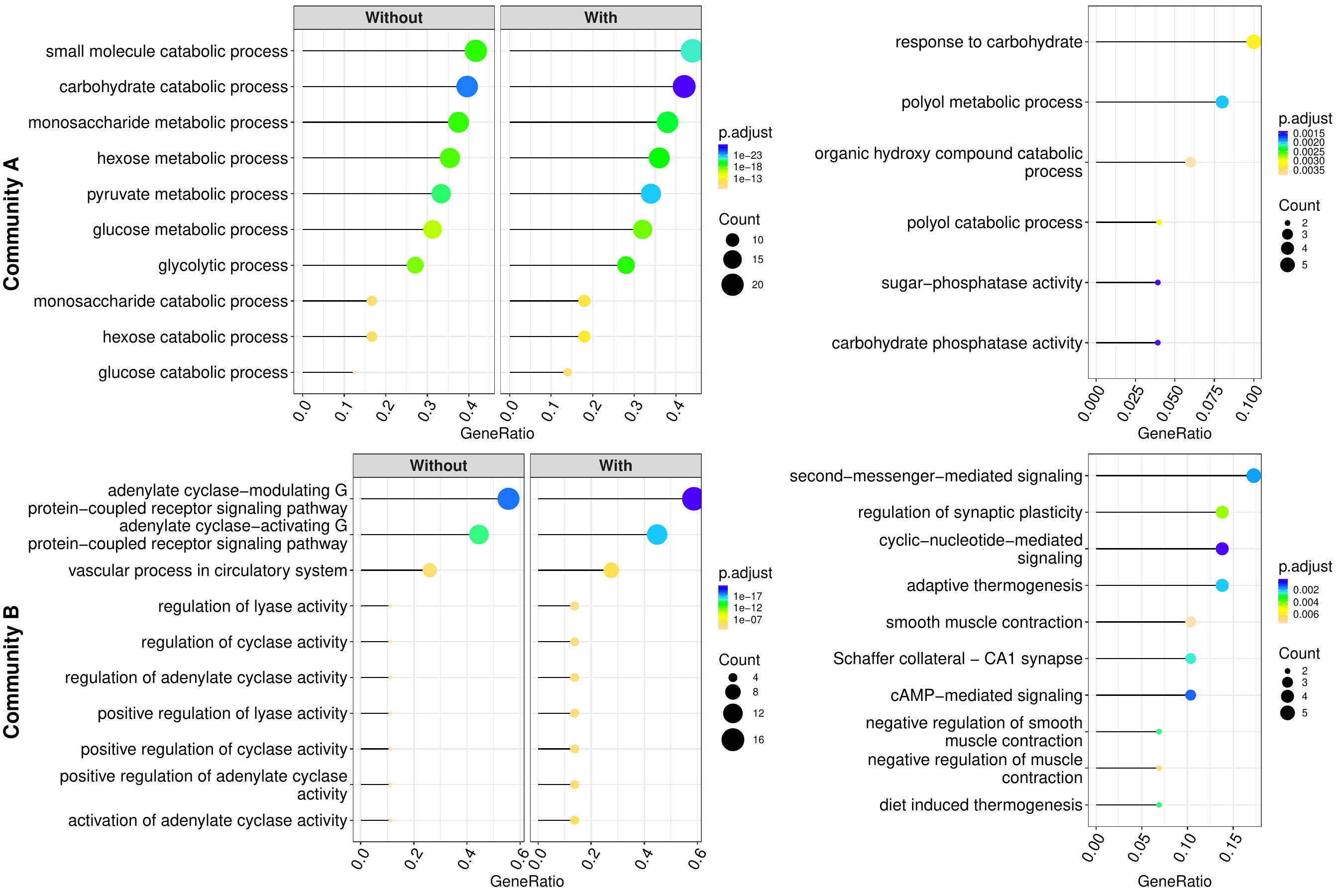}
\caption{ Lollipop plots of GO enrichment analysis for Visium PLC dataset. Left: Comparison results of the analysis without and with the inclusion of proposed Only genes. Right: Newly detected significant GO terms with the proposed Only genes included.
\label{fig:GO_PLC}}
\end{figure}

\section{Data analysis on Xenium PAC dataset}

We apply the proposed approach to the pancreatic cancer (PAC) dataset generated using the 10x Genomics Xenium in situ platform. This dataset profiles the expression of the targeted gene panel across 190,965 cells. The original dataset is publicly available at  
https://www.10xgenomics.com/datasets/pancreatic-cancer-with-xenium-human-multi-tissue-and-cancer-panel-1-standard. Cells with fewer than 50 total counts and genes expressed in less than 10 cells are excluded for quality control. After preprocessing, 164,274 cells with expression profiles across 474 genes are retained for downstream analysis. The median value of expressed counts per cell is 127 with the zero expression proportion being about 86.37\%.

The network is constructed using PPI information from the STRING database. Among the 474 genes, 221 genes form connections, resulting in 21 disconnected sub-networks. Notably, as the Xenium in situ platform provides spatially resolved gene expression at single-cell resolution, no additional correction for cellular composition is performed.

The proposed approach identifies 224 SV genes.
In addition to the proposed approach, analysis is also conducted by alternatives with the upset plot illustrating the number and overlap of SV genes identified by each method shown in Figure \ref{fig:upset_PAC}. It is worth noting that the computational and memory complexity of methods such as SpatialDE, SPARK, and MERINGUE increases quadratically or cubically with the number of sequenced cells. As a result, these methods are not capable of handling datasets at this scale, and their results are therefore not included for comparison. It is observed that among 474 genes, 118 overlap genes are consistently identified by all seven methods. In addition, more than 85\% genes are identified as SV genes by nnSVG and SPARK-X while the proposed approach identified a moderate number of 224 genes.  

\begin{figure}[htb]
\centering
\includegraphics[width=0.8\textwidth]{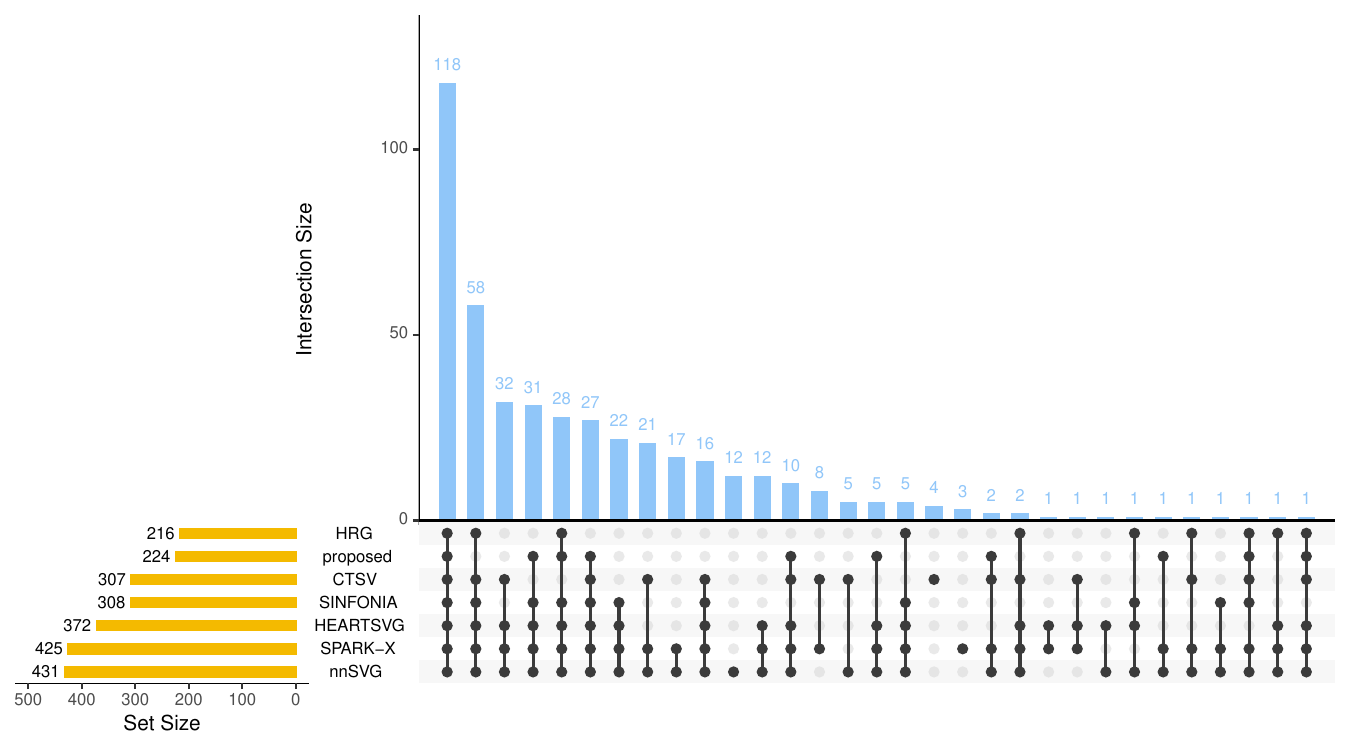}
\caption{ Upset plot of the numbers of SV genes identified by different approaches and their overlaps for Xenium PAC dataset.
\label{fig:upset_PAC}}
\end{figure}

A total of 113 connected SV genes are identified by the proposed approach with their network visualization presented in Figure \ref{fig:network_PAC}. GO enrichment analysis results of the two detected communities are provided in Table \ref{tab:GO_PAC}. The biological implications have been supported by relevant literature. Specifically, the enrichment of the cell killing pathway (GO:0001906) in community A is supported by previous research highlighting the importance of regulated cell death in pancreatic cancer. As discussed in \cite{chen2021cell}, cell death plays a central role in shaping the progression and response of pancreatic cancer treatment, underscoring the biological and clinical relevance of this pathway.
For community B, the enrichment of the MAPK regulation is consistent with previous findings showing that activation or inhibition of this pathway plays a key role in the regulation of pancreatic cancer progression \citep{lin2021trpm2, zhang2022isoliquiritigenin}.

 Within the network, we also highlight a subset of target genes identified by only a few methods. For example, \textit{GZMB} and \textit{NKG7} are detected only by nnSVG, SPARK-X, HEARTSVG, and the proposed approach, while \textit{EGFR} in community B is not selected by SINFONIA or HRG. These genes are integrated within specific communities and may share functional roles with hub genes, suggesting their potential biological relevance. Further validation for these target genes are presented in Figure \ref{fig:GO_PAC} where the inclusion of these genes not only enhances the significance of previously detected GO terms but also leads to the discovery of novel terms, further supporting their functional importance.

\begin{figure}[htb]
\centering
\includegraphics[width=3.5in]{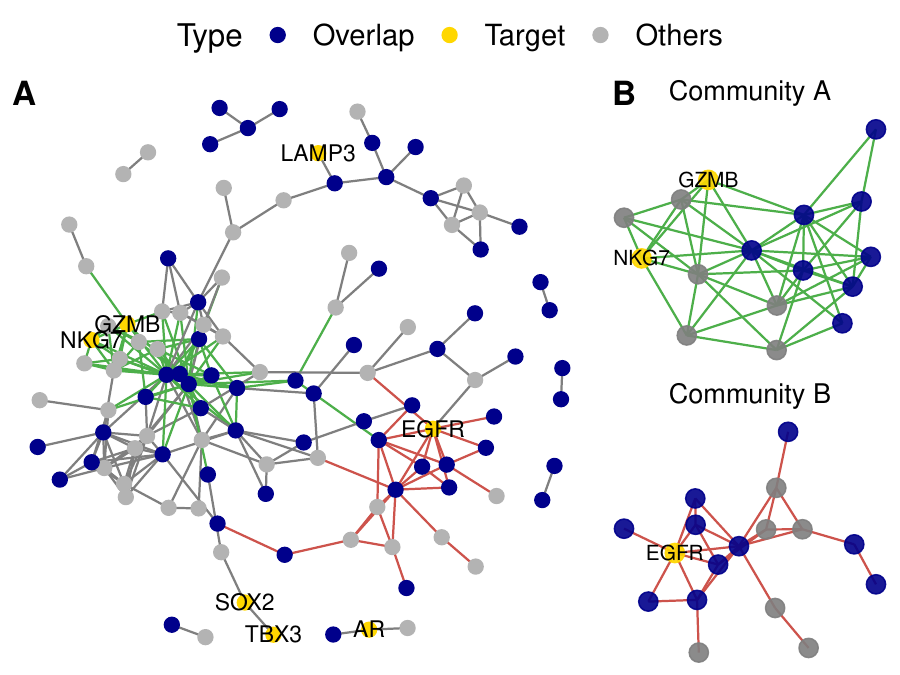}
\caption{ (A) Network of 113 connected SV genes identified by the proposed approach for Xenium PAC dataset. (B) Network of two representative detected communities.
\label{fig:network_PAC}}
\end{figure}

\begin{table}[htb]
\caption{Top five significant GO terms associated with the two representative communities detected in Xenium PAC dataset.\label{tab:GO_PAC}}
\centering
\begin{tabular}{cll}
\hline
\multicolumn{1}{l}{ID} & \multicolumn{1}{l}{adjusted P value} & Description              \\ \hline
\multicolumn{3}{c}{\textit{Community A}}                                                                                                                      \\
GO:0001906            & 1.43$\times 10 ^{-11}$                             & cell killing            \\
GO:0001909           & 1.62$\times 10 ^{-11}$                            &leukocyte mediated cytotoxicity \\
GO:0009897            & 1.04$\times 10^{-10}$                             & external side of plasma membrane \\
GO:0002449           & 4.41$\times 10 ^{-10}$                             & lymphocyte mediated immunity\\
GO:0042267           & 2.02$\times 10 ^{-9}$                             & natural killer cell mediated cytotoxicity \\ \hline
\multicolumn{3}{c}{\textit{Community B}} \\
GO:0043410           & 3.85$\times 10 ^{-6}$                             &  positive regulation of MAPK cascade\\
GO:0005179           & 5.91$\times 10 ^{-6}$                             &hormone activity                                          \\
GO:0004714             & 9.37$\times 10 ^{-6}$                            & transmembrane receptor protein tyrosine kinase activity      \\
GO:0005159           & 1.04$\times 10 ^{-5}$                             & insulin-like growth factor receptor binding \\
GO:0019199            & 1.43$\times 10^{-5}$                             & transmembrane receptor protein kinase activity       
               \\ \hline
\end{tabular}
\end{table}

\begin{figure}[htb]
\centering
\includegraphics[width=0.9\textwidth]{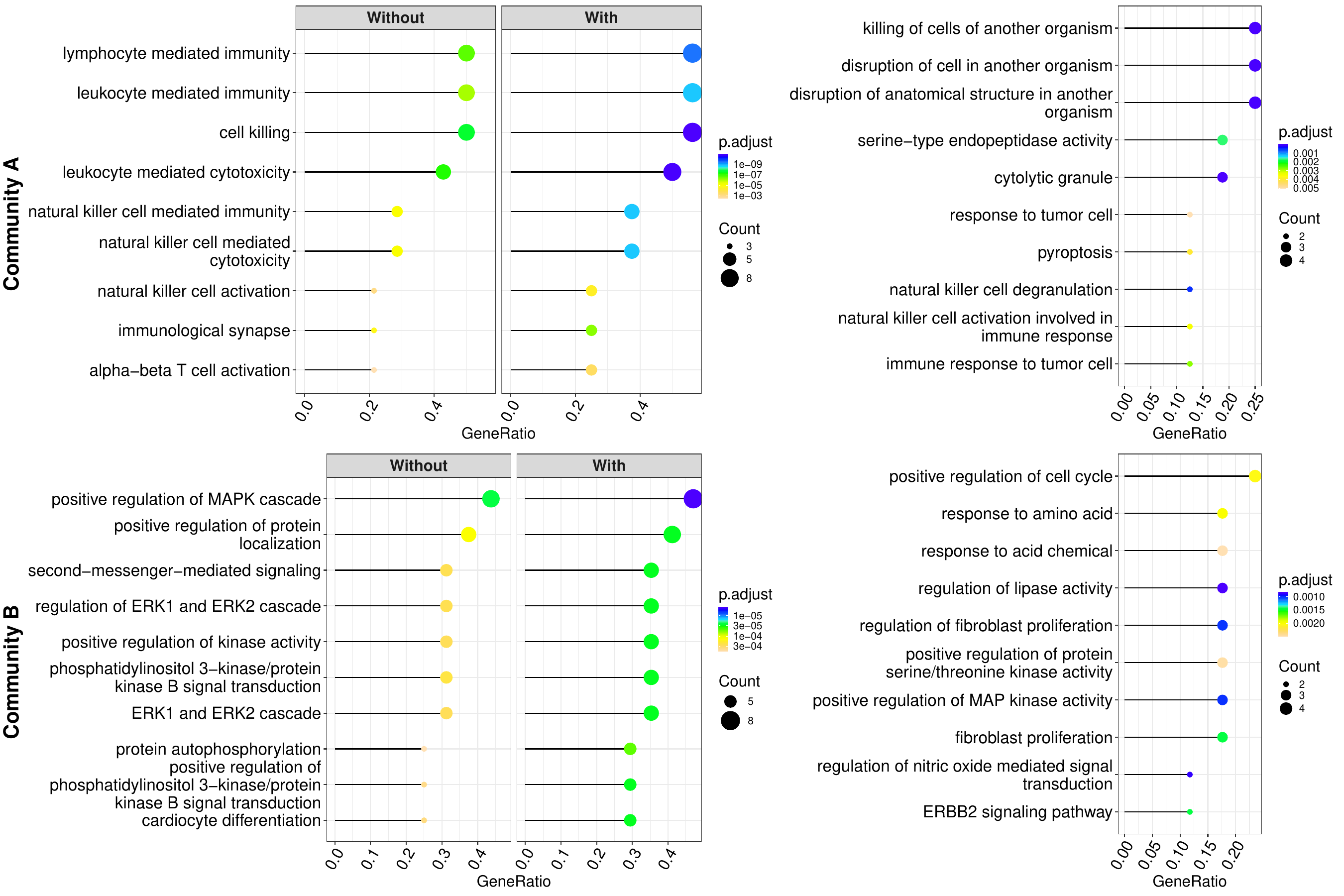}
\caption{Lollipop plots of GO enrichment analysis for Xenium PAC dataset. Left: Comparison results of the analysis without and with the inclusion of Target genes. Right: Newly detected significant GO terms with the Target genes included.
\label{fig:GO_PAC}}
\end{figure}

\newpage
\clearpage
\section{Exploratory analysis on spatial rotation}

We conduct an exploratory analysis to assess the robustness of our method with respect to spatial coordinate rotations. Specifically, four datasets are generated in which the original spatial coordinates are rotated by 0, 30, 60, and 90 degrees, respectively. Figure \ref{fig:venn} presents the Venn diagram of the identified SV gene sets under these different rotation angles. The substantial overlaps observed among the sets suggest that our method is relatively robust to moderate spatial rotations.

\begin{figure}[htb]
\centering
\includegraphics[width=3in]{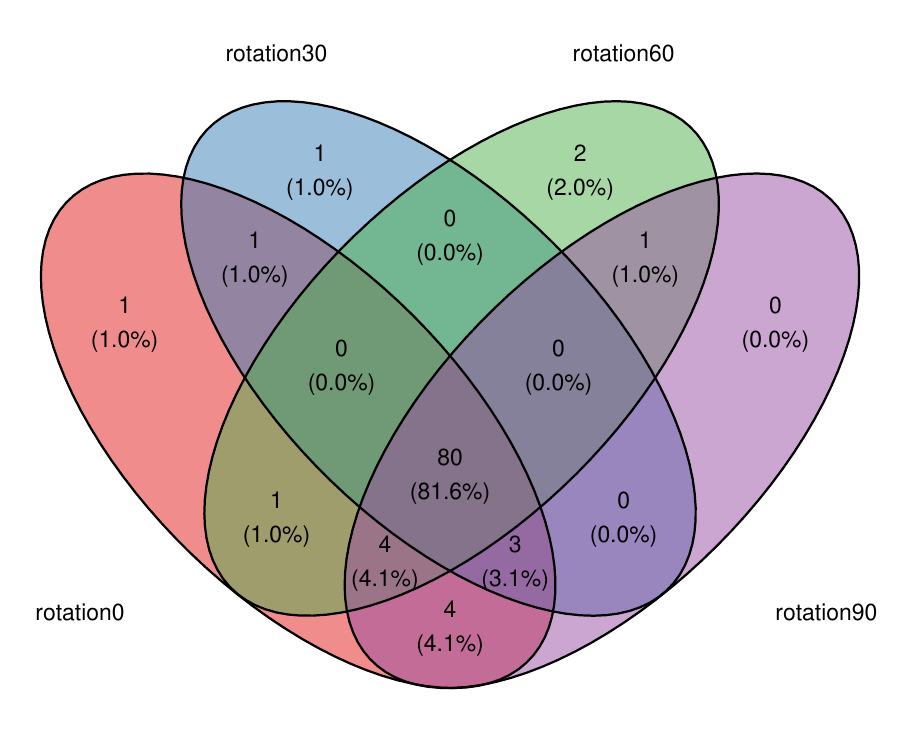}
\caption{Venn plot of the identified SV gene sets under different rotation angles. 
\label{fig:venn}}
\end{figure}
}

\end{document}